\pdfoutput=1
\documentclass[12pt]{article}
\usepackage{amsmath}
\usepackage{amsfonts}
\usepackage{amssymb}
\usepackage{amsthm}
\usepackage{mathtools}
\usepackage{enumerate}
\usepackage{natbib}
\usepackage{url}
\usepackage{bm}
\usepackage[hidelinks]{hyperref}
\usepackage{microtype}

\newcommand{\blind}{1}
\def\linespacing{1.5}

\newcommand{\E}{\mathbb E}
\newcommand{\R}{\mathbb R}
\newcommand{\Prob}{\mathbb P}
\newcommand{\tr}{\operatorname{tr}}
\newcommand{\vect}{\operatorname{vec}}
\newcommand{\rank}{\operatorname{rank}}
\newcommand{\Nor}{\mathcal N}
\newcommand{\MN}{\mathcal{MN}}
\newcommand{\dd}{\,\mathrm d}
\newcommand{\GL}{\mathrm{GL}}
\newcommand{\cG}{\mathcal G}
\newcommand{\cI}{\mathcal I}
\newcommand{\cF}{\mathcal F}
\newcommand{\cH}{\mathcal H}
\newcommand{\cP}{\mathcal P}
\newcommand{\cQ}{\mathcal Q}
\DeclareMathOperator*{\argmin}{arg\,min}

\newcommand{\by}{\bm{y}}
\newcommand{\bx}{\bm{x}}
\newcommand{\bz}{\bm{z}}
\newcommand{\bw}{\bm{w}}

\newcommand{\bb}{\bm{b}}
\newcommand{\bc}{\bm{c}}

\newcommand{\be}{\bm{e}}

\newcommand{\bmv}{\bm{m}}
\newcommand{\bs}{\bm{s}}
\newcommand{\bt}{\bm{t}}
\newcommand{\bu}{\bm{u}}
\newcommand{\bv}{\bm{v}}
\newcommand{\bF}{\bm{F}}
\newcommand{\bgamma}{\bm{\gamma}}
\newcommand{\bbeta}{\bm{\beta}}
\newcommand{\btheta}{\bm{\theta}}
\newcommand{\bmu}{\bm{\mu}}
\newcommand{\btau}{\bm{\tau}}
\newcommand{\bzeta}{\bm{\zeta}}
\newcommand{\boldeta}{\bm{\eta}}
\newcommand{\beps}{\bm{\varepsilon}}
\newcommand{\bpsi}{\bm{\psi}}
\newcommand{\bvarrho}{\bm{\varrho}}

\newcommand{\bTheta}{\bm{\Theta}}
\newcommand{\bXi}{\bm{\Xi}}

\newcommand{\bY}{\bm Y}
\newcommand{\bX}{\bm X}
\newcommand{\bZ}{\bm Z}
\newcommand{\bW}{\bm W}
\newcommand{\bB}{\bm B}
\newcommand{\bA}{\bm A}
\newcommand{\bC}{\bm C}

\newcommand{\bU}{\bm U}
\newcommand{\bV}{\bm V}
\newcommand{\bK}{\bm K}
\newcommand{\bS}{\bm S}
\newcommand{\bT}{\bm T}
\newcommand{\bM}{\bm M}
\newcommand{\bH}{\bm H}
\newcommand{\bI}{\bm I}
\newcommand{\bP}{\bm P}
\newcommand{\bE}{\bm E}
\newcommand{\bR}{\bm R}
\newcommand{\bDelta}{\bm \Delta}
\newcommand{\bSigma}{\bm \Sigma}
\newcommand{\bPhi}{\bm \Phi}
\newcommand{\bOmega}{\bm \Omega}
\newcommand{\bUpsilon}{\bm \Upsilon}
\newcommand{\htheta}{\hat{\btheta}}
\newcommand{\hUpsilon}{\hat{\bUpsilon}}
\newcommand{\bxi}{\bm \xi}
\newcommand{\bdelta}{\bm \delta}
\newcommand{\tY}{\tilde{\bY}}
\newcommand{\tZ}{\tilde{\bZ}}
\newcommand{\hDelta}{\hat{\bDelta}}
\newcommand{\hSigma}{\hat{\bSigma}}
\newcommand{\ECR}{E^{\mathrm{CR}}}
\newcommand{\bO}{\bm O}
\newcommand{\bL}{\bm L}
\newcommand{\bN}{\bm N}
\newcommand{\bJ}{\bm J}
\newcommand{\pv}{\mathrm{p}}

\theoremstyle{plain}
\newtheorem{theorem}{Theorem}
\newtheorem{proposition}{Proposition}
\newtheorem{lemma}{Lemma}
\newtheorem{corollary}{Corollary}
\theoremstyle{definition}
\newtheorem{assumption}{Assumption}
\newtheorem{definition}{Definition}
\theoremstyle{remark}
\newtheorem{remark}{Remark}

\begin{document}

\def\spacingset#1{\renewcommand{\baselinestretch}%
{#1}\small\normalsize} \spacingset{1}

\if1\blind
{
  \title{\bf Anytime-Valid Inference in Linear Models with Clustered Outcomes:\\ Exact Multivariate Tests and Cluster-Robust Confidence Sequences}
  \author{Michael Lindon$^*$}
  \maketitle
} \fi

\if0\blind
{
  \bigskip
  \bigskip
  \bigskip
  \begin{center}
    {\LARGE\bf Anytime-Valid Inference in Linear Models with Clustered Outcomes:\\ Exact Multivariate Tests and Cluster-Robust Confidence Sequences}
\end{center}
  \medskip
} \fi

\bigskip
\begin{abstract}
We extend anytime-valid inference for linear models from independent scalar outcomes to units that contribute a vector of dependent outcomes: multiple metrics per user in an online experiment, repeated measurements per patient in a clinical trial, or every member of a cluster in a cluster-randomized trial. We develop a parametric and a nonparametric treatment. For the Gaussian multivariate linear model with an unstructured error covariance, we derive a closed-form Bayes factor under the right-Haar prior and a matrix-normal mixture alternative and prove that it is an exact $e$-value at every sample size and a test martingale in the invariant filtration. To our knowledge, this Bayes factor is new in the generality considered here: a $d\times q$ coefficient matrix, nuisance regressors and an arbitrary fixed positive-definite prior precision under the right-Haar nuisance prior. It yields sequential Hotelling and MANOVA tests with exact confidence sequences. Under a common within-cluster design, the cluster-robust Wald statistic of a pooled regression is equal to Hotelling's $T^2$ statistic computed from the per-cluster least-squares summaries, giving an exact Gaussian benchmark with $G-1$ degrees of freedom. For general clustered population least-squares projections, the corresponding sandwich statistic is an asymptotic $e$-process, without Gaussian, common-design or linearity assumptions. Each treatment includes growth and first-order expected stopping-time results; under stronger regularity, direct comparisons of expected log evidence identify a fixed mixture precision that maximizes expected log growth through constant order within a specified family, for a specified frequentist alternative. Applied to randomized experiments with several outcomes, the robust procedure gives asymptotically anytime-valid joint tests, confidence ellipsoids and simultaneous confidence sequences for the vector of average treatment effects from ordinary least-squares estimates and cluster-robust standard errors.
\end{abstract}

\noindent%
\if1\blind
\vfill
\footnotesize
\qquad
\\
$^*$Netflix, Inc., 121 Albright Way, Los Gatos, CA 95032, USA. michael.s.lindon@gmail.com
\normalsize
\fi
\newpage
\spacingset{1}
\tableofcontents
\newpage
\spacingset{\linespacing}

\section{Introduction}
\label{sec:intro}

Ordinary least squares with cluster-robust standard errors is one of the most widely used procedures in empirical research. Whenever observations come in groups whose members share unobserved influences, the rows of a regression are not independent, and the textbook standard errors can understate the uncertainty, sometimes by a large factor \citep[Section~II]{cameron2015practitioner}. The remedy of \citet{liang1986longitudinal} keeps the OLS coefficients and replaces the covariance estimate by a sandwich built from the outer products of the estimated scores of each cluster. It is used in cluster-randomized trials, where treatment is assigned to a village, a school or a clinic and outcomes are measured on its members; in surveys with clustered sampling; in longitudinal clinical studies with repeated measurements per patient; and in online experiments, where a user contributes many sessions, page views or purchases and the unit of randomization is the user. \citet{cameron2015practitioner} and \citet{mackinnon2023cluster} survey the practice and its pitfalls, and \citet{abadie2023when} ask when clustering is warranted at all. This paper treats two regression models for such data. The first is the multivariate linear model $\by_g=\bB'\bx_g+\bDelta'\bz_g+\beps_g$: the $g$th observation is a vector of $q$ outcomes, all regressed on the same covariates $\bx_g$ and $\bz_g$, and the $j$th columns of $\bB$ and $\bDelta$ are the coefficients of outcome $j$. It is the classical model of multivariate analysis, in which hypotheses on a vector of coefficients are tested with Hotelling's $T^2$ and the MANOVA tests \citep{anderson2003introduction}. The second is the clustered regression $\by_g=\bW_g\bgamma+\beps_g$ on which cluster-robust standard errors are computed: every row of cluster $g$ carries its own regressors, the rows of $\bW_g$, and all rows share one coefficient vector $\bgamma$. The second contains the first, with $\bW_g=\bI_q\otimes(\bx_g',\bz_g')$ and $\bgamma$ the stacked columns of $\bB$ and $\bDelta$; and under a common within-cluster design, $\bW_g=\bW_0$ for every $g$, the second reduces to the intercept-only case of the first, applied to the per-cluster least-squares estimates (Section~\ref{sec:clustered}). In both models the inferential guarantee of the existing tests is fixed-sample: the test or interval is valid at one predetermined number of clusters, chosen before the data are seen.

In both models the data consist of independent clusters, each contributing an outcome \emph{vector} $\by_g$ whose components are dependent in an unknown way, together with the regressors of its rows. The data arrive one cluster at a time, so that the sample is a sequence of outcome vectors and the sample size is the number of clusters. An online experiment is watched from the day it starts, a trial is inspected at interim looks, a survey is fielded in waves, and in each case the analyst wants to examine the evidence as clusters accrue and to stop as soon as it is decisive, whether because a treatment is harmful, because it is clearly beneficial, or because the study has answered its question. A fixed-sample guarantee does not permit this. Sequential monitoring proceeds across clusters as a sequence of completed cluster outcome vectors is observed. Inference is updated after each completed cluster, with no intermediate updates within a cluster, and the within-cluster covariance is not restricted to a parametric correlation structure.

The tool is the $e$-value. An $e$-value for a null hypothesis is a nonnegative statistic whose expectation under the null is at most one \citep{shafer2021testing,vovk2021evalues,grunwald2024safe}; rejecting when it exceeds $1/\alpha$ is a level-$\alpha$ test by Markov's inequality. On an accumulating sample the object is an $e$-process: a nonnegative process whose value at any stopping time is an $e$-value, of which a test martingale is the leading example. By Ville's inequality the data may then be inspected continuously and collection stopped at any time, for any reason, and inverting a family of $e$-processes gives a confidence sequence with the same time-uniform coverage \citep{howard2021time}. The exact processes constructed in this paper are test martingales in their invariant filtrations, and hence are $e$-values at fixed times and at stopping times of those filtrations (Remark~\ref{rem:design}). Sequential monitoring is the use we emphasize, but $e$-values have further properties that $p$-values lack, and our exact $e$-values inherit them under the stated sampling and stopping conditions. $e$-values can be multiplied across independent studies and averaged across arbitrarily dependent ones without adjustment, which makes them the natural currency for evidence combination and for multiple testing, where the $e$-BH procedure controls the false discovery rate under any dependence \citep{wang2022false}, and they keep a type-I risk guarantee when the decision, including the significance level, is chosen after the data are seen \citep{grunwald2024beyond}. \citet{ramdas2023game} and \citet{ramdas2025hypothesis} give comprehensive treatments. The exact $e$-processes of this paper have a further reading. They are Bayes factors, computed with the right-Haar prior on the nuisance parameters and a proper prior on the coefficients of interest, and under suitable group-invariance and common-prior conditions such Bayes factors are $e$-values \citep{hendriksen2021optional,grunwald2024safe,ramdas2023game}. A Bayesian may therefore use them as they stand, as measures of evidence that enter posterior model probabilities, and a frequentist can use the same Bayes factors to obtain time-uniform type-I error control; the results below are contributions to the literature on Bayesian analysis as much as to that on safe anytime-valid inference.

For a scalar outcome per independent unit, \citet{lindon2026anytime} built the anytime-valid analogues of the linear-model tests. Their construction is the likelihood ratio of an invariantly sufficient statistic, obtained by integrating the Gaussian likelihood against the right-Haar prior on the nuisance parameters and a Gaussian mixture on the standardized coefficients of interest. The resulting $e$-process is a closed-form function of the OLS estimate and residual standard error, gives sequential analogues of the $t$-test, the $F$-test and the confidence ellipsoid, is optimal in the GROW sense of \citet{grunwald2024safe} (Section~\ref{sec:background}) against the mixture alternative, and remains valid asymptotically, through a heteroskedasticity-robust construction, when the Gaussian and homoskedastic assumptions are dropped and even when the linear model is only a population least-squares projection, which is what makes it applicable to average treatment effects in randomized experiments. That theory is restricted to independent scalar outcomes: a single number per independently sampled unit, whose error variance may vary with the unit's covariates but is itself a scalar. \citet{lindon2026anytime} identify dependent outcomes as an important open extension; this paper provides it. Our contributions are as follows.

\begin{enumerate}[(i)]
\item \emph{Parametric inference in Gaussian models} (Section~\ref{sec:exact}). For $\by_g=\bB'\bx_g+\bDelta'\bz_g+\beps_g$ with unknown unstructured error covariance $\bSigma$, Theorem~\ref{thm:main} gives the right-Haar Bayes factor for $H_0:\bDelta=0$ under an arbitrary fixed matrix-normal prior precision. It is an exact $e$-value and a test martingale in the invariant filtration; Corollary~\ref{cor:cs} gives its confidence sequence. The formula covers scalar regression, Hotelling's problem, the two-sample case of \citet{zoh2018powerful}, and MANOVA, with the precedents and distinctions detailed in Section~\ref{sec:special}; we did not find the general formula in earlier work. Validity is verified directly through the matrix angular central Gaussian law of \citet{chikuse1990matrix}, without an amenability assumption on $\GL(q)$. For the clustered regression $\by_g=\bW_g\bgamma+\beps_g$, in which $\by_g\in\R^{n_g}$, the rows of $\bW_g\in\R^{n_g\times k}$ are the regressors of the rows of the cluster, $\bgamma\in\R^k$ is shared by all rows, and the error covariance may differ across clusters, the fit is a single least-squares fit to the rows of all clusters (pooled OLS) with the CR1 covariance estimator. Proposition~\ref{prop:common} shows that under a common within-cluster design, $\bW_g=\bW_0$ for every $g$, the CR1 Wald statistic for a $q$-dimensional contrast is equal to Hotelling's $T^2$ statistic computed from the per-cluster OLS contrasts. This gives the exact Gaussian benchmark for $\ECR_G$ and fixes its finite-sample form at $G-1$ degrees of freedom.
\item \emph{Nonparametric asymptotic inference for clustered projections} (Section~\ref{sec:asymptotic}). Theorem~\ref{thm:asymptotic} proves that $\ECR_G$ is an asymptotic $e$-process for a contrast $\bR\bgamma^\star$ of a clustered population least-squares projection, with no Gaussian, common-design, equal-size or correct-specification assumption. Its inversion gives an asymptotic confidence sequence. The proof uses the multivariate strong approximation of \citet{einmahl1989extensions}, following the template of \citet{waudbysmith2024time} and \citet{lindon2026anytime}, and separately establishes asymptotic late-start type-I error control.
\item \emph{Growth and mixture selection} (Sections~\ref{sec:growth}--\ref{sec:log-growth} and~\ref{sec:cr-growth}--\ref{sec:cr-log-growth}). Each procedure has power one and first-order expected stopping-time limits under its stated assumptions. Under stronger regularity, direct comparisons of expected log evidence give the asymptotically optimal fixed mixture precision within a specified family and the limiting loss from another choice. These comparisons hold the frequentist alternative fixed.

\item \emph{Application to randomized experiments with several outcomes} (Section~\ref{sec:rct}). Under randomization, the population least-squares treatment coefficient is the average treatment effect for each outcome. Stacking the regressions therefore brings the vector of effects within the clustered projection theory, giving joint tests, confidence ellipsoids and simultaneous confidence sequences from ordinary OLS and CR1 estimates. The assumptions allow nonlinear conditional outcome means, non-Gaussian and heteroskedastic errors, and dependence among a unit's outcomes. The exact Gaussian counterpart and the circumstances in which its pooled covariance remains asymptotically appropriate are compared with the robust procedure.
\end{enumerate}
Throughout, the sequential tests and confidence sequences are closed-form expressions of the same OLS estimates and (cluster-robust) covariance estimates that a fixed-sample analysis computes, so they are no harder to implement than the fixed-sample procedures they replace; and the exact ones are Bayes factors, so they can be reported as Bayesian evidence as well.

\paragraph{Related work.} The conjugate matrix-normal--inverse-Wishart integration underlying Theorem~\ref{thm:main} builds on the posterior analysis of multivariate normal parameters and regression by \citet{geisser1963posterior}, \citet{tiao1964bayesian} and \citet{geisser1965bayesian}; \citet{brown1998multivariate,brown2002bayes} use the same matrix-normal prior tied to $\bSigma$ with a proper inverse-Wishart prior for variable selection and model averaging, and the matrix-variate $t$ distributions that arise are those of \citet{dickey1967matricvariate} and \citet{dawid1981some}. \citet{kiefer1965admissible} studied the proper Bayes character of Hotelling's $T^2$ and other fully invariant tests of classical multivariate analysis; in the intercept-only case our Bayes factor is an explicit increasing function of $T^2$ obtained with the right-Haar prior. Their equation~(4.6) also gives the determinant-ratio test recovered by our $g$-prior specialization, but their proper-prior Bayes factor has a different exponent (Section~\ref{sec:special}(d)). \citet{zoh2018powerful} derive the $d=1$ two-sample case of our Bayes factor and note its null distribution, but do not use it sequentially. \citet[Section~5.1]{dickey1971weighted} gives Bayes tests of linear hypotheses under normal--gamma priors for regression with known covariance shape and unknown scalar variance. Bayes factors for the multivariate normal linear model with different priors are developed by \citet{mulder2022bayesian} and implemented in \citet{mulder2021bfpack}; \citet{johnson2005bayes} introduced Bayes factors based on test statistics, of which the $d=1$ case of ours is an instance. The group-integral representation of the likelihood ratio of a maximal invariant is due to Stein and Wijsman \citep[Theorem~4, equation~(3)]{wijsman1967cross}; \citet[Section~4, equation~(19)]{andersson1982distributions} gives a quotient-measure formulation. The right-Haar Bayes-factor identity is given by \citet[Theorem~2.1]{berger1998bayes}; \citet[Section~5.5, Theorem~9]{hendriksen2021optional} establish optional-stopping validity under their stated group-model assumptions for stopping times adapted to the invariant filtration. On the sequential side, the scalar one-sample specialization belongs to the sequential $t$-test literature. \citet[Section~2]{rushton1950sequential} constructs a sequential test from ratios of noncentral $t$ densities at specified standardized effects; \citet{ramdas2023game} trace the right-Haar representation to \citet{cox1952sequential}. For a normal mean with unknown variance, \citet[Section~5, equations~(11)--(12)]{lai1976confidence} uses a flat improper mixture to recover the confidence sequence of \citet[equation~(21), p.~1402]{robbins1970statistical}; \citet{wang2025anytime} discuss this construction in modern terms. This differs from the proper Gaussian mixture in \eqref{eq:hotelling}, whose scalar one-sample test martingale and closed-form confidence sequence appear in \citet{lindon2026anytime} and \citet{wang2025anytime}. \citet[equation~(1.6)]{jackson1961sequential} construct a Wald-type probability-ratio test between two fixed values of the noncentrality, using confluent hypergeometric functions. \citet[Section~I.6]{hall1965relationship} derive invariant sequential $F$-tests with nuisance means and one- and two-sample $T^2$-tests, crediting earlier work by Stein and others. \citet[Sections~2--3]{khan1978confidence} uses a proper Gaussian mixture for a multivariate mean with known covariance and a scale-invariant flat mixture to obtain confidence sequences for the full coefficient vector in scalar-response regression with unknown variance. \citet[Section~4]{khan1978confidence} also constructed confidence sequences for a multivariate normal mean with unknown covariance by mixing the Jackson--Bradley likelihood ratio over an improper measure on the squared standardized effect. \citet[Section~2]{sinha1984invariant} construct invariant likelihood-ratio mixtures for multivariate linear regression with unknown covariance, including inference on a block of coefficients with the remaining coefficients treated as nuisances. Their framework allows general mixing measures; their evaluated confidence sequences use a flat improper mixture and a separately calibrated crossing bound. Section~\ref{sec:special} distinguishes that construction from the explicit proper Gaussian-mixture Bayes factor studied here. \citet{perezortiz2024estatistics} developed $e$-statistics for invariant problems under amenability, including proper mixtures in their Corollary~9 and a lower-triangular multivariate-mean example in Section~4.1; they discuss Hotelling's problem under the nonamenable general linear group in Section~5.1; \citet{chugg2023time} construct confidence spheres for vector means under sub-Gaussian, other tail, boundedness and moment assumptions, including empirical and semi-empirical constructions; these are distinct from the unknown-covariance Gaussian regression Bayes factor studied here. The strong-approximation formulation of asymptotic confidence sequences used here follows \citet{waudbysmith2024time}, building on \citet{howard2021time} and \citet{robbins1970statistical}; \citet{lindon2026anytime} give the heteroskedasticity-robust version for OLS. \citet{bibaut2024} establish asymptotic type-I error calibration for delayed-start sequential tests using strong invariance principles, including settings with dependent observations; their analysis is a further precedent for the late-start guarantee proved here. Earlier scalar delayed-start boundary-crossing limits appear in \citet[Theorem~3]{lai1976confidence}. For fixed-sample cluster-robust inference see \citet{cameron2015practitioner} and \citet{mackinnon2023cluster}; the fixed-$G$ $F(q,G-q)$ Wald reference of \citet{bester2011inference} is recalled in Section~\ref{sec:clustered}, where Proposition~\ref{prop:common} makes it exact under Gaussian cluster contrasts with a common design. \citet{ibragimov2010tstatistic} instead construct a $t$-statistic from per-cluster estimates. General background on $e$-values is in \citet{grunwald2024safe}, \citet{ramdas2023game} and \citet{ramdas2025hypothesis}.

\paragraph{Outline.} Section~\ref{sec:background} gives the definitions. Section~\ref{sec:exact} develops the Gaussian treatment, including the common-design clustered benchmark and stopping-time analysis. Section~\ref{sec:asymptotic} develops the nonparametric cluster-robust treatment, with its validity and stopping-time guarantees. Section~\ref{sec:rct} applies the procedures to randomized experiments with several outcomes. Section~\ref{sec:discussion} concludes. Proofs are grouped by treatment in the appendix, followed by further remarks on optimality and invariance.

\paragraph{Notation.} Vectors are column vectors and use bold lowercase letters, such as $\by_g$, $\bx_g$ and $\btheta$; matrices use bold uppercase letters, such as $\bY_G$, $\bB$ and $\bSigma$. Scalar quantities and individual entries are nonbold, retaining conventional uppercase symbols such as $G$, $E_G$ and $Q_G$. Thus $y_{gj}$ is the $j$th entry of $\by_g$, while the $g$th row of $\bY_G$ is $\by_g'$. This convention applies to random quantities and their realizations alike. Dimensions are specified at first use; scalar specializations are identified explicitly. Descriptive subscripts retain their plain form.

\section{Anytime-Valid Inference}
\label{sec:background}
Throughout, $G$ counts clusters, $n_g$ the rows of cluster $g$, and $n=\sum_{g\leq G}n_g$ the total number of rows. In Sections~\ref{sec:exact} and~\ref{sec:rct} a cluster is a unit with an outcome vector of length $q$, so $n=qG$; when $q=1$ every row is its own cluster and $n=G$ is the sample size of \citet{lindon2026anytime}. This section collects the definitions in the form in which the results below deliver them, following the conventions of \citet{ramdas2023game}; \citet{grunwald2024safe} and \citet{ramdas2025hypothesis} give full accounts.

\paragraph{$e$-values, test martingales and $e$-processes.} Let $\cP$ be a family of distributions for the sequence of clusters and $\cH_0\subset\cP$ a null hypothesis. An \emph{$e$-value} for $\cH_0$ is a nonnegative random variable $E$ with $\E_P[E]\leq1$ for every $P\in\cH_0$; \citet{ramdas2023game} call the random variable an $e$-variable and its realized value the $e$-value, a distinction we do not maintain. The test that rejects when $E\geq\alpha^{-1}$ has level $\alpha$ by Markov's inequality. Let $(\cF_G)_{G\geq0}$ be a filtration and $E_G$ be $\cF_G$-measurable. The process $(E_G)_{G\geq0}$ is a \emph{test supermartingale} for $\cH_0$ if it is a nonnegative supermartingale with $E_0=1$ under every $P\in\cH_0$, and a \emph{test martingale} if it is a nonnegative martingale with $E_0=1$ under every $P\in\cH_0$ \citep{ramdas2023game}; Theorem~\ref{thm:main} delivers a test martingale. Ville's inequality \citep{ville1939etude} gives $\Prob_P[\exists G:E_G\geq\alpha^{-1}]\leq\alpha$ for every $P\in\cH_0$, so the sequential test that rejects the first time $E_G\geq\alpha^{-1}$ has level $\alpha$ whatever the stopping rule; equivalently, $\Prob_P[E_\tau\geq\alpha^{-1}]\leq\alpha$ for every stopping time $\tau$ \citep[Lemma~3]{howard2021time}, and by optional stopping $\E_P[E_\tau]\leq1$, so that $E_\tau$ is an $e$-value. A nonnegative process $(E_G)_{G\geq0}$ is an \emph{$e$-process} for $\cH_0$ if for every $P\in\cH_0$ there is a test martingale $(M_G^P)$ for $P$ with $E_G\leq M_G^P$ for all $G$; equivalently, $\E_P[E_\tau]\leq1$ for every stopping time $\tau$ and every $P\in\cH_0$ \citep{ramdas2023admissible,ramdas2023game}. Every test supermartingale for $\cH_0$ is an $e$-process, an $e$-process need not be a test supermartingale for $\cH_0$, and it has the same time-uniform guarantee; it is the object that Theorem~\ref{thm:asymptotic} delivers in asymptotic form. In either case $\pv_G=\min(1,1/E_G)$ satisfies $\Prob_P[\pv_\tau\leq\alpha]\leq\alpha$ for every stopping time $\tau$, equivalently $\Prob_P[\exists G:\pv_G\leq\alpha]\leq\alpha$, for every $P\in\cH_0$, and is called an \emph{anytime-valid $p$-value} \citep{johari2022always,howard2021time,ramdas2023game}; so is its running minimum $\min_{k\leq G}\pv_k$.

\paragraph{Confidence sequences.} A sequence of sets $(C_G^\alpha)_{G\geq1}$ is a \emph{$(1-\alpha)$ confidence sequence} for a parameter $\btheta$ if $\Prob_P[\btheta\in C_G^\alpha\ \forall G]\geq1-\alpha$ for every $P\in\cP$ \citep{howard2021time}. Given a family $\{E_G^{\btheta_0}\}$ of $e$-processes, one for each null $\btheta=\btheta_0$, the sets $C_G^\alpha=\{\btheta_0:E_G^{\btheta_0}<\alpha^{-1}\}$ form one, by Ville's inequality applied at the true value. Corollary~\ref{cor:cs} uses this construction, and Corollary~\ref{cor:asympcs} gives its asymptotic counterpart. The $d=1$ case of Corollary~\ref{cor:cs} and the sets of Corollary~\ref{cor:asympcs} are ellipsoids in closed form; the general matrix-parameter set in Corollary~\ref{cor:cs} need not be an ellipsoid.

\paragraph{Power and growth.} The sequential test has \emph{power one} against $P\in\cP\setminus\cH_0$ if $\Prob_P[\tau<\infty]=1$ for $\tau=\inf\{G:E_G\geq\alpha^{-1}\}$ \citep{robbins1970statistical}; $G^{-1}\log E_G\to c>0$ almost surely suffices, and Proposition~\ref{prop:growth} establishes it for every fixed alternative. The finer measure of the power of an $e$-value against an alternative $Q$ is its expected logarithmic growth $\E_Q[\log E]$: an $e$-value that maximizes it is \emph{growth-rate optimal} (GRO) for $Q$, and against a composite alternative $\cQ$ the worst-case criteria GROW and REGROW maximize $\inf_{Q\in\cQ}\E_Q[\log E]$ and $\inf_{Q\in\cQ}\E_Q[\log E-\log E^Q]$, where $E^Q$ is the GRO $e$-value for $Q$ \citep{grunwald2024safe,ramdas2023game}. For a simple null and a simple alternative the likelihood ratio is GRO, at every fixed $G$ and at stopping times that are almost surely finite under both laws. Remark~\ref{rem:optimality} states what the exact $e$-process of this paper achieves in this sense.

\paragraph{Asymptotic versions.} Section~\ref{sec:asymptotic} drops the Gaussian model, and its guarantees are asymptotic in the sense introduced by \citet{waudbysmith2024time}, whose arXiv v7, Definition~C.2 is the $e$-process form used by \citet{lindon2026anytime} and here.
\begin{definition}[Asymptotic $e$-process; \citealp{waudbysmith2024time}]\label{def:asymp}
A nonnegative process $(E_G)_{G\geq1}$ is an \emph{asymptotic $e$-process} for $\cH_0\subset\cP$ if there is an $e$-process $(E_G^S)_{G\geq0}$ for $\cH_0$, possibly defined on an enriched probability space, such that $\log E_G/\log E_G^S\to1$ almost surely under every $P\in\cP$.
\end{definition}
The logarithmic equivalence is required under every $P\in\cP$, not only under the null. It is an approximation criterion, not by itself a theorem transferring a fixed threshold-crossing probability. In particular it does not control a triangular sequence of procedures whose tuning changes with the monitoring start. Theorem~\ref{thm:asymptotic} proves the required late-start guarantee separately. The comparison process is defined on a joint extension carrying a coupled Gaussian sequence, with its martingale property asserted in that sequence's invariant filtration, not the filtration containing both the raw data and the Gaussian variables. The coupling law can depend on $P$; the proof below uses the same Hotelling-form comparison statistic for every $P$. We call a family of sets obtained by inverting asymptotic $e$-processes, one for each null value of the parameter, an \emph{asymptotic confidence sequence}; for a scalar parameter, \citet[arXiv v9, Definition~2.1]{waudbysmith2024time} define the notion through intervals whose lower and upper widths are almost-sure relative approximations to those of a nonasymptotic confidence sequence with the same center, and \citet[arXiv v9, Definition~B.6]{waudbysmith2024time} and \citet{lindon2026anytime} give multivariate definitions through the normalized measure of the symmetric difference of the two sets. Remark~\ref{rem:guarantee} discusses what these definitions guarantee and what they do not; the operational statement is the lower bound on the $\liminf$ of simultaneous coverage in \eqref{eq:late_start_coverage}, with the tuning and monitoring regime of Theorem~\ref{thm:asymptotic}, in the sense of \citet[arXiv v9, Definition~2.7]{waudbysmith2024time}.

\section{Parametric Inference in Gaussian Models}
\label{sec:exact}

This section works under a Gaussian linear model with a common error covariance. We first construct exact tests and confidence sequences, then analyze their growth, stopping times and mixture precision under fixed alternatives. The latter results are asymptotic, as $\alpha\downarrow0$ or as $G\to\infty$; they are separate from the finite-sample null guarantees.

\subsection{Model, hypotheses and priors}
\label{sec:model}
For $g=1,2,\ldots$ let $\by_g\in\R^q$ be the outcome vector of unit $g$, with nuisance regressors $\bx_g\in\R^p$ and regressors of interest $\bz_g\in\R^d$, treated as fixed (Remark~\ref{rem:design}). The model is
\begin{equation}\label{eq:model}
\by_g=\bB'\bx_g+\bDelta'\bz_g+\beps_g,\qquad \beps_g\overset{iid}{\sim}\Nor_q(0,\bSigma),
\end{equation}
with $\bB\in\R^{p\times q}$, $\bDelta\in\R^{d\times q}$ and $\bSigma$ positive definite, all unknown. Stacking the first $G$ observations as rows, $\bY_G\in\R^{G\times q}$, $\bX_G\in\R^{G\times p}$, $\bZ_G\in\R^{G\times d}$,
\[
\bY_G=\bX_G\bB+\bZ_G\bDelta+\bE_G,\qquad \bE_G\sim\MN_{G\times q}(0,\bI_G,\bSigma),
\]
where $\MN_{G\times q}(\bM,\bm A,\bm C)$ denotes the matrix-normal distribution with mean $\bM$, row covariance $\bm A$ and column covariance $\bm C$, that is, $\vect(\bE_G)\sim\Nor(0,\bSigma\otimes\bI_G)$. In the notation $\by_g\sim\Nor(\bX_g\bbeta+\bZ_g\bdelta,\bSigma)$ this is
\[
\bX_g=\bx_g'\otimes\bI_q,\quad \bbeta=\vect(\bB')\in\R^{pq},\qquad \bZ_g=\bz_g'\otimes\bI_q,\quad \bdelta=\vect(\bDelta')\in\R^{dq},
\]
that is, every component of $\by_g$ is regressed on the same regressors with its own coefficients. The exact results of this section require this structure; Appendix~\ref{app:mean-structures} explains why and treats other mean structures. The case $q=1$ is the model of \citet{lindon2026anytime}; the case $p=0$, $d=1$, $z_g\equiv1$ is the one-sample multivariate mean problem of Hotelling with $\bmu=\bDelta'$. The sample size $G$ counts \emph{vectors}, so dependence within a vector is unrestricted; in the convention of Section~\ref{sec:background} each unit is a cluster with $q$ rows and $n=qG$.

$\bP_{\bA}$ is the orthogonal projector onto the column space of $\bA$, $\bW_G=[\bX_G,\bZ_G]$, and
\[
\tY_G=(\bI_G-\bP_{\bX_G})\bY_G,\qquad \tZ_G=(\bI_G-\bP_{\bX_G})\bZ_G .
\]
For a positive definite $d\times d$ matrix $\bPhi$ define the two $q\times q$ matrices
\begin{equation}\label{eq:S01}
\bS_{0,G}=\tY_G'\tY_G,\qquad
\bS_{1,G}=\tY_G'\tY_G-\tY_G'\tZ_G\big(\bPhi+\tZ_G'\tZ_G\big)^{-1}\tZ_G'\tY_G .
\end{equation}
$\bS_{0,G}$ is the residual sum-of-squares-and-cross-products (SSCP) matrix of the null regression of $\bY_G$ on $\bX_G$. $\bS_{1,G}$ is the penalized residual SSCP matrix $\min_{\bDelta}\{(\tY_G-\tZ_G\bDelta)'(\tY_G-\tZ_G\bDelta)+\bDelta'\bPhi\bDelta\}$, the minimum being in the Loewner order (Lemma~\ref{lem:square}). When $\bW_G$ has full column rank and $\nu_G\coloneqq G-p-d\geq q$, write
\[
\hDelta_G=(\tZ_G'\tZ_G)^{-1}\tZ_G'\bY_G,\qquad
\hSigma_G=\frac{\bY_G'(\bI_G-\bP_{\bW_G})\bY_G}{\nu_G},\qquad
\bT_G=\hDelta_G\hSigma_G^{-1}\hDelta_G'\in\R^{d\times d},
\]
\[
\bM_G=\tZ_G'\tZ_G-\tZ_G'\tZ_G\big(\bPhi+\tZ_G'\tZ_G\big)^{-1}\tZ_G'\tZ_G=\big(\bPhi^{-1}+(\tZ_G'\tZ_G)^{-1}\big)^{-1}.
\]
$\hDelta_G$ and $\hSigma_G$ are the OLS coefficient matrix and the usual unbiased estimator of $\bSigma$; $\bT_G$ generalizes $\hat{\bxi}_G\hat{\bxi}_G'$ of \citet{lindon2026anytime} ($q=1$) and $T^2/G$ of Hotelling's one-sample problem ($d=1$).

\paragraph{Group and priors.}
\label{sec:group}
The null is $H_0:\bDelta=0$; a nonzero null $\bDelta=\bDelta_0$ is tested by replacing $\bY_G$ with $\bY_G-\bZ_G\bDelta_0$ (Corollary~\ref{cor:cs}). The model is invariant under the group $\cG$ of maps
\[
\bY_G\mapsto\bY_G\bC'+\bX_G\bA,\qquad \bC\in\GL(q),\ \bA\in\R^{p\times q},
\]
which act on the parameters by $(\bB,\bSigma,\bDelta)\mapsto(\bB\bC'+\bA,\ \bC\bSigma\bC',\ \bDelta\bC')$, preserve $H_0$, and act transitively on the nuisance parameter $(\bB,\bSigma)$. The group law is $(\bC_2,\bA_2)(\bC_1,\bA_1)=(\bC_2\bC_1,\ \bA_1\bC_2'+\bA_2)$, its right-Haar measure is $|\det\bC|^{-q}\dd\bC\dd\bA$, and the image of that measure under $(\bC,\bA)\mapsto(\bB,\bSigma)=(\bA,\bC\bC')$ is the right-Haar prior
\begin{equation}\label{eq:prior}
\pi(\bB,\bSigma)\ \propto\ |\bSigma|^{-(q+1)/2},
\end{equation}
which is $1/\sigma^2$ when $q=1$ and the invariant measure on the cone of positive definite matrices when $p=0$. (The left-Haar measure is $|\det\bC|^{-q-p}\dd\bC\dd\bA$ and induces $|\bSigma|^{-(p+q+1)/2}$; see Remark~\ref{rem:prior}.) Under the alternative the coefficients of interest receive the matrix-normal mixture
\begin{equation}\label{eq:mixture}
\bDelta\mid\bSigma\ \sim\ \MN_{d\times q}\big(0,\ \bPhi^{-1},\ \bSigma\big),
\qquad\text{i.e.}\quad \bdelta\mid\bSigma\sim\Nor_{dq}\big(0,\ \bPhi^{-1}\otimes\bSigma\big),
\end{equation}
with density $(2\pi)^{-dq/2}|\bPhi|^{q/2}|\bSigma|^{-d/2}\exp\{-\tfrac12\tr(\bSigma^{-1}\bDelta'\bPhi\bDelta)\}$. Its scale is tied to $\bSigma$, so that it is a fixed $\MN(0,\bPhi^{-1},\bI_q)$ prior on the standardized coefficients $\bDelta\bSigma^{-1/2}$; it is $\bdelta\sim\Nor(0,\sigma^2\bPhi^{-1})$ when $q=1$ and $\bmu\sim\Nor(0,\bSigma/\phi)$ in the one-sample mean problem ($\bPhi=\phi$).

Write $P_{\bB,\bSigma}$ for the law of the sequence $(\by_g)_{g\geq1}$ under \eqref{eq:model} with $\bDelta=0$, and $Q_{\bB,\bSigma}$ for its law when $\bDelta$ is drawn once from \eqref{eq:mixture}. A measurable $f:\R^{G\times q}\to\R$ is \emph{invariant} if $f(\bY\bC'+\bX_G\bA)=f(\bY)$ for all $(\bC,\bA)$, and $\cI_G$ denotes the $\sigma$-field generated by the invariant functions of $\bY_G$.

The role of the group is the classical one \citep{hall1965relationship,perezortiz2024estatistics}. It fixes $H_0$ and acts transitively on the nuisance parameter, so the law of every invariant statistic is free of $(\bB,\bSigma)$, and on $\cI_G$ the composite null $\{P_{\bB,\bSigma}\}$ and the composite alternative $\{Q_{\bB,\bSigma}\}$ are each a single distribution. The likelihood ratio of a maximal invariant is then a test martingale, and Wijsman's representation theorem identifies it with the ratio of the marginal likelihoods under the right-Haar prior \citep{wijsman1967cross,berger1998bayes}, which is why \eqref{eq:prior} is the prior to use; \citet{hall1965relationship} show that the reduction may pass through a sufficient statistic first, and \citet{berger1998bayes} and \citet[Sections~5.2--5.5]{hendriksen2021optional} draw the consequences for Bayes factors and stopping times of the invariant filtration. \citet{perezortiz2024estatistics} prove that this likelihood ratio is GROW optimal when the group is amenable, which is the route by which \citet{lindon2026anytime} obtained optimality for $q=1$, where the group is the affine semidirect product of $\GL(1)$ and $\R^p$. Here $\cG$ contains $\GL(q)$, which is not amenable for $q\geq2$ and is the example those authors single out. Validity does not need amenability: Appendix~\ref{app:evalue} verifies Theorem~\ref{thm:main}(ii) directly from the distribution of the orientation of the whitened residual matrix. Optimality has to be restated (Remark~\ref{rem:optimality}).

\subsection{Bayes factor and exact inference}
\label{sec:mainresults}

\begin{theorem}\label{thm:main}
Suppose $\bX_{p+q}$ has full column rank, $\bPhi\succ0$ is fixed across sample sizes, and let $G\geq p+q$. Set $E_G=1$ before $p+q$.
\begin{enumerate}
\item[\textup{(i)}] \textup{(Bayes factor.)} With the priors \eqref{eq:prior} and \eqref{eq:mixture}, the Bayes factor of $H_1$ against $H_0$ is, almost surely,
\begin{equation}\label{eq:E}
E_G=\left(\frac{|\bPhi|}{|\bPhi+\tZ_G'\tZ_G|}\right)^{q/2}\left(\frac{|\bS_{0,G}|}{|\bS_{1,G}|}\right)^{(G-p)/2}.
\end{equation}
If moreover $\bW_G$ has full column rank and $\nu_G=G-p-d\geq q$, then
\begin{equation}\label{eq:Eols}
E_G=\sqrt{\frac{|\bPhi|^{\,q}}{|\bPhi+\tZ_G'\tZ_G|^{\,q}}}\;
\frac{\left|\bI_d+\bM_G\bT_G/\nu_G\right|^{-\frac{\nu_G+d}{2}}}{\left|\bI_d+\tZ_G'\tZ_G\,\bT_G/\nu_G\right|^{-\frac{\nu_G+d}{2}}}.
\end{equation}
\item[\textup{(ii)}] \textup{(Exact $e$-value.)} For every $(\bB,\bSigma)$ and every bounded invariant $f$,
\[
\E_{Q_{\bB,\bSigma}}\big[f(\bY_G)\big]=\E_{P_{\bB,\bSigma}}\big[f(\bY_G)\,E_G\big].
\]
That is, $E_G$ is the density of $Q_{\bB,\bSigma}$ with respect to $P_{\bB,\bSigma}$ on $\cI_G$. In particular $\E_{P_{\bB,\bSigma}}[E_G]=1$ for every $(\bB,\bSigma)$.
\item[\textup{(iii)}] \textup{(Test martingale.)} $\cI_G\subset\cI_{G+1}$, $E_{p+q}=1$, and $(E_G)_{G\geq p+q}$ is a martingale with respect to $(\cI_G)_{G\geq p+q}$ under every $P_{\bB,\bSigma}$. Hence $P_{\bB,\bSigma}\big[\exists G: E_G\geq\alpha^{-1}\big]\leq\alpha$ for all $(\bB,\bSigma)$.
\end{enumerate}
\end{theorem}
The proof is in Appendix~\ref{app:exact}. Part (i) integrates out $\bB$, $\bDelta$ and $\bSigma$ in turn; part (ii) is a direct verification: after whitening by $\bSigma$, the orientation of the residual matrix is uniform on the Stiefel manifold under the null and has the matrix angular central Gaussian law of \citet[Theorem~2.2]{chikuse1990matrix} under the alternative, and the ratio of the two densities is $E_G$; part (iii) follows from (ii) applied at consecutive times. Expression \eqref{eq:Eols} is the multivariate analogue of the ratio of two multivariate $t$ densities in \citet{lindon2026anytime}: it can be written as a ratio of matrix-variate $t$ density functions with row-scale matrices $\bPhi^{-1}+(\tZ_G'\tZ_G)^{-1}$ and $(\tZ_G'\tZ_G)^{-1}$. These are sampling densities of the symmetrically studentized coefficient matrix at the representative $\bSigma=\bI_q$; the Gram matrix is the nuisance-free invariant, and the oriented matrix need not have that same law at arbitrary $\bSigma$ (Appendix~\ref{app:macg}).

\begin{corollary}[Confidence sequences]\label{cor:cs}
For $\bDelta_0\in\R^{d\times q}$ let $E_G^{\bDelta_0}$ be \eqref{eq:E} evaluated at $\bY_G-\bZ_G\bDelta_0$; this leaves $\hSigma_G$ unchanged and replaces $\hDelta_G$ by $\hDelta_G-\bDelta_0$. Then $(E_G^{\bDelta_0})$ is a test martingale for $H_0:\bDelta=\bDelta_0$, $\pv_G^{\bDelta_0}=\min(1,1/E_G^{\bDelta_0})$ is an anytime-valid $p$-value, and $C_G^\alpha=\{\bDelta_0:E_G^{\bDelta_0}\leq\alpha^{-1}\}$ satisfies $P[\bDelta\in C_G^\alpha\ \forall G\geq p+q]\geq1-\alpha$ for every $(\bB,\bDelta,\bSigma)$. For $d=1$, whenever $\bW_G$ has full column rank and $\nu_G\geq q$, with $\bPhi=\phi>0$, $\hat\bdelta_G=\hDelta_G'\in\R^q$ and $a_G=\phi/(\phi+\|\tZ_G\|_2^2)$, it is the ellipsoid
\begin{equation}\label{eq:ellipsoid}
C_G^\alpha=\left\{\bdelta_0\in\R^q:\ \|\tZ_G\|_2^2\,(\hat\bdelta_G-\bdelta_0)'\hSigma_G^{-1}(\hat\bdelta_G-\bdelta_0)\ \leq\ \nu_G\,\frac{1-u_G}{\max(0,\ u_G-a_G)}\right\},
\end{equation}
where $u_G=(\alpha^2 a_G^{\,q})^{1/(\nu_G+1)}$, with the convention that the right side is $+\infty$ (so $C_G^\alpha=\R^q$) when $u_G\leq a_G$. The fixed-$G$ Hotelling region replaces the right side by $\frac{q\nu_G}{\nu_G-q+1}F_{1-\alpha}(q,\nu_G-q+1)$.
\end{corollary}
Linear hypotheses reduce to Theorem~\ref{thm:main} by reparametrization. For $\bC_0\in\R^{q\times q_0}$ of full column rank, $\bY_G\bC_0$ satisfies \eqref{eq:model} with $(q,\bDelta,\bSigma)$ replaced by $(q_0,\bDelta\bC_0,\bC_0'\bSigma\bC_0)$, which tests $\bDelta\bC_0=0$, a hypothesis on $q_0$ linear combinations of the outcomes. For row contrasts, let $\bL\in\R^{r_0\times d}$ have full row rank and let the columns of $\bN$ span $\ker(\bL)$. Write $\bDelta=\bL'(\bL\bL')^{-1}\bTheta+\bN\bXi$, so $\bTheta=\bL\bDelta$. Testing $\bL\bDelta=0$ therefore uses tested design $\bZ_G\bL'(\bL\bL')^{-1}$ and nuisance design $[\bX_G,\bZ_G\bN]$, with a fixed prior precision for $\bTheta$, as in the contrasts construction of \citet{lindon2026anytime}. In particular a hypothesis on a subset of the rows of $\bDelta$ is tested by moving the other rows into the nuisance block.

\begin{remark}[Starting time and the two forms]\label{rem:start}
Form \eqref{eq:E} needs only $G\geq p+q$, whereas the OLS form \eqref{eq:Eols} needs $G\geq p+d+q$ so that $\hSigma_G$ is invertible. For $p+q\leq G<p+d+q$ the residual matrix $\nu_G\hSigma_G$ is singular but $\bS_{0,G}$ and $\bS_{1,G}$ are not, and $E_G$ is already nontrivial for $G>p+q$. For earlier samples use residual dimension $N_G=G-\rank(\bX_G)$ rather than $G-p$. The rank assumption in Theorem~\ref{thm:main} implies $N_G\leq q$ for $G\leq p+q$, and the full-row-rank residual matrices form a single orbit, so the invariant $\sigma$-field is trivial up to null sets and we set $E_G=1$. If the nuisance design first has full rank at a later design-determined $G_0\geq p+q$, \eqref{eq:E} still has mean one and is a martingale for $G\geq G_0$; Ville's bound applies with initial expectation one. One can extend it by $E_G=1$ and the trivial filtration before $G_0$. With $q=1$ this shows that the process of \citet{lindon2026anytime} is defined from $G=p+1$ through $E_G=(|\bPhi|/|\bPhi+\tZ_G'\tZ_G|)^{1/2}\,(\tY_G'\tY_G/\bS_{1,G})^{(G-p)/2}$, before $s_G^2$ exists.
\end{remark}

\begin{remark}[Dependence on the prior for $\bSigma$]\label{rem:prior}
Provided $G-p+2s-q-1>q-1$ so both marginal integrals are finite, replacing \eqref{eq:prior} by $|\bSigma|^{-s}$ changes the exponent in \eqref{eq:E} from $(G-p)/2$ to $(G-p+2s-q-1)/2$. Because $\bS_{0,G}\succeq\bS_{1,G}$, the ratio $|\bS_{0,G}|/|\bS_{1,G}|$ is at least one and exceeds one with positive probability when $\tZ_G\neq0$. For such a nontrivial design the resulting statistic therefore has null expectation strictly below one for $s<(q+1)/2$ (valid but conservative) and strictly above one for $s>(q+1)/2$ (not an $e$-value). In particular the left-Haar, or Jeffreys-rule, prior $|\bSigma|^{-(p+q+1)/2}$ gives the exponent $G/2$ and is not an $e$-value when $p\geq1$. Within this determinant-power family, only $s=(q+1)/2$ gives the equality in Theorem~\ref{thm:main}(ii) for a nontrivial design. In the one-sample case (b) with $|\bSigma|^{-s}$ the exponent $G/2$ in \eqref{eq:hotelling} becomes $(G+2s-q-1)/2$; for instance $s=q/2$ yields $(G-1)/2$.
\end{remark}

\begin{remark}[Design, filtration and stopping]\label{rem:design}
The exact validity statements treat the regressors as fixed, or condition on a random regressor sequence independent of the errors. This covers covariates and non-adaptive randomized assignment, but not designs in which $(\bx_{G+1},\bz_{G+1})$ depend on past outcomes. The time-uniform bound in Theorem~\ref{thm:main}(iii) is a statement about the supremum of the path, so it controls the type-I error of ``reject the first time $E_G\geq\alpha^{-1}$'' however monitoring is stopped, even by a rule that looks at the raw data. The $e$-process property $\E_P[E_\tau]\leq1$, which the further uses of $e$-values listed in Section~\ref{sec:intro} require at a data-dependent sample size, holds for stopping times of the invariant filtration $(\cI_G)$, for example any rule based on the path $(E_k)_{k\leq G}$, on $(\bT_k)_{k\leq G}$ and on the design. This is the situation that \citet{ramdas2023game} describe for the group-invariant test martingales of \citet{perezortiz2024estatistics}: the process is a test martingale, and hence an $e$-process, in the coarsened filtration rather than in the filtration of the raw data, and in exchange it has the growth property of Remark~\ref{rem:optimality}.
\end{remark}

\begin{remark}[Optimality]\label{rem:optimality}
By Theorem~\ref{thm:main}(ii), $E_G$ is the likelihood ratio of a simple null against a simple alternative on $\cI_G$, so by Gibbs' inequality it maximizes $\E_{Q_{\bB,\bSigma}}[\log T]$, simultaneously for all $(\bB,\bSigma)$, among $e$-variables $T$ that are invariant under $\cG$: the evidence does not depend on the coordinate system chosen for the outcome vector. In the terminology of Section~\ref{sec:background} it is GRO for every $Q_{\bB,\bSigma}$, hence both GROW and REGROW, relative to the invariant $e$-variables, at every $G$ and, by the stopped likelihood-ratio argument, at stopping times of $(\cI_G)$ that are almost surely finite under both laws. When $q\geq2$ and the residual alternative covariance $\bK_G$ is not a scalar multiple of the identity, it is not GROW among all $e$-variables: the lower-triangular subgroup gives an $e$-variable with a strictly larger expected log-growth, the gain being the same for every $Q_{\bB,\bSigma}$ and bounded in $G$. That $e$-variable is the product of the scalar $e$-values of \citet{lindon2026anytime} for the regressions of each outcome on the preceding ones. Degenerate cases such as $\tZ_G=0$ have no strict gain. Appendix~\ref{app:grow} states this precisely, gives the closed form of the triangular Bayes factor, and gives the reasons for working with $E_G$.
\end{remark}

\subsection{Special cases and relation to prior work}
\label{sec:special}
The general Bayes factor \eqref{eq:E} specializes to the following known formulas and test statistics. Its $e$-value, martingale and confidence-sequence properties give explicit proper Gaussian-mixture procedures in the multivariate cases (b)--(d), subject to the fixed-prior condition in (d). The scalar sequential results, including $q=1$ in (b), are due to earlier work.

\begin{corollary}[Special cases]\label{cor:special}
\leavevmode
\begin{enumerate}
\item[\textup{(a)}] $q=1$. Then $\hSigma_G=s_G^2$ is the squared residual standard error, $\hDelta_G=\hat{\bdelta}_G\in\R^d$, $\bT_G=\hat{\bxi}_G\hat{\bxi}_G'$ with $\hat{\bxi}_G=\hat{\bdelta}_G/s_G$, $|\bI_d+\bM_G\hat{\bxi}_G\hat{\bxi}_G'/\nu_G|=1+\hat{\bxi}_G'\bM_G\hat{\bxi}_G/\nu_G$, and \eqref{eq:Eols} is exactly the $E_G$ of \citet{lindon2026anytime}.
\item[\textup{(b)}] $p=0$, $d=1$, $z_g\equiv1$, $\bPhi=\phi$ (the one-sample multivariate mean, $\by_g\sim\Nor_q(\bmu,\bSigma)$, $H_0:\bmu=0$). Then $\tZ_G'\tZ_G=G$, $\bM_G=\phi G/(\phi+G)$, $\nu_G=G-1$, $G\bT_G=T^2=G\bar \by_G'\bS^{-1}\bar \by_G$ is Hotelling's statistic with $\bS$ the sample covariance matrix, and, with $a=\phi/(\phi+G)$,
\begin{equation}\label{eq:hotelling}
E_G=a^{q/2}\left(\frac{1+T^2/(G-1)}{1+a\,T^2/(G-1)}\right)^{G/2},
\end{equation}
an increasing function of $T^2$ alone.
\item[\textup{(c)}] $d=1$ (one regressor of interest, e.g.\ a treatment indicator, $q$ outcomes, $p$ covariates). With $\bPhi=\phi>0$, $a_G=\phi/(\phi+\|\tZ_G\|_2^2)$, $\hat\bdelta_G=\hDelta_G'\in\R^q$ and the regression-adjusted Hotelling statistic $T_G^2=\|\tZ_G\|_2^2\,\hat\bdelta_G'\hSigma_G^{-1}\hat\bdelta_G$,
\begin{equation}\label{eq:d1}
E_G=a_G^{\,q/2}\left(\frac{1+T_G^2/\nu_G}{1+a_G\,T_G^2/\nu_G}\right)^{\frac{\nu_G+1}{2}},
\qquad \frac{\nu_G-q+1}{q\,\nu_G}\,T_G^2\ \overset{H_0}{\sim}\ F(q,\ \nu_G-q+1).
\end{equation}
\item[\textup{(d)}] $\bPhi=\frac{g}{G}\tZ_G'\tZ_G$ (the $g$-prior). Let $\ell_1,\ldots,\ell_s$, $s=\min(d,q)$, be the eigenvalues of $\bH_G(\nu_G\hSigma_G)^{-1}$, padded with zeros if necessary, where $\bH_G=\hDelta_G'\tZ_G'\tZ_G\hDelta_G$ is the hypothesis SSCP matrix. Then
\begin{equation}\label{eq:gprior}
E_G=\left(\frac{g}{g+G}\right)^{\frac{dq}{2}}\prod_{j=1}^{s}\left(\frac{1+\ell_j}{1+\frac{g}{g+G}\,\ell_j}\right)^{\frac{G-p}{2}},
\end{equation}
a function of the classical MANOVA roots; $\prod_j(1+\ell_j)^{-1}$ is Wilks' $\Lambda$. For each fixed $G$ this is an exact $e$-value by Theorem~\ref{thm:main}(ii). As a sequence in $G$, Theorem~\ref{thm:main}(iii) guarantees a test martingale when $\tZ_G'\tZ_G/G$ does not depend on $G$, as in \textup{(b)}. Otherwise the prior changes with $G$, so the fixed-time $e$-value statement alone gives no sequential guarantee. An asymptotic claim requires additional design-stabilization assumptions, as in the scalar construction of \citet{lindon2026anytime}.
\end{enumerate}
\end{corollary}
The verifications are in Appendix~\ref{app:special}. The relevant sequential precedent is \citet{sinha1984invariant}: their equations~(2.3)--(2.5) mix maximal-invariant likelihood ratios over a general measure, using Wijsman's representation on the full general linear group; their Section~2.2 also allows nuisance regression coefficients. A probability mixing measure gives a mean-one martingale within that framework. Their evaluated formulas instead use a flat improper mixing measure, giving the determinant statistic in equation~(2.6) and confidence regions calibrated through their inequality~(1.2). In particular, their equations~(2.10)--(2.11) already give sequential Hotelling ellipsoids. The present formula evaluates a proper Gaussian mixture, yielding a normalized Bayes factor with the threshold $1/\alpha$ and the confidence sets in Corollary~\ref{cor:cs}. Thus the general multivariate invariant-mixture method and the existence of multivariate regression confidence sequences are prior work; the formula and its resulting procedures are the contributions considered here. We now place each case in the literature.

\paragraph{(a) Scalar outcomes.} Case (a) is Theorem~4.1 of \citet{lindon2026anytime}, whose $t$- and $F$-type $e$-processes and confidence sequences are therefore the $q=1$ instance of the present results. The group is then the affine semidirect product of $\GL(1)$ and $\R^p$, which is amenable, and the GROW optimality proved there through \citet{perezortiz2024estatistics} is the one property that does not extend to $q\geq2$ (Appendix~\ref{app:grow}).

\paragraph{(b) Hotelling's one-sample problem.} Case (b) is the sequential analogue of Hotelling's $T^2$ test \citep[p.~364, equation~(4)]{hotelling1931generalization}. That Hotelling's test is Bayes for suitable proper priors was shown by \citet{kiefer1965admissible}, who used it to prove admissibility; \eqref{eq:hotelling} is the explicit Bayes factor for the right-Haar prior and the mixture $\bmu\mid\bSigma\sim\Nor(0,\bSigma/\phi)$, and its exact $e$-value property is the statement that its expectation under the $T^2(q,G-1)$ null law, $\frac{G-q}{q(G-1)}T^2\sim F(q,G-q)$, equals one, whatever $\bSigma$ is. \citet[equations~(1.3)--(1.6)]{jackson1961sequential} construct Wald-type sequential probability-ratio tests between two fixed noncentralities, with the $T^2$ ratio expressed through confluent hypergeometric functions. This is a different alternative specification from the Gaussian mixture used here. \citet[equations~(4.13)--(4.16)]{khan1978confidence} integrate that ratio against Lebesgue measure on the squared standardized effect, obtaining a Gauss hypergeometric statistic with a separately calibrated crossing bound. That improper-mixture confidence sequence also precedes the present proper Gaussian-mixture construction. The present \eqref{eq:hotelling} uses the specified Gaussian mixture over the alternative, has power one against every $\bmu\neq0$ (Proposition~\ref{prop:growth}), and inverts to the closed-form confidence sequence \eqref{eq:ellipsoid}. \citet[Section~4.1]{perezortiz2024estatistics} display the lower-triangular invariant likelihood ratio for a fixed standardized mean as an integral over a Wishart--Cholesky law; the triangular statistic of Appendix~\ref{app:grow}, whose closed form differs from \eqref{eq:hotelling}, has for $q\geq2$ a strictly larger expected log-growth than \eqref{eq:hotelling} under the Gaussian mixture. The case $q=1$ of (b) is the proper Gaussian-mixture sequential $t$-test of \citet{lindon2026anytime} and \citet{wang2025anytime}; the latter mention sequential Hotelling tests as future work.

\paragraph{(c) One regressor of interest.} For $d=1$ in the two-sample design, $z_g$ a group indicator and $\bx_g\equiv1$, \eqref{eq:d1} is exactly Lemma~1 of \citet{zoh2018powerful}, with $\|\tZ_G\|_2^2=G_1G_2/G$ for group sizes $G_1+G_2=G$, $\phi=\tau_0$ and $\nu_G=G-2$; they use the same priors \eqref{eq:prior} and $\bdelta\mid\bSigma\sim\Nor(0,\bSigma/\tau_0)$, note the null $F$ law, and use the Bayes factor at a fixed $G$, after a random projection to lower dimension, as a test in high dimensions, without $e$-values or sequential use. Case (c) extends this to any regression adjustment $\bx_g$, which is the form needed for treatment effects on several outcomes in Section~\ref{sec:rct}. It is a Bayes factor of the test-statistic type of \citet{johnson2005bayes}: under $Q_{\bB,\bSigma}$, $a_GT_G^2$ has the null law of $T_G^2$, so \eqref{eq:d1} is the ratio of the densities of $T_G^2$ under the two hypotheses, which gives a one-line proof of Theorem~\ref{thm:main}(ii) when $d=1$.

\paragraph{(d) MANOVA.} Under the $g$-prior the Bayes factor depends on the data only through the eigenvalues of $\bH_G(\nu_G\hSigma_G)^{-1}$, padded with zeros if necessary, the roots on which every classical MANOVA test is based: Wilks' $\Lambda$ \citep{wilks1932certain}, the Lawley--Hotelling trace, Pillai's trace and Roy's largest root \citep[Section~8.6]{anderson2003introduction}. Since $\prod_j(1+c\,\ell_j)=|\nu_G\hSigma_G+c\bH_G|/|\nu_G\hSigma_G|$, \eqref{eq:gprior} is an increasing function of the determinant ratio $|\nu_G\hSigma_G+\bH_G|/|\nu_G\hSigma_G+\frac{g}{g+G}\bH_G|$. This is the statistic in equation~(4.6) of \citet{kiefer1965admissible}, with their $1-k$ equal to $g/(g+G)$; the reduction of their Lemma~3.1 on which it rests is clarified in \citet{kiefer1972correction}. Their canonical coefficient matrix has $q$ rows and $d$ columns in our notation. Integrating the joint density in their equation~(4.5) over the parameters of their proper prior gives, up to a factor independent of the observations, this ratio to the power $d/2$ (their $r/2$), whereas \eqref{eq:gprior} raises it to $(G-p)/2$. Thus the fixed-sample determinant-ratio test is a precedent, but the Bayes factor differs. They do not study its sequential use. The matrix-normal--inverse-Wishart algebra behind Theorem~\ref{thm:main}(i) is classical (Section~\ref{sec:intro}, related work). After projecting out the flat-prior nuisance regressors, the determinant ratio in \eqref{eq:E} is the formal specialization of the inverse-Wishart marginal-likelihood algebra in \citet[equations~(7)--(8)]{brown2002bayes} to a zero scale matrix and zero Wishart degrees of freedom, that is, $\delta=1-q$ in their shape parametrization. For $q\geq2$ this lies outside the closure of the proper inverse-Wishart range, so it is an algebraic specialization rather than a limit through proper priors, and the exact null expectation and martingale property do not follow from it. The default Bayes factors of \citet{mulder2022bayesian}, implemented in \citet{mulder2021bfpack}, use a fractional prior and are not of the form \eqref{eq:E}. The two features of \eqref{eq:E} that make its sequential use work are the right-Haar prior, which gives the exact null expectation (Remark~\ref{rem:prior}), and the mixture tied to $\bSigma$, which makes $\bU$ matrix normal under the alternative in the proof of (ii).

\subsection{Growth, power and first-order expected stopping time}
\label{sec:growth}
The next result describes performance under fixed alternatives in the Gaussian model. The prior precision is fixed as the evidence threshold increases.

\begin{proposition}[Growth rate, power one and expected stopping time]\label{prop:growth}
Under \eqref{eq:model}, let the design be deterministic or independent of the errors, and fix $\bPhi\succ0$ and $\bDelta$. Suppose $\bW_G'\bW_G/G\to\bOmega_{\bW}\succ0$ almost surely. Define $E_G$ by \eqref{eq:E} when $G\geq p+q$, $\rank(\bX_G)=p$ and $\bS_{0,G},\bS_{1,G}\succ0$, and set $E_G=1$ otherwise, as in Remark~\ref{rem:start}. Write $\bOmega_{\tZ}=\lim_G\tZ_G'\tZ_G/G$ and $\bSigma_0=\bSigma+\bDelta'\bOmega_{\tZ}\bDelta$. Then
\[
\begin{aligned}
\frac{\log E_G}{G}\ \overset{a.s.}{\longrightarrow}\ r_{\bDelta}
&\coloneqq\frac12\log\big|\bI_q+\bSigma^{-1}\bDelta'\bOmega_{\tZ}\bDelta\big|\\
&=\frac12\log\big|\bI_d+\bOmega_{\tZ}\,\bDelta\bSigma^{-1}\bDelta'\big|
=\frac12\log\frac{|\bSigma_0|}{|\bSigma|},
\end{aligned}
\]
which is positive if and only if $\bDelta\neq0$ and equals $\frac12\log(1+\bxi'\bOmega_{\tZ}\bxi)$ when $q=1$, where $\bxi=\bdelta/\sigma$ is the standardized coefficient vector, $\bDelta=\bdelta\in\R^d$ and $\bSigma=\sigma^2$ in that case. Hence the sequential test that rejects when $E_G\geq\alpha^{-1}$ has power one against every $\bDelta\neq0$.

For a fixed $\bDelta\neq0$, define
\[
\tau_\alpha=\inf\{G\geq p+q:E_G\geq\alpha^{-1}\},\qquad 0<\alpha<1.
\]
Keeping $\bPhi\succ0$ fixed as $\alpha\downarrow0$,
\[
\frac{\tau_\alpha}{\log(1/\alpha)}\ \overset{a.s.}{\longrightarrow}\ \frac1{r_{\bDelta}}.
\]
If the design sequence is deterministic, then $\E_{\bB,\bDelta,\bSigma}[\tau_\alpha]<\infty$ for every $\alpha\in(0,1)$, the convergence also holds in $L^1$, and in particular
\begin{equation}\label{eq:expected_stopping}
\E_{\bB,\bDelta,\bSigma}[\tau_\alpha]\sim
\frac{\log(1/\alpha)}{r_{\bDelta}}
\qquad(\alpha\downarrow0).
\end{equation}
The expectation is under the fixed alternative in \eqref{eq:model}. The same conclusions hold unconditionally when the design rows $\bw_g=(\bx_g',\bz_g')'$ are i.i.d., bounded and independent of the errors, with $\bOmega_{\bW}=\E[\bw_g\bw_g']\succ0$.
\end{proposition}

\paragraph{Information interpretation of the growth rate.}
For a fixed positive-definite value of $\bOmega_{\bW}$, choose a law for a design row $\bw_g=(\bx_g',\bz_g')'$ with $\E[\bw_g\bw_g']=\bOmega_{\bW}$. Let $Q^\star$ be the joint law of $(\bw_g,\by_g)$ under \eqref{eq:model} with this design law, and let $\cP_0$ be the collection of joint laws under the same model with $\bDelta=0$, any coefficient matrix $\widetilde{\bB}\in\R^{p\times q}$, any covariance $\bV\succ0$, and the same design law. The \emph{reverse information projection} of $Q^\star$ onto $\cP_0$ is the null law $P^\star$ minimizing $D_{\mathrm{KL}}(Q^\star\,\|\,P)$ over $P\in\cP_0$.

To compute it, write
\[
\bOmega_{\bX}=\E[\bx_g\bx_g'],\qquad
\bm\Pi=\bOmega_{\bX}^{-1}\E[\bx_g\bz_g'],\qquad
\widetilde\bz_g=\bz_g-\bm\Pi'\bx_g.
\]
Since the design law is common to $Q^\star$ and $\cP_0$, the divergence from $Q^\star$ to the law $P\in\cP_0$ with parameters $(\widetilde{\bB},\bV)$ is the expected conditional divergence given the design: with $\varphi_{\bV}$ the density of $\Nor_q(0,\bV)$,
\[
D_{\mathrm{KL}}(Q^\star\,\|\,P)
=\E\!\left[\log\frac{\varphi_{\bSigma}(\by_g-\bB'\bx_g-\bDelta'\bz_g)}
{\varphi_{\bV}(\by_g-\widetilde{\bB}'\bx_g)}\right].
\]
Appendix~\ref{app:growthproof} minimizes this expression over $(\widetilde{\bB},\bV)$. The minimum is attained at the coefficient matrix $\bB_0=\bB+\bm\Pi\bDelta$ and the covariance
\[
\bSigma_0=\E[\boldeta_g\boldeta_g']
=\bSigma+\bDelta'\bOmega_{\tZ}\bDelta,
\qquad \boldeta_g=\by_g-\bB_0'\bx_g=\bDelta'\widetilde\bz_g+\beps_g,
\]
so $P^\star$ is the Gaussian null with parameters $(\bB_0,\bSigma_0)$, and $D_{\mathrm{KL}}(Q^\star\,\|\,P^\star)=\frac12\log(|\bSigma_0|/|\bSigma|)$. Here $\bSigma_0$ is the covariance parameter of $P^\star$ and the uncentered second moment of its residual under $Q^\star$; it need not equal $\operatorname{Cov}(\boldeta_g)$. When $p=0$, take $\widetilde\bz_g=\bz_g$ and $\boldeta_g=\by_g$.

Consequently, the determinant identity in Proposition~\ref{prop:growth} gives
\begin{equation}\label{eq:growth-kl}
r_{\bDelta}
=D_{\mathrm{KL}}(Q^\star\,\|\,P^\star)
=\min_{P\in\cP_0}D_{\mathrm{KL}}(Q^\star\,\|\,P).
\end{equation}
This identity depends on the chosen design law only through $\bOmega_{\bW}$ and does not require independence across clusters. For i.i.d. design rows with finite second moments, choosing their actual distribution makes $Q^\star$ the true single-cluster joint law, so $r_{\bDelta}$ is the KL divergence per cluster to its closest Gaussian null. For deterministic designs, $Q^\star$ is a reference law with the limiting design moments; Appendix~\ref{app:growthproof} also gives the limiting average conditional KL identity.

\begin{remark}[Expected stopping time]\label{rem:expected-stopping}
For other random designs independent of the errors, the expectation statements hold conditionally on almost every design sequence satisfying the Gram-matrix limit, as in Remark~\ref{rem:design}. That almost-sure limit alone does not justify averaging the expected stopping times over the random design; the bounded i.i.d. assumption above supplies the required tail control.

For power-one mixture tests in a one-parameter exponential family, the analogous first-order expected-sample-size result is classical \citep[Lemma~5, p.~1272]{pollak1975approximations}. A modern counterpart for portfolio-based $e$-processes is given by \citet[Theorem~3.3]{waudbysmith2026instance} under sublinear regret and moment assumptions. Section~\ref{sec:log-growth} compares expected log evidence directly across fixed mixture precisions.
\end{remark}

\subsection{Mixture selection for a fixed alternative}
\label{sec:log-growth}

In a Bayesian analysis, $\bPhi$ specifies the prior on the standardized coefficients in \eqref{eq:mixture} and should reflect the analyst's prior beliefs. The following comparison addresses a frequentist choice of mixture: hold the true parameters $(\bB,\bDelta,\bSigma)$ fixed and choose the mixture to maximize expected log evidence under repeated sampling from that alternative. Thus the expectation is under a fixed parameter value, rather than the mixture alternative $Q_{\bB,\bSigma}$ in Remark~\ref{rem:optimality}.

All fixed precisions share the first-order growth rate of Proposition~\ref{prop:growth}. Their differences in expected log evidence nevertheless have nonzero limits, which can be obtained directly from the ratio of Bayes factors.

\begin{proposition}[Relative expected log growth]\label{prop:log-growth}
Under \eqref{eq:model}, suppose that $\bw_g=(\bx_g',\bz_g')'$ are i.i.d., bounded and independent of the Gaussian errors, with $\bOmega_{\bW}=\E[\bw_g\bw_g']\succ0$. Fix $\bDelta\neq0$, $\bSigma\succ0$ and a positive-definite shape $\bPhi_0$. Let $E_G^{(\phi)}$ be the statistic with precision $\bPhi=\phi\bPhi_0$, using the convention at inactive samples in Proposition~\ref{prop:growth}. Define
\begin{equation}\label{eq:optimal-precision}
\mathcal I_0=\tr(\bSigma^{-1}\bDelta'\bPhi_0\bDelta)>0,
\qquad \phi_*=\frac{dq}{\mathcal I_0}.
\end{equation}
For every fixed $\phi>0$, as $G\to\infty$,
\begin{equation}\label{eq:precision-cost}
\E[\log E_G^{(\phi_*)}]-\E[\log E_G^{(\phi)}]
=\frac{dq}{2}\left\{\frac{\phi}{\phi_*}-1-\log\frac{\phi}{\phi_*}\right\}+o(1).
\end{equation}
The expectation averages over design and outcomes under the fixed true parameters.
\end{proposition}

Since $t-1-\log t\geq0$, with equality only at $t=1$, $\phi_*$ maximizes expected log growth through constant order among fixed precisions with the specified shape, for the specified alternative. Appendix~\ref{app:log-growth} proves the comparison by cancelling common terms in the exact Bayes factors.

For the one-sample Gaussian mean in Corollary~\ref{cor:special}(b), $\phi_*=q/c$, where $c=\bmu'\bSigma^{-1}\bmu$. In frequentist use, a specified planning alternative therefore supplies a mixture precision to fix before monitoring. Estimating the alternative from the same accumulating data and repeatedly changing the mixture is not covered by Theorem~\ref{thm:main}(iii).

\subsection{Clustered regression under a common design}
\label{sec:clustered}

Model \eqref{eq:model} has cluster-level regressors and outcome-specific coefficients. The regression that motivates cluster-robust inference is different: for clusters $g=1,\ldots,G$ with outcome vectors $\by_g\in\R^{n_0}$,
\begin{equation}\label{eq:clustered}
\by_g=\bW_g\bgamma+\beps_g,\qquad \beps_g\sim\Nor_{n_0}(0,\bSigma_{\varepsilon})\ \text{independently across clusters},
\end{equation}
where the rows of $\bW_g\in\R^{n_0\times k}$ vary \emph{within} the cluster, $\bgamma\in\R^k$ is shared by all rows, and $\bSigma_\varepsilon$ is a common within-cluster error covariance. The exact reduction below requires a common design as well; Section~\ref{sec:asymptotic} allows unequal cluster sizes, heterogeneous designs and covariances, and non-Gaussian outcomes. The shared-coefficient structure is generally not preserved by arbitrary transformations of the outcome vector, so Theorem~\ref{thm:main} does not apply directly; Appendix~\ref{app:mean-structures} gives the details and other reductions. Under a common design, compression to per-cluster OLS contrasts gives the following exact reduction and fixes the degrees of freedom. The target is $\btheta=\bR\bgamma$ for a fixed $\bR\in\R^{q\times k}$ of full row rank, and $H_0:\btheta=\btheta_0$.

\paragraph{Cluster-robust quantities.} Let $\bA_G=\sum_g \bW_g'\bW_g$, let $\hat\bgamma_G=\bA_G^{-1}\sum_g\bW_g'\by_g$ be pooled OLS, and let $\hat \bs_g=\bW_g'(\by_g-\bW_g\hat\bgamma_G)$ be the estimated cluster scores. The cluster-robust variance estimator of \citet{liang1986longitudinal} with the finite-sample factor $G/(G-1)$ \citep{cameron2015practitioner}, commonly labelled CR1, and the Wald statistic are
\begin{equation}\label{eq:CR1}
\hat \bV_G=\frac{G}{G-1}\,\bA_G^{-1}\Big(\sum_{g=1}^G\hat \bs_g\hat \bs_g'\Big)\bA_G^{-1},
\qquad
Q_G(\btheta_0)=(\bR\hat\bgamma_G-\btheta_0)'\big(\bR\hat \bV_G\bR'\big)^{-1}(\bR\hat\bgamma_G-\btheta_0).
\end{equation}
For $\phi>0$ define
\begin{equation}\label{eq:ECR}
\ECR_G(\btheta_0)=\left(\frac{\phi}{\phi+G}\right)^{q/2}
\left(\frac{1+Q_G(\btheta_0)/(G-1)}{1+\frac{\phi}{\phi+G}\,Q_G(\btheta_0)/(G-1)}\right)^{G/2},\qquad G>q,
\end{equation}
and $\ECR_G(\btheta_0)=1$ for $G\leq q$. More generally, at any sample where $\bA_G$ or $\bR\hat \bV_G\bR'$ is singular, set $\ECR_G(\btheta_0)=1$ for every $\btheta_0$ and take its confidence set to be $\R^q$. These conventions affect at most finitely many samples almost surely under the assumptions of Section~\ref{sec:asymptotic}. This is \eqref{eq:hotelling} with $T^2$ replaced by the cluster-robust Wald statistic. For $q=1$, a two-sided $t_{G-1}$ reference is common \citep[Section~4.2]{mackinnon2023cluster}. For joint tests, \citet{bester2011inference}, as reviewed by \citet[Section~4.3.3]{mackinnon2023cluster}, obtain the scaled $F(q,G-q)$ reference asymptotically with fixed $G$ under their conditions of equal cluster sizes and limited within-cluster dependence. In the common-design Gaussian benchmark of Proposition~\ref{prop:common}, this reference is exact: $\frac{G-q}{q(G-1)}Q_G(\btheta_0)\sim F(q,G-q)$; $\ECR_G$ is an exact or asymptotic sequential replacement under the corresponding assumptions below.

\begin{proposition}[Common within-cluster design]\label{prop:common}
Suppose $\bW_g=\bW_0$ for every $g$, with $\rank(\bW_0)=k$. Let $\hat\bgamma_g=(\bW_0'\bW_0)^{-1}\bW_0'\by_g$ be the OLS fit of cluster $g$ alone and let $\bb_g=\bR\hat\bgamma_g\in\R^q$.
\begin{enumerate}
\item[\textup{(a)}] \textup{(Algebra; no distributional assumption.)} $\hat\bgamma_G=\frac1G\sum_g\hat\bgamma_g$, $\hat \bs_g=\bW_0'\bW_0(\hat\bgamma_g-\hat\bgamma_G)$, and $\hat \bV_G=\frac1G\bS_{\hat\bgamma}$, where $\bS_{\hat\bgamma}$ is the sample covariance matrix of $\hat\bgamma_1,\ldots,\hat\bgamma_G$ with divisor $G-1$. Hence, with $\bar \bb$ and $\bS_b=\bR\bS_{\hat\bgamma}\bR'$ the sample mean and covariance of $\bb_1,\ldots,\bb_G$,
\[
Q_G(\btheta_0)=G\,(\bar \bb-\btheta_0)'\bS_b^{-1}(\bar \bb-\btheta_0),
\]
whenever $\bS_b$ is invertible, which is the algebraic form of Hotelling's $T^2$ statistic for the sample $\bb_1,\ldots,\bb_G$.
\item[\textup{(b)}] \textup{(Exactness.)} Under \eqref{eq:clustered} with positive definite compressed covariance $\bSigma_R$, or more generally whenever $\bb_1,\bb_2,\ldots$ are i.i.d.\ $\Nor_q(\btheta,\bSigma_R)$ with $\bSigma_R\succ0$, the vectors $\bb_g-\btheta_0$ follow model \eqref{eq:model} with $p=0$, $d=1$, $z_g\equiv1$, $\bDelta'=\btheta-\btheta_0$ and covariance $\bSigma_R=\bR(\bW_0'\bW_0)^{-1}\bW_0'\bSigma_\varepsilon \bW_0(\bW_0'\bW_0)^{-1}\bR'$. With $\bPhi=\phi$, the Bayes factor $E_G$ of Theorem~\ref{thm:main} equals $\ECR_G(\btheta_0)$. Consequently $\ECR_G(\btheta_0)$ is an exact $e$-value at every $G$ and a test martingale in the invariant filtration of the compressed vectors under $H_0$, and
\[
\Big\{\btheta_0:\ Q_G(\btheta_0)\leq (G-1)\,\frac{1-u_G}{\max(0,u_G-a_G)}\Big\},\qquad a_G=\frac{\phi}{\phi+G},\quad u_G=\big(\alpha^2a_G^{\,q}\big)^{1/G},
\]
is an exact $(1-\alpha)$ confidence sequence for $\btheta$.
\end{enumerate}
\end{proposition}
The proof is in Appendix~\ref{app:clustered}. Thus, under a common design, cluster-robust inference \emph{is} one-sample multivariate inference on the compressed vectors $\bb_g$: the CR1 contrast covariance is their sample covariance divided by $G$, and estimating the $k$ regression coefficients costs a single degree of freedom, the centering $\hat\bgamma_g-\hat\bgamma_G$. The number $k$ of within-cluster regressors sets the dimension of $\hat\bgamma_g$ and does not enter $\nu=G-1$. For $q=1$ the compression is that of \citet{ibragimov2010tstatistic}, who apply a $t_{G-1}$ test to per-cluster estimates; Proposition~\ref{prop:common}(a) shows that with a common design this is the same as the CR1 $t$-statistic of the pooled regression, and (b) gives its anytime-valid version.

\begin{remark}[The rule for the degrees of freedom]\label{rem:dfrule}
In every exact case above, $\nu$ equals the number of clusters minus the rank of the \emph{cluster-level} design, the matrix with one row per cluster: the constant in Proposition~\ref{prop:common} ($G-1$, whatever $k$ is), $[\bx_g',\bz_g']$ in Theorem~\ref{thm:main} ($G-p-d$, each outcome vector being a cluster), and, in \citet{lindon2026anytime}, where every observation is its own cluster, all $p+d$ regressors ($G-p-d$). Degrees of freedom count directions of $\R^G$ consumed by estimation, not parameters estimated: estimating the mean of a $k$-vector costs one, whatever $k$ is, and the dimension $q$ of the contrast does not enter.

The $G-q$ in the fixed-sample $F(q,G-q)$ reference is a denominator degree of freedom of Hotelling's $T^2$, not the residual divisor in \eqref{eq:ECR}. Replacing $Q_G/(G-1)$ by $Q_G/(G-q)$ makes the statistic larger and invalidates its exact $e$-value property for $q>1$. Appendix~\ref{app:clustered} gives the comparison and its null-expectation limit.
\end{remark}

\begin{remark}[What compression requires and what it gives up]\label{rem:compression}
Proposition~\ref{prop:common} needs balanced clusters with a common design, a common error covariance and Gaussian $\bb_g$. It uses OLS rather than GLS within the cluster, so it ignores what the $n_0-k$ residual directions say about $\bSigma_\varepsilon$. For cluster-specific error covariances $\bSigma_{\varepsilon,g}$, write $\bSigma_{R,g}=\bR(\bW_0'\bW_0)^{-1}\bW_0'\bSigma_{\varepsilon,g}\bW_0(\bW_0'\bW_0)^{-1}\bR'$. If these compressed covariances differ, the i.i.d. Gaussian reduction fails. Differing designs can also make pooled OLS differ from the unweighted average of the $\hat\bgamma_g$; differing error covariances alone do not change that algebra under a common design. When $k>n_0$ the $\hat\bgamma_g$ do not exist, and their role is taken by the influence contributions $\bR\bH^{-1}\bW_g'(\by_g-\bW_g\bgamma)$ with $\bH=\E[\bW_g'\bW_g]$, which reduce to $\bR(\hat\bgamma_g-\bgamma)$ under a common design. In those cases, and for non-Gaussian outcomes, $\ECR_G$ is an asymptotic rather than an exact $e$-process (Section~\ref{sec:asymptotic}).
\end{remark}

Section~\ref{sec:asymptotic} keeps the formula \eqref{eq:ECR} while replacing the Gaussian model by a population least-squares projection. The benchmark above fixes its finite-sample form, which logarithmic asymptotic equivalence alone cannot select.

\section{Nonparametric Asymptotic Inference for Clustered Projections}
\label{sec:asymptotic}

We now replace the Gaussian model by a population least-squares projection and use the sandwich statistic $\ECR_G$ of \eqref{eq:ECR}. It has the exact benchmark of Proposition~\ref{prop:common}, and this section establishes its asymptotic validity for general clusters. Outcomes need not be Gaussian, clusters may differ in size and in design, dependence within a cluster is unrestricted, and the linear model need not hold: the target is the population least-squares projection. The result extends the heteroskedasticity-robust construction of \citet{lindon2026anytime} from independent observations to independent clusters.

\subsection{Population projection and cluster-robust statistic}

Let $(\by,\bW)\sim P$ denote a generic cluster, consisting of an outcome vector $\by$ and a design matrix $\bW$ with $k$ columns and as many rows as $\by$; the number of rows is part of the draw and need not be the same for every cluster. The observed clusters $(\by_g,\bW_g)$, $g=1,2,\ldots$, are independent draws from $P$, with $\by_g\in\R^{n_g}$ and $\bW_g\in\R^{n_g\times k}$. Define the population least-squares target
\begin{equation}\label{eq:projection}
\bgamma^\star=\argmin_{\bb\in\R^k}\E_P\big[(\by-\bW \bb)'(\by-\bW \bb)\big],
\qquad\text{so that}\qquad \E_P\{\bW'(\by-\bW\bgamma^\star)\}=0 ,
\end{equation}
which is unique when $\E_P(\bW'\bW)$ is positive definite and the second moments exist. No model for $\E(\by\mid \bW)$ is assumed. Throughout this section $\|\cdot\|$ is the Euclidean norm of a vector and the spectral norm of a matrix. The parameter of interest is $\btheta^\star=\bR\bgamma^\star$ for a fixed $\bR\in\R^{q\times k}$ of full row rank, and $H_0:\btheta^\star=\btheta_0$. The population cluster scores are $\bpsi_g=\bW_g'(\by_g-\bW_g\bgamma^\star)$, with partial sums $\bs_G=\sum_{g\leq G}\bpsi_g$, and
\[
\bH=\E(\bW_g'\bW_g),\qquad \bOmega=\E(\bpsi_g\bpsi_g'),\qquad \bSigma_R=\bR\bH^{-1}\bOmega\bH^{-1}\bR'.
\]
Under the common design of Proposition~\ref{prop:common}, $\bSigma_R$ is the covariance of the $\bb_g$ defined there, which is why the same symbol is used. The quantities $\bA_G$, $\hat\bgamma_G$, $\hat \bs_g$, $\hat \bV_G$, $Q_G(\btheta_0)$ and $\ECR_G(\btheta_0)$ are those of Section~\ref{sec:clustered}, whose definitions apply verbatim to clusters of unequal size; write $\hat \bM_G=\sum_{g\leq G}\hat \bs_g\hat \bs_g'$ for the cluster meat. Pooled OLS satisfies the exact linearization
\begin{equation}\label{eq:linearization}
\hat\bgamma_G-\bgamma^\star=\bA_G^{-1}\bs_G ,
\end{equation}
the cluster analogue of the observation-level score representation in \citet{lindon2026anytime}.

\subsection{Assumptions and asymptotic validity}
The guarantee is asymptotic in the sense of Definition~\ref{def:asymp}.

\begin{assumption}[Cluster-score regularity]\label{ass:scores}
The clusters $\{(\by_g,\bW_g)\}_{g\geq1}$ are i.i.d., $\bH$ is positive definite, and $\bSigma_R$ is positive definite. On a sufficiently rich probability space there are i.i.d.\ $\bz_g\sim\Nor_k(0,\bOmega)$ with
\[
\Big\|\sum_{g=1}^G\bpsi_g-\sum_{g=1}^G\bz_g\Big\|=O_{a.s.}(G^{\vartheta})\qquad\text{for some }0<\vartheta<1/2 .
\]
Moreover $\bA_G/G\to\bH$ and $\hat \bM_G/G\to\bOmega$ almost surely.
\end{assumption}

Assumption~\ref{ass:scores} states the cluster-level requirements directly: the independent cluster scores admit a multivariate strong Gaussian approximation, the design stabilizes at the cluster level, and the covariance of the fitted cluster scores consistently estimates that of the true ones. Extensions to triangular arrays or non-identically distributed clusters require corresponding assumptions and are not asserted here. The next two lemmas give primitive sufficient conditions.

\begin{lemma}[Multivariate strong Gaussian approximation; {\citealp[Theorem~1]{einmahl1989extensions}}]\label{lem:kmt}
Let $(\bu_g)_{g\geq1}$ be i.i.d.\ random vectors in $\R^k$ with $\E \bu_g=0$, covariance $\bOmega$ and $\E\|\bu_g\|^s<\infty$ for some $s>3$. On a sufficiently rich probability space there exist i.i.d.\ $\bz_g\sim\Nor_k(0,\bOmega)$ with $\|\sum_{g\leq G}\bu_g-\sum_{g\leq G}\bz_g\|=O_{a.s.}(G^\vartheta)$ for some $0<\vartheta<1/2$.
\end{lemma}

\citet[Theorem~1, p.~22]{einmahl1989extensions} gives the stronger coupling error $o_{a.s.}(G^{1/s})$ for $s>3$, so $\vartheta=1/s$ suffices here. The result is used only at a polynomial rate strictly smaller than $G^{1/2}$. If $\bOmega$ is singular, project onto its range, apply the nondegenerate multivariate approximation there, and map back; every centered component in its null space is zero almost surely.

\begin{lemma}[Primitive conditions]\label{lem:primitive}
Let the clusters be i.i.d.\ with $\bH\succ0$ and $\bSigma_R\succ0$, and write $\bA_g=\bW_g'\bW_g$, so that $\bA_G=\sum_{g\leq G}\bA_g$. If $\E\|\bpsi_g\|^s<\infty$ for some $s>3$ and $\E\|\bA_g\|^2<\infty$, then Assumption~\ref{ass:scores} holds. In particular, subject to the stated positive-definiteness conditions, it holds whenever cluster sizes, regressors and outcomes are bounded.
\end{lemma}

\begin{theorem}[Cluster-robust asymptotic $e$-process]\label{thm:asymptotic}
Let $\cP$ be the family of distributions of the cluster sequence that satisfy Assumption~\ref{ass:scores}, and let $\phi>0$ and $\btheta_0\in\R^q$. The process
\[
\ECR_G(\btheta_0)=\left(\frac{\phi}{\phi+G}\right)^{q/2}
\left(\frac{1+Q_G(\btheta_0)/\nu_G}{1+\frac{\phi}{\phi+G}\,Q_G(\btheta_0)/\nu_G}\right)^{(\nu_G+1)/2},\qquad \nu_G=G-1,\quad G>q,
\]
of \eqref{eq:ECR}, with the singular-sample convention stated there, is an asymptotic $e$-process for $H_0:\bR\bgamma^\star=\btheta_0$.

Moreover, fix $\rho>0$ and $\alpha\in(0,1)$. For each monitoring start $m$, use $\phi_m=\rho m$ at every $G\geq m$ and denote the resulting process by $\ECR_{G,m}$. For each fixed $P\in\cH_0$,
\begin{equation}\label{eq:late_start}
\limsup_{m\to\infty}P\!\left(\exists G\geq m:\ \ECR_{G,m}(\btheta_0)\geq\alpha^{-1}\right)\leq\alpha.
\end{equation}
Equivalently, define $C_{G,m}^{\alpha}=\{\btheta\in\R^q:\ECR_{G,m}(\btheta)<\alpha^{-1}\}$. For each fixed $P\in\cP$, with $\btheta^\star=\bR\bgamma^\star$,
\begin{equation}\label{eq:late_start_coverage}
\liminf_{m\to\infty}P\!\left(\btheta^\star\in C_{G,m}^{\alpha}\ \text{for all }G\geq m\right)\geq1-\alpha.
\end{equation}

\end{theorem}

\subsection{Confidence sequences and reporting}

\begin{corollary}[Asymptotic anytime-valid $p$-values]\label{cor:pvalue}
Under the conditions of Theorem~\ref{thm:asymptotic}, the untruncated reciprocal $\pv_G(\btheta_0)=1/\ECR_G(\btheta_0)$ is an asymptotic anytime-valid $p$-value in the logarithmic sense of \citet[arXiv v7, Definition~C.3]{waudbysmith2024time}. One may report $\min(1,\pv_G)$ without changing any rejection decision at $\alpha\in(0,1)$. With $\phi_m=\rho m$, both reporting conventions satisfy the operational bound \eqref{eq:late_start}.
\end{corollary}

\begin{corollary}[Asymptotic cluster-robust confidence sequence]\label{cor:asympcs}
Under the conditions of Theorem~\ref{thm:asymptotic}, $C_G^\alpha=\{\btheta\in\R^q:\ECR_G(\btheta)<\alpha^{-1}\}$, $G>q$, is an asymptotic $(1-\alpha)$ confidence sequence for $\bR\bgamma^\star$ in the inversion sense of Section~\ref{sec:background}. With
\begin{equation}\label{eq:kappa}
u_G(\phi,q,\alpha)=\left(\frac{\alpha^{2/q}\phi}{\phi+G}\right)^{q/(\nu_G+1)},
\qquad
\kappa_G(\phi,q,\alpha)=\frac{\nu_G}{q}\,\frac{1-u_G(\phi,q,\alpha)}{\max\{0,\ u_G(\phi,q,\alpha)-\frac{\phi}{\phi+G}\}},
\end{equation}
and the convention $\kappa_G=+\infty$ (so $C_G^\alpha=\R^q$) when the denominator vanishes, it is the open Wald ellipsoid
\[
C_G^\alpha=\Big\{\btheta\in\R^q:\ Q_G(\btheta)/q<\kappa_G(\phi,q,\alpha)\Big\}.
\]
For a scalar contrast ($q=1$, $\bR$ a row vector) it is the interval $\bR\hat\bgamma_G\pm\sqrt{\bR\hat \bV_G\bR'}\,\sqrt{\kappa_G(\phi,1,\alpha)}$, the conventional cluster-robust interval with the fixed-sample critical value replaced by $\sqrt{\kappa_G(\phi,1,\alpha)}$. Using $\phi_m=\rho m$ in these formulas gives $C_{G,m}^{\alpha}$, whose simultaneous coverage over $G\geq m$ satisfies \eqref{eq:late_start_coverage}.
\end{corollary}
The proofs are in Appendix~\ref{app:asymptotic}. The first five steps of Theorem~\ref{thm:asymptotic} establish logarithmic equivalence: a strong Gaussian approximation of $\bR(\hat\bgamma_G-\bgamma^\star)$ through \eqref{eq:linearization}; the exact Gaussian mixture martingale of the coupled Gaussian vectors, which is the known-covariance counterpart of Theorem~\ref{thm:main} in the intercept-only case with the same mixture $\bmu\sim\Nor(0,\bSigma_R/\phi)$; the replacement of $\bSigma_R$ by its cluster-robust estimate; the replacement of the Gaussian form $\exp\{\frac12\frac{G}{\phi+G}Q_G\}$ by the Student-type form \eqref{eq:ECR}, which is where $\nu_G=G-1$ enters; and, under an alternative, a comparison with the exact Bayes factor of Theorem~\ref{thm:main} for the coupled Gaussian vectors, which grows at the same linear rate. Step~6 separately proves the late-start crossing bound. Corollary~\ref{cor:cr-growth} below gives the growth and stopping-time conclusions, with its expectation statement proved separately by complete convergence.

\begin{remark}[Relation to the exact results]\label{rem:asymp_exact}
Proposition~\ref{prop:common} supplies the exact common-design Gaussian benchmark for $\ECR_G$. Theorem~\ref{thm:asymptotic} uses the same formula for a population projection, with no shared-coefficient conditional-mean model required. When the coefficients also depend on cluster-level covariates, the compressed vectors $\bb_g$ follow \eqref{eq:model} with those covariates as regressors, and Theorem~\ref{thm:main} gives a different exact benchmark, with $\nu$ equal to $G$ minus the rank of the cluster-level design and a pooled residual covariance in place of the sandwich; the cluster-robust statistic with $G-1$ is then only asymptotically equivalent to it, and Theorem~\ref{thm:asymptotic} still applies. When every cluster is a single observation ($n_g\equiv1$), $\hat \bV_G$ is the heteroskedasticity-consistent sandwich of \citet{white1980heteroskedasticity} with the factor $G/(G-1)$, and Theorem~\ref{thm:asymptotic} reduces to the heteroskedasticity-robust theorem of \citet{lindon2026anytime} up to finite-sample conventions: that paper scales the sandwich by $G/(G-p-d)$ and uses $\nu_G=G-p-d$ with exponent $(\nu_G+q)/2$, against $G/(G-1)$, $\nu_G=G-1$ and exponent $G/2$ here. Steps~4 and~5 of the proof show that these differences are immaterial asymptotically.
\end{remark}

\begin{remark}[What the asymptotic guarantee does and does not say]\label{rem:guarantee}
Definition~\ref{def:asymp} is pathwise and logarithmic. It does not imply that $\E_P[\ECR_G]$ is close to one, nor finite-sample error control for a fixed $\phi$ with monitoring from the first clusters. Type-I error is governed by crossing probabilities. Equation~\eqref{eq:late_start} proves a concrete late-start guarantee, in the coverage sense formalized by \citet[arXiv v9, Definition~2.7]{waudbysmith2024time}: take $\phi_m=\rho m$ with fixed $\rho>0$ and monitor all $G\geq m$ as $m\to\infty$. The equivalent coverage bound is the $\liminf$ statement \eqref{eq:late_start_coverage}. Both bounds are pointwise in $P$ and assert neither convergence of the probabilities nor uniformity over the model class. For fixed $\phi$, the null LIL instead gives $\ECR_G\to0$ almost surely, so the probability of a crossing after $m$ tends to zero; that conservative limit does not justify retuning $\phi$ with $m$. In that regime the finite-$G$ differences discussed in Remark~\ref{rem:dfrule} vanish. In practice $\phi$ should target a realistic horizon and monitoring should not begin at very small $G$; the Student-type form helps, since $C_G^\alpha=\R^q$ until $u_G>\phi/(\phi+G)$.
\end{remark}

\subsection{Growth, power and first-order expected stopping time}
\label{sec:cr-growth}
The following result concerns fixed alternatives, with the precision held fixed as $\alpha\downarrow0$.

\begin{corollary}[Cluster-robust growth and expected stopping time]\label{cor:cr-growth}
Under Assumption~\ref{ass:scores}, fix $\phi>0$.
Under a fixed alternative $\btheta^\star=\bR\bgamma^\star\neq\btheta_0$, put
\[
c=(\btheta^\star-\btheta_0)'\bSigma_R^{-1}(\btheta^\star-\btheta_0),
\qquad r_{\mathrm{CR}}=\tfrac12\log(1+c)>0.
\]
Then
\[
\frac{\log\ECR_G(\btheta_0)}G\longrightarrow r_{\mathrm{CR}}
\quad\text{almost surely},
\]
so the test has power one. Define
\[
\tau_\alpha^{\mathrm{CR}}=\inf\{G>q:\ECR_G(\btheta_0)\geq\alpha^{-1}\}.
\]
As $\alpha\downarrow0$, $\tau_\alpha^{\mathrm{CR}}/\log(1/\alpha)\to1/r_{\mathrm{CR}}$ almost surely. If in addition
\begin{equation}\label{eq:stopping_moments}
\E_P\|\bpsi_g\|^4<\infty,\qquad
\E_P\|\bA_g\|^4<\infty,
\end{equation}
then $\E_P[\tau_\alpha^{\mathrm{CR}}]<\infty$ for every $\alpha\in(0,1)$, the convergence holds in $L^1(P)$, and
\begin{equation}\label{eq:cr_expected_stopping}
\E_P[\tau_\alpha^{\mathrm{CR}}]\sim
\frac{\log(1/\alpha)}{r_{\mathrm{CR}}}
\qquad(\alpha\downarrow0).
\end{equation}
This expectation averages over both designs and outcomes.
\end{corollary}

Condition \eqref{eq:stopping_moments} holds if cluster sizes and regressor entries are bounded and $\E_P\|\by_g\|^4<\infty$, in particular for bounded outcomes. Cluster sizes may vary and may be smaller than the number of regressors: only the population matrix $\bH$ and contrast covariance $\bSigma_R$ must be positive definite. The stopping-time conclusion uses fixed $\phi$ and $\alpha\downarrow0$ under a fixed alternative; it does not change the null calibration in \eqref{eq:late_start}, which uses $\phi_m=\rho m$ and $m\to\infty$ at fixed $\alpha$.

\subsection{Mixture selection for a fixed alternative}
\label{sec:cr-log-growth}

The same frequentist criterion applies to the observable cluster-robust statistic. Fix an alternative distribution $P$ and recall $c=(\btheta^\star-\btheta_0)'\bSigma_R^{-1}(\btheta^\star-\btheta_0)>0$.

\begin{proposition}[Relative cluster-robust log growth]\label{prop:cr-log-growth}
In the setup of Section~\ref{sec:asymptotic}, suppose $\bH\succ0$, $\bSigma_R\succ0$ and $\|\bA_g\|+\|\bpsi_g\|\leq K$ almost surely for some finite $K$. Suppose $\btheta^\star\neq\btheta_0$ and put $\phi_*=q/c$. For every fixed $\phi>0$, as $G\to\infty$,
\begin{equation}\label{eq:cr-precision-cost}
\E_P[\log E_G^{\mathrm{CR},(\phi_*)}]-\E_P[\log E_G^{\mathrm{CR},(\phi)}]
=\frac q2\left\{\frac{\phi}{\phi_*}-1-\log\frac{\phi}{\phi_*}\right\}+o(1).
\end{equation}
\end{proposition}

Thus $\phi_*$ maximizes expected log growth through constant order among fixed precisions for the specified $P$. Bounded cluster sizes, regressors and outcomes imply the boundedness condition. Individual $\bA_g$ may be singular, and neither a linear conditional mean nor a common within-cluster design is required. Appendix~\ref{app:cr-log-growth} proves the comparison directly from \eqref{eq:ECR}. This tuning criterion is separate from the late-start null calibration \eqref{eq:late_start}.

\begin{remark}[Choice of $\phi$]\label{rem:lambda}
$\phi$ plays the role of a planned sample size. Minimizing $\kappa_G(\phi,q,\alpha)$ over $\phi$ at a target $G=G^\star$ gives the narrowest confidence sequence at $G^\star$; in the exact Gaussian benchmark validity is unaffected by the actual stopping time, although the width changes. Outside that benchmark the guarantees remain asymptotic. $\phi$ must be fixed before monitoring begins. In the Gaussian benchmark of Proposition~\ref{prop:common}, $\phi$ is the prior precision of the standardized contrast, $\btheta\mid\bSigma_R\sim\Nor(\btheta_0,\bSigma_R/\phi)$. The $e$-process is log-optimal among invariant $e$-variables against that mixture alternative (Remark~\ref{rem:optimality}). For a specified fixed frequentist alternative, maximizing expected log growth gives a separate tuning criterion: \eqref{eq:cr-precision-cost} yields $\phi_*=q/c$ through constant order, whereas minimizing the radius here targets width at the chosen horizon.
\end{remark}

\section{Application: Randomized Experiments with Several Outcomes}
\label{sec:rct}

We now apply Theorem~\ref{thm:asymptotic} to a randomized experiment in which each unit contributes a vector of outcomes. Under randomization, the population least-squares coefficient of the treatment indicator is the average treatment effect even when the outcome depends nonlinearly on the covariates, a standard fact that \citet{lindon2026anytime} use for a scalar outcome to bring causal estimands within the robust $e$-process; here the same holds outcome by outcome, and the $q$ regressions together form one clustered regression with the units as clusters. Fixed-sample inference from covariate-adjusted OLS with sandwich standard errors is established prior work; see \citet[Section~5, Theorem~2]{lin2013agnostic} for the finite-population randomized setting. Our formulation below instead assumes i.i.d.\ sampling from a superpopulation and derives sequential guarantees. Remark~\ref{rem:rctcompare} relates the procedure to the exact results of Section~\ref{sec:exact}.

\paragraph{Potential outcomes.} We use potential outcomes and assume SUTVA (no interference between units and no hidden versions of treatment), as in \citet[Section~6]{lindon2026anytime}, so that unit $i$ has potential outcome vectors $\by_i(0),\by_i(1)\in\R^q$ and observed outcome $\by_i=D_i\by_i(1)+(1-D_i)\by_i(0)$, where $D_i\sim\mathrm{Bernoulli}(\pi)$, $\pi\in(0,1)$, is the treatment indicator. Randomization means $D_i\perp(\by_i(0),\by_i(1),\bmv_i)$, where $\bmv_i\in\R^r$ are pre-treatment covariates with mean $\bmu_m$ and positive definite covariance $\bOmega_m$. The data $\{(\by_i(0),\by_i(1),D_i,\bmv_i)\}_{i\geq1}$ are i.i.d.\ draws from a joint distribution and arrive sequentially. The estimand is the vector of average treatment effects
\[
\bdelta^\star=\E[\by_i(1)-\by_i(0)]\in\R^q,
\]
one per outcome. When interference is present, outcomes must be allowed to depend on other units' treatment assignments. \citet[Section~2.1]{hudgens2008toward} develop such a framework under partial interference, allowing interference within groups but assuming none between groups.

\paragraph{The regression.} For each outcome $j=1,\ldots,q$ consider the regression on treatment and pre-treatment covariates without interactions,
\begin{equation}\label{eq:rctreg}
y_{ij}=\mu_j+\bmv_i'\bzeta_j+(D_i-\pi)\tau_j+\varepsilon_{ij},
\end{equation}
with regressor vector $\bw_i=(1,\bmv_i',D_i-\pi)'\in\R^k$, $k=r+2$, and coefficient vector $\bgamma_j=(\mu_j,\bzeta_j',\tau_j)'$. The covariate means need not be known. The regression is a population least-squares projection; neither linear conditional outcome means nor constant treatment effects are assumed. The $q$ regressions share the regressors and have outcome-specific coefficients, so together they are the mean structure of model \eqref{eq:model} with $\bx_i=(1,\bmv_i')'$, $z_i=D_i-\pi$, $\bB'=[\bmu,\bm\Lambda_\zeta']$, where $\bmu=(\mu_1,\ldots,\mu_q)'$ and $\bm\Lambda_\zeta=[\bzeta_1,\ldots,\bzeta_q]\in\R^{r\times q}$, and $\bDelta'=\btau$, where $\btau=(\tau_1,\ldots,\tau_q)'$. Equally, they are one instance of the clustered regression of Section~\ref{sec:asymptotic} with the units as clusters,
\begin{equation}\label{eq:rctcluster}
\by_i\in\R^q,\qquad \bW_i=\bI_q\otimes \bw_i'\in\R^{q\times qk},\qquad \bgamma=(\bgamma_1',\ldots,\bgamma_q')'\in\R^{qk},
\end{equation}
since $\bW_i\bgamma=(\bw_i'\bgamma_1,\ldots,\bw_i'\bgamma_q)'$. Pooled OLS for \eqref{eq:rctcluster} is OLS outcome by outcome, because $\bA_G=\bI_q\otimes\bW_G'\bW_G$, where $\bW_G$ is the $G\times k$ matrix with rows $\bw_i'$; it is the matrix $[\bX_G,\bZ_G]$ of Section~\ref{sec:exact}, since $\bw_i=(\bx_i',z_i')'$. Let $\bR_\tau\in\R^{q\times qk}$ select the $\tau$-coordinates of $\bgamma$, so that $\bR_\tau\bgamma=(\tau_1,\ldots,\tau_q)'$.

\begin{proposition}[Average treatment effects as population least-squares coefficients]\label{prop:rct}
Assume SUTVA, randomization, $\E\|\by_i(t)\|^2<\infty$ for $t=0,1$, and $\E\|\bmv_i\|^2<\infty$. Then $\bOmega_{\bW}=\E(\bw_i\bw_i')$ is positive definite, the population least-squares projection \eqref{eq:projection} of the clustered regression \eqref{eq:rctcluster} is $\bgamma^\star=(\bgamma_1^{\star\prime},\ldots,\bgamma_q^{\star\prime})'$ with $\bgamma_j^\star=\bOmega_{\bW}^{-1}\E(\bw_iy_{ij})$, and for every outcome $j$,
\begin{gather*}
\tau_j^\star=\E[y_{ij}(1)-y_{ij}(0)],\qquad
\bzeta_j^\star=\bOmega_m^{-1}\E[(\bmv_i-\bmu_m)y_{ij}],\\
\mu_j^\star=\E[y_{ij}]-\bmu_m'\bzeta_j^\star.
\end{gather*} Hence $\bR_\tau\bgamma^\star=\bdelta^\star$. Moreover $\bH=\E(\bW_i'\bW_i)=\bI_q\otimes\bOmega_{\bW}\succ0$, and the population cluster scores $\bpsi_i=\bW_i'(\by_i-\bW_i\bgamma^\star)=\beps_i\otimes \bw_i$, with $\beps_i=\by_i-(\bI_q\otimes \bw_i')\bgamma^\star$, have mean zero. If in addition $\E\|\beps_i\otimes \bw_i\|^2<\infty$, their covariance is $\bOmega=\E(\beps_i\beps_i'\otimes \bw_i\bw_i')$; this extra integrability does not follow from the preceding second-moment conditions alone.
\end{proposition}
The proof (Appendix~\ref{app:rct}) applies the identification argument of \citet{lindon2026anytime} to each outcome; nothing about the outcomes' joint distribution is used beyond second moments. Centering the covariates at any common vector, including their sample mean, leaves the treatment estimates and their CR1 covariance block unchanged. Thus the procedure uses the ordinary OLS estimates and CR1 covariance without an additional variance correction. Appendix~\ref{app:rct_interactions} gives the corresponding adjustment for fully interacted regression with sample-mean centering.

\begin{corollary}[Asymptotic anytime-valid inference for the vector of average treatment effects]\label{cor:rct}
Under the assumptions of Proposition~\ref{prop:rct}, let $\hat\bgamma_G$ be the pooled OLS estimate of \eqref{eq:rctcluster} from the first $G$ units, $\hat\bdelta_G=\bR_\tau\hat\bgamma_G$ the vector of estimated treatment effects, $\hat \bV_G$ the CR1 covariance \eqref{eq:CR1} with the units as clusters, and $Q_G(\bdelta_0)=(\hat\bdelta_G-\bdelta_0)'(\bR_\tau\hat \bV_G\bR_\tau')^{-1}(\hat\bdelta_G-\bdelta_0)$.
\begin{enumerate}
\item[\textup{(i)}] \textup{(Joint test and confidence ellipsoid.)} Suppose in addition that $\E\|\bmv_i\|^4<\infty$, $\E\|\beps_i\otimes \bw_i\|^s<\infty$ for some $s>3$, and that $\bSigma_{R_\tau}=\bR_\tau\bH^{-1}\bOmega\bH^{-1}\bR_\tau'$ is positive definite. Then $\ECR_G(\bdelta_0)$ of \eqref{eq:ECR}, computed from $Q_G(\bdelta_0)$ with the units as clusters and $\nu_G=G-1$, is an asymptotic $e$-process for $H_0:\bdelta^\star=\bdelta_0$; the reciprocal $1/\ECR_G(\bdelta_0)$ has the logarithmic asymptotic $p$-value property, and its reported truncation $\pv_G(\bdelta_0)=\min\{1,1/\ECR_G(\bdelta_0)\}$ has the same rejection events; and the ellipsoid
\[
C_G^\alpha=\big\{\bdelta\in\R^q:\ Q_G(\bdelta)/q<\kappa_G(\phi,q,\alpha)\big\}
\]
of Corollary~\ref{cor:asympcs} is an asymptotic $(1-\alpha)$ confidence sequence for $\bdelta^\star$.
\item[\textup{(ii)}] \textup{(Simultaneous confidence sequences.)} On the event $\{\bdelta^\star\in C_G^\alpha\ \forall G\}$, for every nonzero $\bc\in\R^q$ and every $G$ with a finite ellipsoid simultaneously,
\[
\big|\bc'(\hat\bdelta_G-\bdelta^\star)\big|<\sqrt{q\,\kappa_G(\phi,q,\alpha)\;\bc'\bR_\tau\hat \bV_G\bR_\tau'\bc}\,.
\]
For $\bc=0$ use the singleton $\{0\}$, and before the ellipsoid is finite use $\R$ for nonzero $\bc$. Hence the intervals $\bc'\hat\bdelta_G\pm\sqrt{q\kappa_G(\phi,q,\alpha)\,\bc'\bR_\tau\hat \bV_G\bR_\tau'\bc}$, over all $\bc$ and all $G$, are simultaneous asymptotic confidence sequences for all linear combinations of the treatment effects; with $\bc=\be_j$, the $j$th standard basis vector of $\R^q$, they give simultaneous asymptotic confidence sequences for the per-outcome effects $\delta_j^\star$.
\end{enumerate}
With $\phi_m=\rho m$ and monitoring over $G\geq m$, the ellipsoids satisfy \eqref{eq:late_start_coverage} with $\btheta^\star=\bdelta^\star$. The simultaneous coverage probability of the intervals over all $\bc$ and all $G\geq m$ has the same $\liminf$ lower bound $1-\alpha$, for each fixed law satisfying the assumptions in \textup{(i)}.
\end{corollary}

\begin{remark}[Inference through the parametric model]\label{rem:rctcompare}
The $q$ regressions \eqref{eq:rctreg} share the regressors $\bw_i$, each with its own coefficients, which is the form of model \eqref{eq:model} (Appendix~\ref{app:mean-structures}(c)). One may therefore ask whether inference could proceed through that parametric model, with randomization supplying the asymptotic justification, the question that \citet{lindon2026anytime} ask of linear regression with a scalar outcome. Suppose first that \eqref{eq:rctreg} holds as a model, with conditional mean linear in $\bw_i$ and errors $\beps_i=(\varepsilon_{i1},\ldots,\varepsilon_{iq})'\overset{iid}{\sim}\Nor_q(0,\bSigma)$ independent of $(D_i,\bmv_i)$; for example, this observed-data model follows from the stronger potential-outcome model $\by_i(t)=\bmu+\bm\Lambda_\zeta'\bmv_i+(t-\pi)\btau+\beps_i$ and $\bdelta^\star=\btau$. Then, conditionally on the design, $E_G$ of Corollary~\ref{cor:special}\textup{(c)} with $z_i=D_i-\pi$, $\bx_i=(1,\bmv_i')'$ (so $p=1+r$, $d=1$, $\nu_G=G-r-2$) and $\bPhi=\phi$, evaluated at $\bY_G-\bZ_G\bdelta_0'$, is an exact $e$-process for $H_0:\bdelta^\star=\bdelta_0$, and the ellipsoid \eqref{eq:ellipsoid}, with $T_G^2$ the regression-adjusted Hotelling statistic, is an exact $(1-\alpha)$ confidence sequence for $\bdelta^\star$; the time-uniform bounds hold unconditionally because the design is independent of the errors (Remark~\ref{rem:design}). Since the design $\bI_q\otimes \bw_i'$ varies across units, this benchmark is Theorem~\ref{thm:main} directly rather than the compression of Proposition~\ref{prop:common}. This exact version and Corollary~\ref{cor:rct}(i) use the same coefficient estimate $\hat\bdelta_G$ and differ in the covariance estimate and the degrees of freedom. The exact version uses the pooled residual covariance $\hSigma_G$ of the $q$ regressions, scaled by $\|\tZ_G\|_2^{-2}$, with $\nu_G=G-p-1$; the robust version uses $\bR_\tau\hat \bV_G\bR_\tau'=\frac{G}{G-1}\sum_ih_{iG}^2\hat\beps_{iG}\hat\beps_{iG}'$, where $h_{iG}=\bw_i'(\bW_G'\bW_G)^{-1}\be_\tau$, $\be_\tau\in\R^k$ is the standard basis vector at the position of $D_i-\pi$ in $\bw_i$, and $\hat\beps_{iG}$ is the $q$-vector of residuals of unit $i$, with $\nu_G=G-1$; this is the cluster sandwich of \citet[Section~2, Theorem~1]{liang1986longitudinal}, applied to the stacked system with each unit's $q$ residuals as a cluster, and reduces to the heteroskedasticity-consistent covariance of \citet{white1980heteroskedasticity} when $q=1$.

Now drop the model and keep only randomization. The proof of Theorem~\ref{thm:asymptotic} uses the covariance estimate only through its almost-sure limit (Steps~3 and~6), so the exact statistic is an asymptotic $e$-process under the assumptions of Corollary~\ref{cor:rct}, by the same proof and with matching prior scales, that is, with Corollary~\ref{cor:rct}(i) tuned to $\phi/\lim_G(\|\tZ_G\|_2^2/G)$, whenever its covariance estimate, multiplied by $G$, has the same limit as the sandwich. Let $\beps_i$ be the projection residual of Proposition~\ref{prop:rct} and $\bV_t=\E[\beps_i\beps_i'\mid D_i=t]$ the residual covariance in arm $t$, so that $\E[\beps_i\beps_i']=\pi\bV_1+(1-\pi)\bV_0$. Because $D_i$ is independent of $\bmv_i$, $\bOmega_{\bW}$ is block diagonal with $\pi(1-\pi)$ in the cell of $D_i-\pi$, $\bR_\tau\bH^{-1}\bpsi_i=\beps_i(D_i-\pi)/\{\pi(1-\pi)\}$, and $\|\tZ_G\|_2^2/G\to\pi(1-\pi)$; hence, almost surely,
\[
G\,\bR_\tau\hat \bV_G\bR_\tau'\longrightarrow\frac{\bV_1}{\pi}+\frac{\bV_0}{1-\pi},
\qquad
\frac{G\,\hSigma_G}{\|\tZ_G\|_2^2}\longrightarrow\frac{\bV_1}{1-\pi}+\frac{\bV_0}{\pi}.
\]
Nonlinearity of $\E[\by_i\mid \bmv_i,D_i]$ in $\bmv_i$, non-Gaussian outcomes, dependence among the $q$ outcomes of a unit and heteroskedasticity as a function of $\bmv_i$ enter only through $\bV_1$ and $\bV_0$; the two limits differ by $(\bV_1-\bV_0)(1-2\pi)/\{\pi(1-\pi)\}$. When the allocation is balanced, $\pi=1/2$, or the residual covariance is the same in both arms, the two limits coincide: the parametric procedure is then justified asymptotically by randomization alone, and under the Gaussian homoskedastic model both limits equal $\bSigma/\{\pi(1-\pi)\}$. Otherwise the covariance limits differ, and asymptotic validity is not guaranteed in general: the pooled covariance can lead to conservative or anticonservative inference. For $q=1$ the Wald statistic of the exact version is asymptotically $\varkappa$ times the robust one, with
\[
\varkappa=\frac{V_1/\pi+V_0/(1-\pi)}{V_1/(1-\pi)+V_0/\pi},
\]
which exceeds one, the anticonservative case, exactly when the smaller arm has the larger residual variance. For $q\geq2$ the scalar $\varkappa$ is replaced by a matrix, and since $\bV_1-\bV_0$ need not be definite, the exact statistic can be anticonservative in some directions of $\R^q$ and conservative in others. This is why Corollary~\ref{cor:rct} is stated through Theorem~\ref{thm:asymptotic}, which is correct at every allocation, and it is the relation between $E_G$ and the robust $e$-process in \citet{lindon2026anytime}, now with a $q\times q$ covariance.
\end{remark}

\begin{remark}[Coordinatewise evidence and the joint statistic]\label{rem:bonferroni}
The joint hypothesis $H_0:\bdelta^\star=\bdelta_0$ may also be examined using the $q$ scalar asymptotic $e$-processes $E_{j,G}(\delta_{0j})$ of Corollary~\ref{cor:rct}(i) with $q=1$, one per outcome, each with a fixed positive tuning parameter $\phi_j$. For genuine $e$-values, arithmetic averaging preserves validity under arbitrary dependence, whereas multiplication need not \citep[Sections~3--4]{vovk2021evalues}. Here the coordinatewise guarantees are asymptotic, and we compare their averaged evidence
\[
\bar E_G=\frac1q\sum_{j=1}^qE_{j,G}(\delta_{0j})
\]
with the joint statistic through its rejection rule and growth rate.

Since $\bar E_G\geq\max_jE_{j,G}/q$, the rule that rejects when $\bar E_G\geq1/\alpha$ rejects whenever the coordinatewise Bonferroni rule $\max_jE_{j,G}\geq q/\alpha$ does. The mean grows at the rate of its largest term:
\[
\max_j\log E_{j,G}-\log q\leq\log\bar E_G\leq\max_j\log E_{j,G}.
\]
Thus the mean and the Bonferroni statistic $\max_jE_{j,G}/q$ have the same asymptotic growth rate.

Suppose $\bdelta^\star\neq\bdelta_0$. Step~5 of the proof of Theorem~\ref{thm:asymptotic}, together with the null calculation for coordinates with $\delta^\star_j=\delta_{0j}$, gives
\[
\frac1G\log\bar E_G\ \overset{a.s.}{\longrightarrow}\ \max_j\frac12\log\Big(1+\frac{(\delta^\star_j-\delta_{0j})^2}{(\bSigma_{R_\tau})_{jj}}\Big),
\]
whereas
\[
\frac1G\log\ECR_G(\bdelta_0)\ \overset{a.s.}{\longrightarrow}\ \frac12\log\big\{1+(\bdelta^\star-\bdelta_0)'\bSigma_{R_\tau}^{-1}(\bdelta^\star-\bdelta_0)\big\}.
\]
By the Cauchy--Schwarz inequality,
\[
(\delta^\star_j-\delta_{0j})^2\leq(\bSigma_{R_\tau})_{jj}\,(\bdelta^\star-\bdelta_0)'\bSigma_{R_\tau}^{-1}(\bdelta^\star-\bdelta_0)
\]
for every $j$. Consequently, the joint growth rate is never smaller than the best coordinate rate. Equality holds if and only if $\bdelta^\star-\bdelta_0$ is proportional to a column of $\bSigma_{R_\tau}$. In particular, when only outcome $j$ is affected, the joint growth rate is strictly larger if $(\bSigma_{R_\tau})_{jk}\neq0$ for some $k\neq j$: the unaffected outcomes then provide information about the estimation noise in the affected outcome.
\end{remark}

\section{Discussion}
\label{sec:discussion}
We have extended anytime-valid inference for linear models from independent scalar outcomes to clustered outcomes. The first contribution is the general right-Haar Bayes factor for a matrix of coefficients of interest, nuisance regressors and arbitrary fixed matrix-normal prior precision. The formula extends the known scalar and two-sample Bayes factors and differs from the proper-prior Bayes factors behind the classical determinant-ratio tests. For the Gaussian multivariate linear model it is an exact $e$-value at every sample size and a test martingale in the invariant filtration, and it is a closed-form function of the OLS coefficient matrix and the residual covariance, so that Hotelling's $T^2$, the MANOVA tests and the fixed-sample confidence ellipsoid all acquire sequential analogues. For clustered regression with shared coefficients and a common within-cluster design the extension goes through a compression to per-cluster summaries, which shows that the cluster-robust Wald statistic is a Hotelling statistic and fixes the degrees of freedom at $G-1$. For general clustered least-squares projections the same statistic is an asymptotic $e$-process, and its inversion an asymptotic confidence sequence, with the exact result fixing the finite-sample form of the statistic. In randomized experiments with several outcomes this gives joint tests, confidence ellipsoids and simultaneous confidence sequences for the vector of average treatment effects from the least-squares estimates and cluster-robust standard errors of a fixed-sample analysis; any setting in which the estimand is a linear contrast of a clustered population least-squares projection is another application.

The growth analysis accompanies each procedure: Gaussian-model results for $E_G$ and nonparametric results for $\ECR_G$. Power one and the first-order expected stopping-time limits connect sequential detection to the growth rate. Under stronger regularity, direct comparisons of expected log evidence give the asymptotically optimal fixed mixture precision within the specified family and the limiting loss from another choice. These are frequentist comparisons at a specified fixed alternative.

Several questions remain. The first concerns optimality for $q\geq2$. Nonamenability of $\GL(q)$ prevents a direct appeal to the scalar optimality theorem; the Bayes factor is log-optimal among invariant $e$-variables, but Appendix~\ref{app:grow} proves that it is not GROW among all $e$-variables in the nontrivial cases, because the triangular-subgroup Bayes factors grow faster, at the price of depending on the ordering of the outcome coordinates; Appendix~\ref{app:grow} gives the reasons for working with $E_G$ nonetheless. The unrestricted minimax status of the triangular statistics is left for a separate analysis. The second concerns the scope of the exact results. Theorem~\ref{thm:main} covers regressors common to the outcome coordinates, cluster-level covariates included, and Proposition~\ref{prop:common} a common within-cluster design with shared coefficients. Heterogeneous within-cluster designs and unequal cluster sizes, for which no i.i.d.\ Gaussian summary reduction is available, are covered only asymptotically; a general exact reduction for them under unknown covariance remains outside the results proved here.

\subsection*{Acknowledgements}
The author developed the theoretical results, building on his earlier work \citep{lindon2026anytime}. Claude (Anthropic) and ChatGPT (OpenAI) assisted with drafting and editing text, restructuring the manuscript, and providing technical review and feedback. The author takes full responsibility for the content.

\subsection*{Disclosure Statement}
The author is employed by Netflix, Inc.

\bibliographystyle{agsm}
\bibliography{theory}

\appendix
\newpage
\begin{center}
{\large\bf SUPPLEMENTARY MATERIAL}
\end{center}

\section{Proofs for Section~\ref{sec:exact}}
\label{app:exact}

\subsection{The Bayes factor}
\label{app:bf}
Throughout this subsection $G$ is fixed and the subscript is dropped.

\begin{lemma}[Decomposition]\label{lem:decomp}
For all $\bB$ and $\bDelta$, with $\tilde\bB_{\bDelta}=(\bX'\bX)^{-1}\bX'(\bY-\bZ\bDelta)$,
\[
(\bY-\bX\bB-\bZ\bDelta)'(\bY-\bX\bB-\bZ\bDelta)
=(\bB-\tilde\bB_{\bDelta})'\bX'\bX(\bB-\tilde\bB_{\bDelta})+(\tY-\tZ\bDelta)'(\tY-\tZ\bDelta).
\]
\end{lemma}
\begin{proof}
Let $\bm R=\bY-\bX\bB-\bZ\bDelta$. Then $\bm R'\bm R=\bm R'\bP_{\bX}\bm R+\bm R'(\bI-\bP_{\bX})\bm R$, with $\bP_{\bX}\bm R=\bX(\tilde\bB_{\bDelta}-\bB)$ and $(\bI-\bP_{\bX})\bm R=\tY-\tZ\bDelta$.
\end{proof}

\begin{lemma}[Completing the square]\label{lem:square}
With $\tilde\bDelta=(\bPhi+\tZ'\tZ)^{-1}\tZ'\tY$ and $\bS_1$ as in \eqref{eq:S01},
\[
(\tY-\tZ\bDelta)'(\tY-\tZ\bDelta)+\bDelta'\bPhi\bDelta
=(\bDelta-\tilde\bDelta)'(\bPhi+\tZ'\tZ)(\bDelta-\tilde\bDelta)+\bS_1 .
\]
If $\bW$ has full column rank, then $\tZ'\tY=\tZ'\tZ\hDelta$, $\bS_0=\nu\hSigma+\hDelta'\tZ'\tZ\hDelta$ and $\bS_1=\nu\hSigma+\hDelta'\bM\hDelta$.
\end{lemma}
\begin{proof}
Both sides of the display equal $\tY'\tY-\tY'\tZ\bDelta-\bDelta'\tZ'\tY+\bDelta'(\bPhi+\tZ'\tZ)\bDelta$. For the second statement, $\bP_{\bW}-\bP_{\bX}=\bP_{\tZ}$, so $\tY=\tZ\hDelta+(\bI-\bP_{\bW})\bY$ with orthogonal summands, which gives $\tZ'\tY=\tZ'\tZ\hDelta$ and $\tY'\tY=\hDelta'\tZ'\tZ\hDelta+\nu\hSigma$; substitute into \eqref{eq:S01}.
\end{proof}

\begin{lemma}[Matrix-normal integral]\label{lem:mn}
For positive definite $\bOmega\in\R^{k\times k}$ and $\bSigma\in\R^{q\times q}$, and any $\bM_0\in\R^{k\times q}$,
\[
\int_{\R^{k\times q}}\exp\!\Big\{-\tfrac12\tr\big[\bSigma^{-1}(\bM-\bM_0)'\bOmega(\bM-\bM_0)\big]\Big\}\dd\bM=(2\pi)^{kq/2}|\bOmega|^{-q/2}|\bSigma|^{k/2}.
\]
\end{lemma}
\begin{proof}
The integrand is the kernel of the $\MN_{k\times q}(\bM_0,\bOmega^{-1},\bSigma)$ density. Equivalently, $\vect(\bM)$ is Gaussian with mean $\vect(\bM_0)$ and covariance $\bSigma\otimes\bOmega^{-1}$, whose determinant is $|\bSigma|^{k}|\bOmega|^{-q}$.
\end{proof}

\begin{lemma}[Inverse-Wishart integral]\label{lem:iw}
For positive definite $\bA\in\R^{q\times q}$ and $\nu>q-1$,
\[
\int_{\bSigma\succ0}|\bSigma|^{-(\nu+q+1)/2}\exp\!\big\{-\tfrac12\tr(\bSigma^{-1}\bA)\big\}\dd\bSigma=2^{\nu q/2}\,\Gamma_q\!\big(\tfrac\nu2\big)\,|\bA|^{-\nu/2},
\]
where $\Gamma_q$ is the multivariate gamma function and $\dd\bSigma$ is Lebesgue measure on the $q(q+1)/2$ free entries.
\end{lemma}
\begin{proof}
This is the statement that the inverse-Wishart density $\mathrm{IW}(\bA,\nu)$,
\[
\frac{|\bA|^{\nu/2}}{2^{\nu q/2}\,\Gamma_q(\nu/2)}\,|\bSigma|^{-(\nu+q+1)/2}\exp\!\big\{-\tfrac12\tr(\bSigma^{-1}\bA)\big\},
\]
integrates to one; equivalently it is the Wishart normalizing constant after the change of variables $\bSigma\mapsto\bSigma^{-1}$, whose Jacobian is $|\bSigma|^{-(q+1)}$ \citep[Corollary~7.2.4 and Theorem~7.7.1]{anderson2003introduction}.
\end{proof}
\noindent In the one-sample case the matrix determinant lemma $|\bW+c\,\bu \bu'|=|\bW|(1+c\,\bu'\bW^{-1}\bu)$ converts the determinant ratio into a function of $T^2$; in general it is replaced by Sylvester's identity, $|\bI_q+\bA\bC|=|\bI_d+\bC\bA|$ for $\bA\in\R^{q\times d}$ and $\bC\in\R^{d\times q}$.

\begin{proof}[Proof of Theorem~\ref{thm:main}\textup{(i)}]
\emph{Step 1: integrate out $\bB$.} The likelihood is
\[
L(\bB,\bDelta,\bSigma)=(2\pi)^{-Gq/2}|\bSigma|^{-G/2}\exp\!\Big\{-\tfrac12\tr\big[\bSigma^{-1}(\bY-\bX\bB-\bZ\bDelta)'(\bY-\bX\bB-\bZ\bDelta)\big]\Big\}.
\]
By Lemma~\ref{lem:decomp} and Lemma~\ref{lem:mn} with $k=p$ and $\bOmega=\bX'\bX$, and writing $c_{\bX}=|\bX'\bX|^{-q/2}$,
\begin{equation}\label{eq:afterB}
\int L\dd\bB=c_{\bX}\,(2\pi)^{-(G-p)q/2}\,|\bSigma|^{-(G-p)/2}\exp\!\Big\{-\tfrac12\tr\big[\bSigma^{-1}(\tY-\tZ\bDelta)'(\tY-\tZ\bDelta)\big]\Big\}.
\end{equation}
Under $H_0$ this is \eqref{eq:afterB} at $\bDelta=0$, whose exponent is $-\tfrac12\tr(\bSigma^{-1}\bS_0)$.

\emph{Step 2: integrate out $\bDelta$ under $H_1$.} Multiply \eqref{eq:afterB} by the density of \eqref{eq:mixture} and use Lemma~\ref{lem:square} and then Lemma~\ref{lem:mn} with $k=d$, $\bOmega=\bPhi+\tZ'\tZ$. The powers of $2\pi$ and of $|\bSigma|$ contributed by the prior and by the Gaussian integral cancel, leaving
\[
c_{\bX}\left(\frac{|\bPhi|}{|\bPhi+\tZ'\tZ|}\right)^{q/2}(2\pi)^{-(G-p)q/2}\,|\bSigma|^{-(G-p)/2}\exp\!\big\{-\tfrac12\tr(\bSigma^{-1}\bS_1)\big\}.
\]
Integrating out the coefficients of interest leaves the null form with $\bS_0$ replaced by $\bS_1$ and one extra constant.

\emph{Step 3: integrate out $\bSigma$.} Multiplied by the prior density $|\bSigma|^{-(q+1)/2}$ of \eqref{eq:prior}, the null expression and the expression above are, as functions of $\bSigma$, inverse-Wishart kernels with $G-p$ degrees of freedom and scale matrices $\bS_0$ and $\bS_1$, so each integral is the reciprocal of an inverse-Wishart normalizing constant, given by Lemma~\ref{lem:iw} with $\nu=G-p$. The lemma needs $G-p>q-1$, which holds because $G\geq p+q$, and positive definite scale matrices. For the latter, write $\bS_0=\bU'\bU$ and $\bS_1=\bU'\bK^{-1}\bU$ with $\bU\in\R^{(G-p)\times q}$ and $\bK\succ0$ as in Appendix~\ref{app:evalue}. Since $\bU$ is Gaussian with nonsingular column covariance and $G-p\geq q$, it has rank $q$ almost surely, and then both matrices are positive definite. Lemma~\ref{lem:iw} therefore gives marginal likelihoods proportional, with the same constant, to $|\bS_0|^{-(G-p)/2}$ and $(|\bPhi|/|\bPhi+\tZ'\tZ|)^{q/2}|\bS_1|^{-(G-p)/2}$. Their ratio is \eqref{eq:E}; the arbitrary constant of the improper prior cancels because the prior is common to both hypotheses.

\emph{Step 4: OLS form.} By Lemma~\ref{lem:square} and Sylvester's identity, for $\bm N\in\{\tZ'\tZ,\bM\}$,
$|\nu\hSigma+\hDelta'\bm N\hDelta|=|\nu\hSigma|\,|\bI_d+\bm N\hDelta(\nu\hSigma)^{-1}\hDelta'|=|\nu\hSigma|\,|\bI_d+\bm N\bT/\nu|$, and $G-p=\nu+d$. This gives \eqref{eq:Eols}.
\end{proof}

\subsection{The \texorpdfstring{$e$}{e}-value and martingale properties}
\label{app:evalue}

The proof of Theorem~\ref{thm:main}(ii) is a direct verification. After whitening by the true $\bSigma$, the orientation of the residual matrix is uniform on the Stiefel manifold under $P_{\bB,\bSigma}$ and has the matrix angular central Gaussian law of \citet{chikuse1990matrix} under $Q_{\bB,\bSigma}$; the ratio of the two densities is $E_G$, and every invariant function is a function of the orientation.

\paragraph{Canonical form.} Fix $G\geq p+q$, let $N=G-p$, and let $\bO_G\in\R^{N\times G}$ satisfy $\bO_G\bO_G'=\bI_N$ and $\bO_G'\bO_G=\bI_G-\bP_{\bX_G}$. Put $\bU=\bO_G\bY_G\in\R^{N\times q}$, $\bV=\bO_G\bZ_G\in\R^{N\times d}$ and $\bK=\bI_N+\bV\bPhi^{-1}\bV'$. Then $\tY'\tY=\bU'\bU$, $\tY'\tZ=\bU'\bV$, $\tZ'\tZ=\bV'\bV$, and by the Woodbury identity $\bK^{-1}=\bI_N-\bV(\bPhi+\bV'\bV)^{-1}\bV'$ and $|\bK|=|\bPhi+\bV'\bV|/|\bPhi|$. Therefore
\begin{equation}\label{eq:hK}
\bS_{0,G}=\bU'\bU,\qquad \bS_{1,G}=\bU'\bK^{-1}\bU,\qquad
E_G=h_{\bK}(\bU)\coloneqq|\bK|^{-q/2}\left(\frac{|\bU'\bK^{-1}\bU|}{|\bU'\bU|}\right)^{-N/2}.
\end{equation}
Since $\bO_G\bX_G=0$, $\bU\sim\MN(0,\bI_N,\bSigma)$ under $P_{\bB,\bSigma}$. Under $Q_{\bB,\bSigma}$, $\bU=\bV\bDelta+\bO_G\bE_G$ is a sum of independent matrix normals with the same column covariance, so $\bU\sim\MN(0,\bK,\bSigma)$; this is where tying the scale of \eqref{eq:mixture} to $\bSigma$ is used. Since $N\geq q$ and $\bSigma$ is nonsingular, $\bU$ has rank $q$ almost surely under both laws. An invariant $f$ satisfies $f(\bY_G)=f(\tY_G)=f(\bO_G'\bU)\eqqcolon\varphi(\bU)$, by invariance under the translations $\bY_G\mapsto\bY_G+\bX_G\bA$, and $\varphi(\bU\bC')=\varphi(\bU)$ for all $\bC\in\GL(q)$, by invariance under $\bY_G\mapsto\bY_G\bC'$. Note also $h_{\bK}(\bU\bC')=h_{\bK}(\bU)$.

\begin{proof}[Proof of Theorem~\ref{thm:main}\textup{(ii)}]
Fix $(\bB,\bSigma)$ and a bounded invariant $f$. Whiten the columns of $\bU$ and then orthonormalize them:
\[
\bU_*=\bU\bSigma^{-1/2},\qquad \bU_{\mathrm{pol}}=\bU_*(\bU_*'\bU_*)^{-1/2},
\]
with symmetric square roots. Both steps multiply $\bU$ on the right by an invertible matrix, so
\[
\varphi(\bU_{\mathrm{pol}})=\varphi(\bU),\qquad h_{\bK}(\bU_{\mathrm{pol}})=h_{\bK}(\bU)=E_G .
\]
Whitening is used only in the proof; the identities above ensure that the resulting expectation identity involves observable quantities.

Suppose first that $N>q$. Let $\nu$ be the uniform probability measure on the Stiefel manifold $\{\bH\in\R^{N\times q}:\bH'\bH=\bI_q\}$. Under $P_{\bB,\bSigma}$, $\bU_*\sim\MN(0,\bI_N,\bI_q)$; under $Q_{\bB,\bSigma}$, $\bU_*\sim\MN(0,\bK,\bI_q)$, that is, its columns are i.i.d.\ $\Nor_N(0,\bK)$. By Theorem~2.2 of \citet{chikuse1990matrix}, the orientation of a matrix with i.i.d.\ $\Nor_N(0,\bK)$ columns has density $g_{\bK}(\bH)=|\bK|^{-q/2}|\bH'\bK^{-1}\bH|^{-N/2}$ with respect to $\nu$, and $\bK=\bI_N$ gives the uniform law. Hence $\bU_{\mathrm{pol}}$ has density $1$ under $P_{\bB,\bSigma}$ and density $g_{\bK}(\bU_{\mathrm{pol}})$ under $Q_{\bB,\bSigma}$, both with respect to $\nu$. Since $\bH'\bH=\bI_q$ on the Stiefel manifold, $g_{\bK}(\bH)=h_{\bK}(\bH)$ there, and therefore
\begin{align*}
\E_{Q_{\bB,\bSigma}}\big[f(\bY_G)\big]
&=\int\varphi(\bH)\,g_{\bK}(\bH)\,\nu(\mathrm d\bH)
=\E_{P_{\bB,\bSigma}}\big[\varphi(\bU_{\mathrm{pol}})\,h_{\bK}(\bU_{\mathrm{pol}})\big]
=\E_{P_{\bB,\bSigma}}\big[f(\bY_G)\,E_G\big].
\end{align*}
When $N=q$, all nonsingular $\bU$ differ by right multiplication by an invertible matrix, so $\varphi(\bU)$ is almost surely constant, and $h_{\bK}(\bU)=|\bK|^{-q/2}|\bK|^{q/2}=1$; the same identity follows. Taking $f\equiv1$ gives $\E_{P_{\bB,\bSigma}}[E_G]=1$.
\end{proof}

\begin{proof}[Proof of Theorem~\ref{thm:main}\textup{(iii)}]
If $F\subset\R^{G\times q}$ is invariant under $\bY_G\mapsto\bY_G\bC'+\bX_G\bA$, then $F\times\R^{1\times q}$ is invariant under $\bY_{G+1}\mapsto\bY_{G+1}\bC'+\bX_{G+1}\bA$, because the group acts row by row; hence $\cI_G\subset\cI_{G+1}$. $E_G$ is $\cI_G$-measurable because $h_{\bK}$ is invariant. For $F\in\cI_G$, applying (ii) at times $G+1$ and $G$ gives $\E_P[E_{G+1}\mathbf 1_F]=Q(F)=\E_P[E_G\mathbf 1_F]$, where $P$ and $Q$ are the laws of the whole sequence and $Q(F)$ is the probability of the event $\{\bY_G\in F\}$; so $\E_P[E_{G+1}\mid\cI_G]=E_G$. At $G=p+q$ we have $N=q$ and $E_{p+q}=1$ as shown above. The last claim is Ville's inequality.
\end{proof}

\subsection{Proof of Corollary~\ref{cor:cs}}
The first claim is Theorem~\ref{thm:main} applied to $\bY_G-\bZ_G\bDelta_0$, which satisfies \eqref{eq:model} with $\bDelta-\bDelta_0$ in place of $\bDelta$, plus Ville's inequality; the $p$-value statement is the same inequality. For $d=1$, by \eqref{eq:d1} with $t=T_G^2(\bdelta_0)/\nu_G$, where $T_G^2(\bdelta_0)$ is the statistic of Corollary~\ref{cor:special}(c) computed from $\bY_G-\bZ_G\bdelta_0'$, $E_G^{\bdelta_0}\leq\alpha^{-1}$ iff $(1+t)/(1+a_Gt)\leq u_G^{-1}$ iff $t(u_G-a_G)\leq1-u_G$, and $u_G<1$. \qed

\subsection{Proof of Corollary~\ref{cor:special}}
\label{app:special}
(a) With $q=1$, $\hSigma_G=s_G^2$, $\hDelta_G=\hat{\bdelta}_G\in\R^{d}$ (as a column) and $\bT_G=\hat{\bdelta}_G\hat{\bdelta}_G'/s_G^2=\hat{\bxi}_G\hat{\bxi}_G'$. Sylvester's identity gives $|\bI_d+\bm N\hat{\bxi}\hat{\bxi}'/\nu|=1+\hat{\bxi}'\bm N\hat{\bxi}/\nu$ for $\bm N\in\{\bM_G,\tZ_G'\tZ_G\}$, and $\bM_G=\tZ_G'\tZ_G-\tZ_G'\tZ_G(\bPhi+\tZ_G'\tZ_G)^{-1}\tZ_G'\tZ_G$ is the matrix in the numerator of Theorem~4.1 of \citet{lindon2026anytime}.

(b) With $p=0$ there is no projection, $\tZ_G=\mathbf 1_G$, $\tZ_G'\tZ_G=G$, $\bM_G=(\phi^{-1}+G^{-1})^{-1}=\phi G/(\phi+G)=aG$, $\nu_G=G-1$, $\hDelta_G'=\bar \by_G$ and $\nu_G\hSigma_G=\sum_g(\by_g-\bar \by_G)(\by_g-\bar \by_G)'=(G-1)\bS$. Hence $\bT_G=\bar \by_G'\bS^{-1}\bar \by_G=T^2/G$, $|\bI_1+\bM_G\bT_G/\nu_G|=1+aT^2/(G-1)$, $|\bI_1+G\bT_G/\nu_G|=1+T^2/(G-1)$, $(|\bPhi|/|\bPhi+G|)^{q/2}=a^{q/2}$ and $(\nu_G+d)/2=G/2$, which is \eqref{eq:hotelling}. Monotonicity in $T^2$: $t\mapsto(1+t)/(1+at)$ is increasing for $a<1$.

(c) With $d=1$, $\tZ_G'\tZ_G=\|\tZ_G\|_2^2$, $\bM_G=\|\tZ_G\|_2^2a_G$ and $\bT_G=\hat\bdelta_G'\hSigma_G^{-1}\hat\bdelta_G$, so $|\bI_1+\bM_G\bT_G/\nu_G|=1+a_GT_G^2/\nu_G$ and $|\bI_1+\tZ_G'\tZ_G\bT_G/\nu_G|=1+T_G^2/\nu_G$, with $(\nu_G+1)/2$ as the exponent. The null law: $\|\tZ_G\|_2\hat\bdelta_G\sim\Nor_q(0,\bSigma)$ independently of $\nu_G\hSigma_G\sim\mathrm{Wishart}_q(\nu_G,\bSigma)$, so $T_G^2$ has Hotelling's $T^2(q,\nu_G)$ distribution \citep[Theorem~5.2.2]{anderson2003introduction}.

(d) With $\bPhi=\frac gG\tZ_G'\tZ_G$, $\bPhi+\tZ_G'\tZ_G=\frac{g+G}{G}\tZ_G'\tZ_G$, so the first factor of \eqref{eq:Eols} is $(g/(g+G))^{dq/2}$, and $\bM_G=(\frac Gg(\tZ_G'\tZ_G)^{-1}+(\tZ_G'\tZ_G)^{-1})^{-1}=\frac{g}{g+G}\tZ_G'\tZ_G$. By Sylvester's identity, for $c\in\{\frac g{g+G},1\}$, $|\bI_d+c\,\tZ_G'\tZ_G\hDelta_G\hSigma_G^{-1}\hDelta_G'/\nu_G|=|\bI_q+c\,(\nu_G\hSigma_G)^{-1}\bH_G|=\prod_{j}(1+c\,\ell_j)$, the product running over the eigenvalues of $\bH_G(\nu_G\hSigma_G)^{-1}$, of which at most $s=\min(d,q)$ are nonzero since $\rank\bH_G\leq d$. With $(\nu_G+d)/2=(G-p)/2$ this is \eqref{eq:gprior}, and $\prod_j(1+\ell_j)^{-1}=|\nu_G\hSigma_G|/|\nu_G\hSigma_G+\bH_G|$ is Wilks' $\Lambda$. The martingale statement: \eqref{eq:gprior} is \eqref{eq:E} with a prior that depends on $G$ through $\tZ_G'\tZ_G/G$; Theorem~\ref{thm:main}(ii) applies at each fixed $G$, and (iii) gives the claimed martingale guarantee when the prior is the same at every $G$. \qed

\subsection{Proof of Proposition~\ref{prop:growth}}
\label{app:growthproof}
Condition on the fixed design (or on an independent random design). Since $\bW_G'\bW_G=O(G)$, Gaussian tail bounds and Borel--Cantelli give $\bW_G'\bE_G/G\to0$ almost surely, entrywise. The strong law also gives $\bE_G'\bE_G/G\to\bSigma$. Thus OLS is strongly consistent, $\nu_G\hSigma_G/G\to\bSigma$, and Lemma~\ref{lem:square} gives $\bS_{0,G}/G\to\bSigma+\bDelta'\bOmega_{\tZ}\bDelta$ almost surely. These are the multivariate counterparts of the consistency argument in \citet{lindon2026anytime}. Since $0\preceq\bM_G\preceq\bPhi$, also $\frac1G\bS_{1,G}=\frac{\nu_G}{G}\hSigma_G+\frac1G\hDelta_G'\bM_G\hDelta_G\to\bSigma$, and $\frac1G\log(|\bPhi+\tZ_G'\tZ_G|/|\bPhi|)\to0$ because the determinant grows polynomially in $G$. Hence, from \eqref{eq:E},
\[
\frac{\log E_G}{G}=\frac{G-p}{2G}\log\frac{|\bS_{0,G}/G|}{|\bS_{1,G}/G|}+o(1)\ \overset{a.s.}{\longrightarrow}\ \frac12\log\big|\bI_q+\bSigma^{-1}\bDelta'\bOmega_{\tZ}\bDelta\big|,
\]
and Sylvester's identity gives the $d\times d$ determinant form, and $|\bSigma_0|=|\bSigma|\,|\bI_q+\bSigma^{-1}\bDelta'\bOmega_{\tZ}\bDelta|$ gives the determinant-ratio form. The eigenvalues of $\bOmega_{\tZ}\bDelta\bSigma^{-1}\bDelta'$ are nonnegative, and not all zero when $\bDelta\neq0$ because $\bOmega_{\tZ}\succ0$ and $\bSigma\succ0$; hence the limit is positive iff $\bDelta\neq0$. Power one follows since $\log E_G\to\infty$ almost surely exceeds $\log\alpha^{-1}$ in finite time.

\paragraph{Stopping time.}
Write $r=r_{\bDelta}>0$ and $b=\log(1/\alpha)$. For any $0<\eta<r$, almost surely
$(r-\eta)G\leq\log E_G\leq(r+\eta)G$ for all sufficiently large $G$.
The maximum over the finitely many earlier values is finite, so for all sufficiently large $b$,
\[
\frac{b}{r+\eta}\leq\tau_{e^{-b}}\leq
\left\lceil\frac{b}{r-\eta}\right\rceil .
\]
Letting $b\to\infty$ and then $\eta\downarrow0$ proves the almost-sure limit.

\paragraph{Expected stopping time.}
To pass from the almost-sure limit to expectations, we must control the contribution of unusually late crossings. We do this by showing that the lower-tail deviations of $G^{-1}\log E_G$ from $r$ are summable. For a deterministic design, Gaussian concentration gives the stronger bound: for every $\eta>0$ there are constants $C_\eta,c_\eta>0$ and a deterministic $G_\eta$ such that
\begin{equation}\label{eq:growth_exp_tail}
\Prob_{\bB,\bDelta,\bSigma}\!\left(\left|\frac{\log E_G}{G}-r\right|>\eta\right)
\leq C_\eta e^{-c_\eta G},\qquad G\geq G_\eta .
\end{equation}
To verify this, put $\bOmega_{\bW,G}=\bW_G'\bW_G/G$, $\bC_G=\bW_G'\bE_G/G$ and $\bV_G=\bE_G'\bE_G/G$. Since $\bOmega_{\bW,G}\to\bOmega_{\bW}\succ0$, each entry of $\bC_G$ is centered Gaussian with variance $O(G^{-1})$. Each entry of $\bV_G-\bSigma$ is an average of centered products of Gaussian coordinates, whose moment-generating functions are finite near zero. Gaussian tail bounds, the exponential Markov inequality and a union bound therefore give, for every fixed $\epsilon>0$,
\[
\Prob\big(\|\bC_G\|+\|\bV_G-\bSigma\|>\epsilon\big)
\leq C_\epsilon e^{-c_\epsilon G}
\]
for all sufficiently large $G$, where $\|\cdot\|$ is the spectral norm. The OLS identities give
\[
\frac{\nu_G\hSigma_G}{G}=\bV_G-\bC_G'\bOmega_{\bW,G}^{-1}\bC_G,
\qquad
\hDelta_G-\bDelta=\text{the last $d$ rows of }\bOmega_{\bW,G}^{-1}\bC_G,
\]
and Lemma~\ref{lem:square} gives
\[
\frac{\bS_{0,G}}G=\frac{\nu_G\hSigma_G}G+
\hDelta_G'\frac{\tZ_G'\tZ_G}G\hDelta_G,\qquad
\frac{\bS_{1,G}}G=\frac{\nu_G\hSigma_G}G+
\frac1G\hDelta_G'\bM_G\hDelta_G .
\]
For sufficiently small $\epsilon$ and large $G$, these identities keep $\bS_{0,G}/G$ and $\bS_{1,G}/G$ in fixed neighborhoods of $\bSigma_0$ and $\bSigma$, respectively, on the event $\|\bC_G\|+\|\bV_G-\bSigma\|\leq\epsilon$. Choose these neighborhoods with compact closures contained in the positive definite cone. Their eigenvalues are then bounded away from zero, and the log determinant is uniformly continuous on their closures. The deterministic design approximation, the bounded prior contribution $0\preceq\bM_G\preceq\bPhi$, and the $O(\log G/G)$ prior-determinant term can all be controlled by taking $G$ sufficiently large. Consequently, for every $\eta>0$ we can choose $\epsilon>0$ and a deterministic $G_\eta$ so that, for $G\geq G_\eta$,
\[
\|\bC_G\|+\|\bV_G-\bSigma\|\leq\epsilon
\quad\Longrightarrow\quad
\left|\frac{\log E_G}{G}-r\right|\leq\eta.
\]
Taking complements and applying the preceding exponential bound proves \eqref{eq:growth_exp_tail}. This requires only eventual closeness of the deterministic design matrices, with no specified convergence rate.

We use the first-moment tail-sum argument of \citet[arXiv v1, proof of Theorem~4.1, equations~(33)--(35)]{fellouris2017multichannel}; an earlier summable-tail argument appears in \citet[proof of Lemma~3, equation~(15)]{hagwood1982expansion}. The same passage-time argument is used for portfolio-based $e$-processes by \citet[arXiv v2, Appendix~A.4]{waudbysmith2026instance}. Only summability of the lower-tail deviations is needed. Fix $0<\eta<r$ and put $m_b=\max\{p+q,G_\eta,\lceil b/(r-\eta)\rceil\}$. For $G\geq m_b$, the event $\{\tau_{e^{-b}}>G\}$ implies $\log E_G<b\leq(r-\eta)G$. Thus
\begin{equation}\label{eq:stopping_tail_sum}
\E[\tau_{e^{-b}}]
=\sum_{G=0}^\infty\Prob(\tau_{e^{-b}}>G)
\leq m_b+\sum_{G=m_b}^\infty
\Prob\!\left(\frac{\log E_G}{G}<r-\eta\right).
\end{equation}
The series is finite by \eqref{eq:growth_exp_tail}, and its tail tends to zero as $b\to\infty$. Combining this upper bound with Fatou's lemma and the almost-sure limit gives
\[
\frac1r\leq\liminf_{b\to\infty}\frac{\E[\tau_{e^{-b}}]}b
\leq\limsup_{b\to\infty}\frac{\E[\tau_{e^{-b}}]}b\leq\frac1{r-\eta}.
\]
Letting $\eta\downarrow0$ proves \eqref{eq:expected_stopping}. For the $L^1$ conclusion, apply Scheff\'e's lemma. Explicitly, nonnegativity gives
\[
\E\left|\frac{\tau_{e^{-b}}}{b}-\frac1r\right|
=\frac{\E[\tau_{e^{-b}}]}{b}+\frac1r
-2\E\left[\min\left\{\frac{\tau_{e^{-b}}}{b},\frac1r\right\}\right]
\longrightarrow0,
\]
because the first term tends to $1/r$ and the expectation of the minimum tends to $1/r$ by bounded convergence. For an independent random design, the same argument gives the expectation and $L^1$ limits conditional on almost every design sequence satisfying the Gram-matrix assumption. Unconditional limits require the additional tail control verified next.

For bounded i.i.d. design rows independent of the errors, $\bOmega_{\bW,G}$ itself has exponentially small fixed-size deviations from $\bOmega_{\bW}$. The entries of $\bw_g\beps_g'$ are centered i.i.d. variables with moment-generating functions finite near zero, since $\bw_g$ is bounded and $\beps_g$ is Gaussian. Exponential Markov bounds therefore give the same joint deviation control for $(\bOmega_{\bW,G},\bC_G,\bV_G)$ unconditionally. On this concentration event, $\bOmega_{\bW,G}\succ0$ for $G\geq G_\eta$, so $\bW_G$ and $\bX_G$ have full column rank; samples before nuisance rank is attained are covered by the same exponentially small exceptional probability. The preceding continuity and tail-sum arguments therefore apply to the extended process, proving the unconditional result.

\paragraph{KL interpretation.}
For a fixed design with $\bX_G$ of full column rank, let
\[
\begin{split}
K_G(\widetilde{\bB},\bV)=\frac1G\sum_{g=1}^G
D_{\mathrm{KL}}\!\left(
\Nor_q(\bB'\bx_g+\bDelta'\bz_g,\bSigma)
\,\middle\|\,\Nor_q(\widetilde{\bB}'\bx_g,\bV)\right).
\end{split}
\]
Set $\bm\Pi_G=(\bX_G'\bX_G)^{-1}\bX_G'\bZ_G$, $\bB_{0,G}=\bB+\bm\Pi_G\bDelta$ and $\bSigma_{0,G}=\bSigma+\bDelta'(\tZ_G'\tZ_G/G)\bDelta$. Orthogonality of the sample residualized design gives
\[
\begin{split}
2K_G(\widetilde{\bB},\bV)
={}&\log\frac{|\bV|}{|\bSigma|}-q\\
&+\tr\!\left[\bV^{-1}\left\{\bSigma_{0,G}
+(\widetilde{\bB}-\bB_{0,G})'\frac{\bX_G'\bX_G}{G}(\widetilde{\bB}-\bB_{0,G})\right\}\right].
\end{split}
\]
For each $\bV\succ0$, the last quadratic term is minimized at $\widetilde{\bB}=\bB_{0,G}$; the remaining expression is minimized at $\bV=\bSigma_{0,G}$. Consequently
\[
\min_{\widetilde{\bB},\,\bV\succ0}K_G(\widetilde{\bB},\bV)
=\frac12\log\frac{|\bSigma_{0,G}|}{|\bSigma|}
\longrightarrow r_{\bDelta}.
\]
The same identity applies conditionally on an independent random design.

For the reference-law formulation, fix $\bOmega_{\bW}\succ0$ and any design law with $\E[\bw_g\bw_g']=\bOmega_{\bW}$, and define $Q^\star$ and $\cP_0$ as in Section~\ref{sec:growth}. Population projection orthogonality, $\E[\bx_g\widetilde\bz_g']=0$, and $\E[\widetilde\bz_g\widetilde\bz_g']=\bOmega_{\tZ}$ give, for the null law $P$ with parameters $(\widetilde{\bB},\bV)$,
\[
\begin{split}
2D_{\mathrm{KL}}(Q^\star\,\|\,P)
={}&\log\frac{|\bV|}{|\bSigma|}-q\\
&+\tr\!\left[\bV^{-1}\left\{\bSigma_0
+(\widetilde{\bB}-\bB_0)'\bOmega_{\bX}(\widetilde{\bB}-\bB_0)\right\}\right].
\end{split}
\]
For each $\bV\succ0$, the quadratic term is minimized at $\widetilde{\bB}=\bB_0$. The remaining covariance term is minimized at $\bV=\bSigma_0$: its excess over the value at $\bSigma_0$ is
\[
\tr(\bV^{-1}\bSigma_0)-\log|\bV^{-1}\bSigma_0|-q\geq0,
\]
with equality only when $\bV=\bSigma_0$. Since $\bOmega_{\bX}\succ0$ and $\bSigma_0\succ0$, the minimum is attained at the unique null law $P^\star$ with parameters $(\bB_0,\bSigma_0)$, and
\[
\min_{P\in\cP_0}D_{\mathrm{KL}}(Q^\star\,\|\,P)
=D_{\mathrm{KL}}(Q^\star\,\|\,P^\star)
=\frac12\log\frac{|\bSigma_0|}{|\bSigma|}
=r_{\bDelta}.
\]
When $p=0$, omit the nuisance-mean terms and use $\widetilde\bz_g=\bz_g$ and $\tZ_G=\bZ_G$. \qed

\subsection{Proof of Proposition~\ref{prop:log-growth}}
\label{app:log-growth}
Fix $\phi_1,\phi_2>0$. On samples with full design rank, let
\[
\bM_{j,G}=\big\{(\phi_j\bPhi_0)^{-1}+(\tZ_G'\tZ_G)^{-1}\big\}^{-1},
\qquad j=1,2,
\]
and write $\widetilde{\bSigma}_G=\bY_G'(\bI_G-\bP_{\bW_G})\bY_G/G$ for the full-model maximum-likelihood covariance estimator. Lemma~\ref{lem:square} gives
$\bS_{1,G}^{(j)}/G=\widetilde{\bSigma}_G+G^{-1}\hDelta_G'\bM_{j,G}\hDelta_G$.
The null residual determinant cancels in \eqref{eq:E}, so
\begin{equation}\label{eq:precision-ratio}
\begin{split}
\log\frac{E_G^{(\phi_1)}}{E_G^{(\phi_2)}}
={}&\frac{dq}{2}\log\frac{\phi_1}{\phi_2}
-\frac q2\log\frac{|\phi_1\bPhi_0+\tZ_G'\tZ_G|}
{|\phi_2\bPhi_0+\tZ_G'\tZ_G|}\\
&-\frac{G-p}{2}\log
\frac{|\widetilde{\bSigma}_G+G^{-1}\hDelta_G'\bM_{1,G}\hDelta_G|}
{|\widetilde{\bSigma}_G+G^{-1}\hDelta_G'\bM_{2,G}\hDelta_G|}.
\end{split}
\end{equation}
The consistency limits in Appendix~\ref{app:growthproof} give
$\tZ_G'\tZ_G/G\to\bOmega_{\tZ}\succ0$,
$\hDelta_G\to\bDelta$ and $\widetilde{\bSigma}_G\to\bSigma\succ0$ almost surely.
Thus $\bM_{j,G}\to\phi_j\bPhi_0$ and the second term in \eqref{eq:precision-ratio} tends to zero. Expanding only the remaining determinant ratio to first order gives
\[
\log\frac{E_G^{(\phi_1)}}{E_G^{(\phi_2)}}
\longrightarrow
\frac{dq}{2}\log\frac{\phi_1}{\phi_2}
-\frac{\phi_1-\phi_2}{2}\mathcal I_0
\qquad\text{almost surely}.
\]

To pass to expectations, restrict the empirical moments to a fixed small neighborhood of their population values, on which the design and residual covariance matrices are uniformly nonsingular. The expansion above is uniform there and the log ratio is bounded by a constant independent of $G$. The probability of leaving the neighborhood is at most $C_0e^{-c_0G}$ by Appendix~\ref{app:growthproof}. On every active sample, bounded design rows give
\[
(1+C G)^{-1}\bS_{0,G}\preceq\bS_{1,G}^{(j)}\preceq\bS_{0,G},
\qquad |\log E_G^{(\phi_j)}|\leq C G\log(1+G),
\]
for a constant depending on the two fixed precisions. The latter bound also holds at inactive samples. Bounded convergence on the neighborhood and this bound on its complement therefore yield
\[
\E\!\left[\log\frac{E_G^{(\phi_1)}}{E_G^{(\phi_2)}}\right]
=\frac{dq}{2}\log\frac{\phi_1}{\phi_2}
-\frac{\phi_1-\phi_2}{2}\mathcal I_0+o(1).
\]
The function $(dq/2)\log\phi-\phi\mathcal I_0/2$ is uniquely maximized at $\phi_*=dq/\mathcal I_0$. Taking $(\phi_1,\phi_2)=(\phi_*,\phi)$ proves \eqref{eq:precision-cost}. \qed

\subsection{Common-design reduction and degrees of freedom}
\label{app:clustered}

\begin{proof}[Proof of Proposition~\ref{prop:common}]
(a) Since $\bA_G=G\bW_0'\bW_0$ and $\sum_g\bW_g'\by_g=\bW_0'\sum_g\by_g$, pooled OLS is the average of the $\hat\bgamma_g$. Then $\hat \bs_g=\bW_0'\by_g-\bW_0'\bW_0\hat\bgamma_G=\bW_0'\bW_0(\hat\bgamma_g-\hat\bgamma_G)$, so $\bA_G^{-1}\hat \bs_g=(\hat\bgamma_g-\hat\bgamma_G)/G$ and $\hat \bV_G=\frac{G}{G-1}\frac1{G^2}\sum_g(\hat\bgamma_g-\hat\bgamma_G)(\hat\bgamma_g-\hat\bgamma_G)'=\bS_{\hat\bgamma}/G$. Finally $\bR\hat\bgamma_G=\bar \bb$ and $\bR\hat \bV_G\bR'=\bS_b/G$.

(b) $\hat\bgamma_g\sim\Nor_k(\bgamma,(\bW_0'\bW_0)^{-1}\bW_0'\bSigma_\varepsilon \bW_0(\bW_0'\bW_0)^{-1})$ independently, which gives the law of $\bb_g$; $\bSigma_R\succ0$ when $\bSigma_\varepsilon\succ0$, although $\bSigma_\varepsilon$ itself need not be nonsingular. The identification with $E_G$ is Corollary~\ref{cor:special}(b) with $T^2=Q_G(\btheta_0)$, by (a), and $E_q=1$ by Theorem~\ref{thm:main}(iii). The confidence sequence is Corollary~\ref{cor:cs} with $d=1$, $\nu_G=G-1$ and $\|\tZ_G\|_2^2=G$.
\end{proof}

\paragraph{The alternative degrees-of-freedom convention.}
The rule needs stating because two other conventions suggest themselves for a $q$-dimensional contrast. The fixed-sample $F(q,G-q)$ reference for the cluster-robust Wald statistic, recalled before Proposition~\ref{prop:common}, carries $G-q$; and the scalar rule $\nu=G-p-d$ of \citet{lindon2026anytime}, read with the $q$ coordinates of $\btheta$ as $q$ tested regressors of a single outcome, also gives $G-q$. Either reading would put $Q_G/(G-q)$ in place of $Q_G/(G-1)$ in \eqref{eq:ECR}, with the exponent unchanged. For $q=1$ nothing is at stake. For $q>1$ the two conventions differ, and the exact benchmark decides between them. The $G-q$ in the $F$ reference is a denominator degree of freedom of Hotelling's $T^2$, not the $\nu$ of the Bayes factor: in Proposition~\ref{prop:common}, $\frac{G-q}{q(G-1)}Q_G\sim F(q,G-q)$ exactly, while the Bayes factor uses $Q_G/(G-1)$, the factor $(G-1)/(G-q)$ being the usual Hotelling correction. The correct reading of the scalar rule is $q$ outcomes and $d=1$ tested regressor (the intercept of the compressed vectors), which gives $\nu=G-1$ and exponent $G/2$. The version with $Q_G/(G-q)$ is pointwise larger than \eqref{eq:ECR}, and in the Gaussian model of Proposition~\ref{prop:common} its null expectation converges, as $G\to\infty$ with $\phi$ fixed, to $\E\exp\{\frac{q-1}{2}\,w/(1+w)\}$ with $w\sim\chi^2_q/\phi$, which is $1+q(q-1)/(2\phi)+O(\phi^{-2})$; it is not an $e$-value. Both versions satisfy the almost-sure logarithmic definition of an asymptotic $e$-process, which cannot distinguish them; the exact benchmark does.

\paragraph{Expectation limit.}
Let $\widetilde E_G$ replace $G-1$ by $G-q$ in both occurrences inside the ratio of \eqref{eq:ECR}, retaining exponent $G/2$, and put $r_G=\widetilde E_G/E_G$. Both statistics are invariant, so Theorem~\ref{thm:main}(ii) gives $\E_P\widetilde E_G=\E_Q r_G$. Under the Gaussian mixture alternative, $Q_G/G\to w=\btheta'\bSigma_R^{-1}\btheta$ almost surely and $w\sim\chi_q^2/\phi$. Expanding the log-ratio of the two versions gives
\[
\log r_G\longrightarrow\frac{q-1}{2}\frac{w}{1+w},\qquad
0\leq\log r_G\leq\frac G2\log\frac{G-1}{G-q}.
\]
The upper bound is uniformly bounded for $G\geq2q+1$. Dominated convergence proves the expectation limit; expansion in $1/\phi$ and $\E\chi_q^2=q$ gives $1+q(q-1)/(2\phi)+O(\phi^{-2})$.

\subsection{Mean structures and reductions}
\label{app:mean-structures}

Model \eqref{eq:model} regresses every component of $\by_g$ on the same regressors $(\bx_g,\bz_g)$, each component with its own coefficients. Theorem~\ref{thm:main} rests on this structure. Its proof transforms the outcome vector by $\by_g\mapsto\bC \by_g$ with $\bC\in\GL(q)$, and under \eqref{eq:model} such a transformation only relabels the parameters, $\bB\mapsto\bB\bC'$, $\bDelta\mapsto\bDelta\bC'$ and $\bSigma\mapsto\bC\bSigma\bC'$, because $\bC\bB'\bx_g=(\bB\bC')'\bx_g$. The model and $H_0$ are preserved, and every $\bSigma$ can be reached from every other, which is what removes $\bSigma$ from the null law of the invariants.

The regression that motivates cluster-robust inference has a different structure. There the rows of a cluster have different regressors and share one coefficient vector: $\by_g\sim\Nor(\bX_g\bbeta+\bZ_g\bdelta,\bSigma)$ with $\bX_g\in\R^{q\times p}$, $\bZ_g\in\R^{q\times d}$, for instance $q$ units in a cluster with unit-level covariates and common slopes. The transformation $\by_g\mapsto\bC \by_g$ now mixes rows with different regressors, and $\bC \bX_g\bbeta$ is not of the form $\bX_g\bbeta^\ast$: with $q=2$, $p=1$ and $\bX_g=(a_{g1},a_{g2})'$, the mean $(a_{g1}\beta,a_{g2}\beta)'$ is carried by $\bC=\left(\begin{smallmatrix}1&0\\1&1\end{smallmatrix}\right)$ to $(a_{g1}\beta,(a_{g1}+a_{g2})\beta)'$, which is of the form $(a_{g1}\beta^\ast,a_{g2}\beta^\ast)'$ only when $a_{g1}\beta=0$. Only the subgroup of $\GL(q)$ that preserves the shared-coefficient structure is available, and an invariant reduction under a smaller group can retain dependence on the shape of $\bSigma$. Theorem~\ref{thm:main} therefore does not apply directly. Sections~\ref{sec:clustered} and~\ref{sec:asymptotic} give its common-design Gaussian and general projection treatments. This is the boundary between the multivariate linear model and the seemingly-unrelated-regression structures of \citet{zellner1962efficient}: the exact reduction of Proposition~\ref{prop:common} is available under a common design, while a general exact treatment of \eqref{eq:clustered} with unknown covariance is not claimed. Three routes lead back to \eqref{eq:model} and keep the exact results.
\begin{enumerate}
\item[(a)] \emph{Embedding.} A model with shared coefficients is a submodel of \eqref{eq:model} with the stacked regressors $\bx_g^\ast=\vect(\bX_g')$ and $\bz_g^\ast=\vect(\bZ_g')$, after deleting redundant columns. Its null is contained in the null of the larger model, since $\bdelta=0$ implies $\bDelta^\ast=0$, so Theorem~\ref{thm:main} and Corollary~\ref{cor:cs} applied to the larger model remain exactly valid. The price is $pq$ nuisance regressors and $dq^2$ tested coefficients in place of $p$ and $d$, which is sensible only for small $q$.
\item[(b)] \emph{Fixed linear summaries.} For a fixed $\bM\in\R^{q^\ast\times q}$ of full row rank, $\bM \by_g\sim\Nor(\bM \bX_g\bbeta+\bM \bZ_g\bdelta,\bM\bSigma\bM')$. Two cases return to \eqref{eq:model}. A scalar summary, $q^\ast=1$ and $\bM=\bc'$, gives $\bc'\by_g=(\bX_g'\bc)'\bbeta+(\bZ_g'\bc)'\bdelta+\bc'\beps_g$, the scalar model of \citet{lindon2026anytime} with regressors $\bX_g'\bc$, $\bZ_g'\bc$ and variance $\bc'\bSigma \bc$, whatever the designs $\bX_g$ and $\bZ_g$ are; with $\bc=\mathbf 1_q/q$ this is the regression of the cluster mean $\bar y_g$ on the within-cluster averages $\bar \bx_g=\bX_g'\mathbf 1_q/q$ and $\bar \bz_g=\bZ_g'\mathbf 1_q/q$, with i.i.d.\ homoskedastic Gaussian errors, so the exact $E_G$ of that paper applies verbatim with $G$ the number of clusters. (Under \eqref{eq:model} itself, $\bc'\by_g$ has coefficient $\bDelta \bc$ and variance $\bc'\bSigma \bc$.) A vector summary, $q^\ast>1$, returns to \eqref{eq:model} when the within-cluster design is the same in every cluster, $\bX_g=\bX_0$ and $\bZ_g=\bZ_0$: the mean $\bM(\bX_0\bbeta+\bZ_0\bdelta)$ is then the same vector for every cluster, so $\bM \by_1,\bM \by_2,\ldots$ are i.i.d.\ Gaussian and follow \eqref{eq:model} with an intercept as the only regressor, $(p,d)=(0,1)$ and $z_g\equiv1$. Section~\ref{sec:clustered} takes $\bM=\bR(\bW_0'\bW_0)^{-1}\bW_0'$ with $\bW_0=[\bX_0,\bZ_0]$, for which $\bM \by_g=\bR\hat\bgamma_g$ is a contrast of the per-cluster OLS estimate with mean $\bR\bgamma$; the nuisance coefficients are removed by the choice of $\bR$. Under a common within-cluster design the cluster-robust Wald statistic is then the one-sample Hotelling statistic of these summaries, and Theorem~\ref{thm:main} applies with $\nu=G-1$ (Proposition~\ref{prop:common}).
\item[(c)] \emph{Cluster-level regressors.} No reduction is needed when the regressors of interest are cluster-level and effects are allowed to differ across components, since that is \eqref{eq:model} itself: a cluster-randomized treatment with $q$ outcomes per cluster, $q$ units per cluster that all share the cluster's regressors, or a unit with $q$ outcome metrics and unit-level covariates and treatment, which is the setting of Section~\ref{sec:rct}.
\end{enumerate}
Unequal cluster sizes are outside \eqref{eq:model} when the full cluster vector is used, since its dimension and covariance change from cluster to cluster. A fixed-dimensional summary as in (b), such as $\bR\hat\bgamma_g$, still gives an exact reduction when it is Gaussian with a common covariance; otherwise Section~\ref{sec:asymptotic} provides the cluster-robust asymptotic route.

\section{Proofs for Section~\ref{sec:asymptotic}}
\label{app:asymptotic}

\begin{proof}[Proof of Lemma~\ref{lem:primitive}]
The first-order condition gives $\E\bpsi_g=0$, and $\operatorname{Cov}(\bpsi_g)=\bOmega$, so Lemma~\ref{lem:kmt} with $\bu_g=\bpsi_g$ gives the strong approximation. The strong law gives $\bA_G/G=G^{-1}\sum_g\bA_g\to\bH$ and $G^{-1}\bs_G\to0$ almost surely, hence, by \eqref{eq:linearization}, $\hat\bgamma_G-\bgamma^\star=(\bA_G/G)^{-1}G^{-1}\bs_G\to0$ almost surely. The fitted scores are $\hat \bs_g=\bpsi_g-\bA_g(\hat\bgamma_G-\bgamma^\star)$, so
\[
\Big\|\frac{\hat \bM_G}{G}-\frac1G\sum_{g=1}^G\bpsi_g\bpsi_g'\Big\|
\leq2\|\hat\bgamma_G-\bgamma^\star\|\,\frac1G\sum_{g=1}^G\|\bpsi_g\|\|\bA_g\|+\|\hat\bgamma_G-\bgamma^\star\|^2\,\frac1G\sum_{g=1}^G\|\bA_g\|^2\longrightarrow0\quad a.s.,
\]
because the averages on the right have finite limits by the strong law and the Cauchy--Schwarz inequality. Finally $G^{-1}\sum_g\bpsi_g\bpsi_g'\to\bOmega$ almost surely since $\E\|\bpsi_g\|^2<\infty$. Bounded cluster sizes, regressors and outcomes make $\bA_g$ and $\bpsi_g$ bounded.
\end{proof}

\begin{proof}[Proof of Theorem~\ref{thm:asymptotic}]
The proof follows the heteroskedastic asymptotic construction of \citet{lindon2026anytime}, with observation-level scores replaced by cluster scores. Steps~1--4 concern the null $\btheta_0=\bR\bgamma^\star$; Step~5 treats alternatives and Step~6 proves \eqref{eq:late_start}.

\emph{Step 1: Gaussian approximation of the estimator.} By Assumption~\ref{ass:scores} there are i.i.d.\ $\bz_g\sim\Nor_k(0,\bOmega)$ with $\|\bs_G-\sum_{g\leq G}\bz_g\|=O_{a.s.}(G^\vartheta)$, and $\bA_G^{-1}=G^{-1}\bH^{-1}+o_{a.s.}(G^{-1})$. By \eqref{eq:linearization},
\[
\bR(\hat\bgamma_G-\bgamma^\star)=\frac1G\bR\bH^{-1}\sum_{g=1}^G\bz_g+\bR\{\bA_G^{-1}-G^{-1}\bH^{-1}\}\sum_{g=1}^G\bz_g+\bR \bA_G^{-1}\Big(\bs_G-\sum_{g=1}^G\bz_g\Big).
\]
The multivariate law of the iterated logarithm gives $\|\sum_{g\leq G}\bz_g\|=O_{a.s.}(\sqrt{G\log\log G})$, so the second term is $o_{a.s.}(\sqrt{\log\log G/G})$ and the third is $O_{a.s.}(G^{\vartheta-1})=o_{a.s.}(\sqrt{\log\log G/G})$. Hence, with $\tilde \bz_g=\bR\bH^{-1}\bz_g\overset{iid}{\sim}\Nor_q(0,\bSigma_R)$ and $\bar \bz_G=G^{-1}\sum_{g\leq G}\tilde \bz_g$,
\[
\htheta_G-\btheta^\star=\bar \bz_G+\bvarrho_G,\qquad \bvarrho_G=o_{a.s.}\big(\sqrt{\log\log G/G}\big),
\]
where $\htheta_G\coloneqq \bR\hat\bgamma_G$ estimates $\btheta^\star=\bR\bgamma^\star$.
Since $\|\bar \bz_G\|=O_{a.s.}(\sqrt{\log\log G/G})$, we get $G\{2\bar \bz_G'\bSigma_R^{-1}\bvarrho_G+\bvarrho_G'\bSigma_R^{-1}\bvarrho_G\}=o_{a.s.}(\log\log G)$, and therefore
\begin{equation}\label{eq:step1}
G\,(\htheta_G-\btheta^\star)'\bSigma_R^{-1}(\htheta_G-\btheta^\star)=G\,\bar \bz_G'\bSigma_R^{-1}\bar \bz_G+o_{a.s.}(\log\log G).
\end{equation}

\emph{Step 2: the exact Gaussian mixture martingale.} Let $\cF_G^z=\sigma(\tilde \bz_1,\ldots,\tilde \bz_G)$. For fixed $\bmu\in\R^q$ the Gaussian likelihood ratio $L_G(\bmu)=\exp\{\bmu'\bSigma_R^{-1}\sum_{g\leq G}\tilde \bz_g-\frac G2\bmu'\bSigma_R^{-1}\bmu\}$ is a nonnegative martingale with respect to $(\cF_G^z)$, because $\E[\exp\{\bmu'\bSigma_R^{-1}\tilde \bz_G-\frac12\bmu'\bSigma_R^{-1}\bmu\}\mid\cF_{G-1}^z]=1$. Let $\Pi=\Nor_q(0,\phi^{-1}\bSigma_R)$. By Tonelli's theorem mixing preserves the martingale property, and Gaussian integration gives
\[
E_G^S=\int L_G(\bmu)\dd\Pi(\bmu)=\left(\frac{\phi}{\phi+G}\right)^{q/2}\exp\left\{\frac12\,\frac{G}{\phi+G}\,G\,\bar \bz_G'\bSigma_R^{-1}\bar \bz_G\right\},\qquad E_0^S=1 .
\]
For $q=1$ this is the Gaussian mixture $e$-process of \citet[arXiv v7, Proposition~C.3]{waudbysmith2024time}, there with an estimated variance; in vector form it is Lemma~5.1 of \citet{lindon2026anytime}. It is the known-covariance counterpart of Theorem~\ref{thm:main} in the intercept-only case, with the same mixture $\bmu\sim\Nor(0,\bSigma_R/\phi)$; indeed \eqref{eq:hotelling} tends to it as $\nu\to\infty$ at fixed $a$ and $T^2$, since $(1+t/\nu)^{(\nu+1)/2}(1+at/\nu)^{-(\nu+1)/2}\to\exp\{\frac12(1-a)t\}$ and $1-a=G/(\phi+G)$.

\emph{Step 3: the feasible Gaussian form.} Assumption~\ref{ass:scores} and $G/(G-1)\to1$ give
\[
\hat\bSigma_{R,G}\coloneqq G\,\bR\hat \bV_G\bR'=\frac{G}{G-1}\,\bR(\bA_G/G)^{-1}(\hat \bM_G/G)(\bA_G/G)^{-1}\bR'\longrightarrow\bSigma_R\quad a.s.,
\]
so $\hat\bSigma_{R,G}^{-1}-\bSigma_R^{-1}=o_{a.s.}(1)$. By Step~1, $\|\htheta_G-\btheta^\star\|=O_{a.s.}(\sqrt{\log\log G/G})$, hence
\[
Q_G(\bR\bgamma^\star)-G\,(\htheta_G-\btheta^\star)'\bSigma_R^{-1}(\htheta_G-\btheta^\star)=G\,(\htheta_G-\btheta^\star)'\{\hat\bSigma_{R,G}^{-1}-\bSigma_R^{-1}\}(\htheta_G-\btheta^\star)=o_{a.s.}(\log\log G),
\]
and with \eqref{eq:step1}, $Q_G(\bR\bgamma^\star)=G\bar \bz_G'\bSigma_R^{-1}\bar \bz_G+o_{a.s.}(\log\log G)$. Define
\[
E_G^{\chi^2}=\left(\frac{\phi}{\phi+G}\right)^{q/2}\exp\left\{\frac12\,\frac{G}{\phi+G}\,Q_G(\bR\bgamma^\star)\right\}.
\]
Then $\log E_G^{\chi^2}-\log E_G^S=o_{a.s.}(\log\log G)$. The law of the iterated logarithm gives
\[
G\bar \bz_G'\bSigma_R^{-1}\bar \bz_G=O_{a.s.}(\log\log G),\qquad\text{so that}\qquad \log E_G^S=-\tfrac q2\log G+O_{a.s.}(\log\log G),
\]
and therefore $\log E_G^{\chi^2}/\log E_G^S\to1$ almost surely.

\emph{Step 4: the Student-type form.} Write $Q_G=Q_G(\bR\bgamma^\star)$, $x_G=Q_G/\nu_G$ and $a_G=\phi/(\phi+G)$. Since $Q_G=O_{a.s.}(\log\log G)$ and $\nu_G=G-1$, $x_G\to0$ almost surely, and $\log\{(1+a_Gx_G)/(1+x_G)\}=(a_G-1)x_G+O_{a.s.}(x_G^2)$. Using $\nu_G+1=G$,
\begin{align*}
\log \ECR_G(\bR\bgamma^\star)-\log E_G^{\chi^2}
&=-\frac G2\log\left(\frac{1+a_Gx_G}{1+x_G}\right)-\frac12\,\frac{G}{\phi+G}\,Q_G\\
&=\frac12\,\frac{G}{\phi+G}\,Q_G\Big(\frac{G}{\nu_G}-1\Big)+O_{a.s.}\Big(\frac{Q_G^2}{G}\Big)=o_{a.s.}(1).
\end{align*}
Since $\log E_G^{\chi^2}=-\frac q2\log G+O_{a.s.}(\log\log G)$, this gives $\log \ECR_G/\log E_G^{\chi^2}\to1$, and combining with Step~3, $\log \ECR_G/\log E_G^S\to1$ almost surely. The same computation applies to any $\nu_G$ with $G/\nu_G\to1$ and any exponent $\nu_G'/2$ with $\nu_G'/G\to1$, which is the under-determination discussed in Section~\ref{sec:clustered}. For a single comparison statistic that also works under alternatives, set $\tilde \bz_g^\star=\bR\bH^{-1}\bz_g+\btheta^\star-\btheta_0$ and let $E_G^{S\star}$ be \eqref{eq:hotelling} for these vectors, extended by one for $G\leq q$. Under every null law these vectors are i.i.d. zero-mean Gaussian, so Theorem~\ref{thm:main} makes this statistic a test martingale in their invariant filtration. Under the null their sample covariance converges to $\bSigma_R$ almost surely; the same argument as Steps~3--4 gives
\[
\log E_G^{S\star}=-\tfrac q2\log G+O_{a.s.}(\log\log G),\qquad
\frac{\log\ECR_G(\btheta_0)}{\log E_G^{S\star}}\longrightarrow1\quad a.s.
\]
This proves the null part for the comparison statistic used under all laws in the extended model.

\emph{Step 5: alternatives.} Let $P\in\cP$ with $\btheta^\star=\bR\bgamma^\star\neq\btheta_0$, and put $c=(\btheta^\star-\btheta_0)'\bSigma_R^{-1}(\btheta^\star-\btheta_0)>0$. Steps~1 and~3 give $\htheta_G-\btheta_0=(\htheta_G-\btheta^\star)+\btheta^\star-\btheta_0\to\btheta^\star-\btheta_0$ and $\hat\bSigma_{R,G}\to\bSigma_R$ almost surely, so $x_G=Q_G(\btheta_0)/\nu_G\to c$ almost surely, while $Ga_Gx_G\to\phi c$. Hence
\[
\log\ECR_G(\btheta_0)=\frac q2\log a_G+\frac G2\Big\{\log(1+x_G)-\log(1+a_Gx_G)\Big\}=\frac G2\log(1+c)+o_{a.s.}(G).
\]
Use the same statistic $E_G^{S\star}$ just defined, now evaluated at the coupled Gaussian vectors with mean $\btheta^\star-\btheta_0$. Its null martingale property is unchanged; it is not claimed to be a martingale under the alternative. Under $P$ the strong law gives $\bar \bz_G^\star\to\btheta^\star-\btheta_0$ and $\bS_G^\star\to\bSigma_R$, so $T_G^2/\nu_G\to c$ almost surely and the computation above gives $\log E_G^{S\star}=\frac G2\log(1+c)+o_{a.s.}(G)$. Therefore $\log\ECR_G(\btheta_0)/\log E_G^{S\star}\to1$ almost surely, which completes the fixed-$\phi$ logarithmic approximation. The Gaussian mixture $E_G^S$ of Step~2 would not serve here: shifted by $\btheta^\star-\btheta_0$, its logarithm grows at rate $c/2$ per cluster, whereas the Student-type form grows at rate $\frac12\log(1+c)$, the rate of the exact Bayes factor in Proposition~\ref{prop:growth}.

\emph{Step 6: late-start crossing probabilities.} An early scalar precedent is \citet[Theorem~3]{lai1976confidence}, which proves delayed-start convergence of Student-statistic crossing probabilities to Wiener-process crossing probabilities, with additional conditions for the infinite horizon. We adapt the strong-approximation and Brownian boundary-crossing arguments of \citet[arXiv v9, Section~2.5, Theorem~2.8 and Appendix~A.7]{waudbysmith2024time} and \citet[arXiv v5, Section~3, Theorem~1 and Appendix~B]{bibaut2024} to the multivariate CR1 Wald statistic and the Student-type boundary. With the tuning $\phi_m=\rho m$ used here, we retain the fixed threshold $1/\alpha$ and bound the $\limsup$ of the crossing probabilities by $\alpha$. \citet[arXiv v5, Theorem~1]{bibaut2024} allow arbitrary positive tuning sequences, including this regime, but use a threshold calibrated to the scalar Wiener crossing probability to obtain asymptotically exact type-I error. The sharp calibration of \citeauthor{waudbysmith2024time} uses a different tuning regime. This is a separate coverage conclusion, in addition to the logarithmic equivalence of Definition~\ref{def:asymp}.

Work under a fixed null law. Let $\bOmega^{1/2}$ be the symmetric nonnegative square root. Enrich the coupling with Brownian bridges and, if needed, independent Gaussian coordinates in the kernel of $\bOmega$, so that $\sum_{g\leq G}\bz_g=\bOmega^{1/2}\bb_k(G)$ for a standard $k$-dimensional Brownian motion $\bb_k$. Define $\bb_R(t)=\bSigma_R^{-1/2}\bR\bH^{-1}\bOmega^{1/2}\bb_k(t)$. Since $\bSigma_R=\bR\bH^{-1}\bOmega\bH^{-1}\bR'$, $\bb_R$ is standard $q$-dimensional Brownian motion. Put
\[
\mathfrak b_\rho(t)=(1+\rho/t)\{q\log(1+t/\rho)+2\log(1/\alpha)\}.
\]
The argument has six parts. Parts (a) and (b) bound $\sqrt{Q_G(\btheta_0)}$ by a Brownian quantity, with one tolerance for all $G\geq m$; part (c) turns a crossing of $\ECR_{G,m}$ into a crossing of a boundary by that Brownian quantity; parts (d) and (e) remove $m$ and the tolerance; part (f) applies Ville's inequality.

\emph{(a) The statistic as a matrix times the coupled path.} Under the null, $\btheta_0=\btheta^\star$, so $\htheta_G-\btheta_0=\bR(\hat\bgamma_G-\bgamma^\star)=\bR \bA_G^{-1}\bs_G$ by \eqref{eq:linearization}, and $Q_G(\btheta_0)=G\,(\htheta_G-\btheta_0)'\hat\bSigma_{R,G}^{-1}(\htheta_G-\btheta_0)$ with $\hat\bSigma_{R,G}$ of Step~3. Hence, at every nonsingular sample,
\[
\sqrt{Q_G(\btheta_0)}=\Big\|\hat\bSigma_{R,G}^{-1/2}\,\sqrt G\,(\htheta_G-\btheta_0)\Big\|
=\Big\|\hUpsilon_G\,\frac{\bs_G}{\sqrt G}\Big\|,
\qquad\text{where}\qquad
\hUpsilon_G\coloneqq\hat\bSigma_{R,G}^{-1/2}\bR(\bA_G/G)^{-1},
\]
and put $\bUpsilon\coloneqq\bSigma_R^{-1/2}\bR\bH^{-1}$ for its limit.
Set $\hUpsilon_G=\bUpsilon$ at singular samples; there are almost surely only finitely many of them, and they cannot cross because the process is set to one there. By construction $\sum_{g\leq G}\bz_g=\bOmega^{1/2}\bb_k(G)$ and $\bUpsilon\bOmega^{1/2}\bb_k(G)=\bb_R(G)$. Substituting
\[
\bs_G=\bOmega^{1/2}\bb_k(G)+\Big(\bs_G-\sum_{g\leq G}\bz_g\Big),\qquad
\hUpsilon_G=\bUpsilon+(\hUpsilon_G-\bUpsilon),
\]
and using the triangle inequality with $\|\hUpsilon_G\bv\|\leq\|\hUpsilon_G\|\|\bv\|$ gives, at every nonsingular sample,
\begin{equation}\label{eq:late_triangle}
\sqrt{Q_G(\btheta_0)}
\leq\frac{\|\bb_R(G)\|}{\sqrt G}
+\big\|(\hUpsilon_G-\bUpsilon)\bOmega^{1/2}\big\|\,\frac{\|\bb_k(G)\|}{\sqrt G}
+\|\hUpsilon_G\|\,\frac{\|\bs_G-\sum_{g\leq G}\bz_g\|}{\sqrt G}.
\end{equation}
The first term is the Brownian quantity of the limit. The second is the error of the estimated standardization; it is proportional to the $k$-dimensional path $\bb_k$, not to $\bb_R$, because the estimation error acts before the projection to $q$ dimensions. The third is the coupling error.

\emph{(b) One tolerance for all $G\geq m$.} The event in \eqref{eq:late_start} is a union over $G\geq m$, so the two error terms in \eqref{eq:late_triangle} must be controlled by a single random quantity for all $G\geq m$ at once. Assumption~\ref{ass:scores} gives $\bA_G/G\to\bH\succ0$, Step~3 gives $\hat\bSigma_{R,G}\to\bSigma_R\succ0$, and inversion and the symmetric square root are continuous on positive definite matrices, so $\hUpsilon_G\to\bUpsilon$ almost surely. The coupling error is $O_{a.s.}(G^\vartheta)$, that is, $\sup_{G\geq1}\|\bs_G-\sum_{g\leq G}\bz_g\|/G^\vartheta<\infty$ almost surely. Define
\begin{equation}\label{eq:omega_m}
\omega_m\coloneqq\max\Big\{\sup_{G\geq m}\big\|(\hUpsilon_G-\bUpsilon)\bOmega^{1/2}\big\|,\ \
m^{\vartheta-1/2}\,\sup_{G\geq m}\|\hUpsilon_G\|\,\sup_{G\geq1}\frac{\|\bs_G-\sum_{g\leq G}\bz_g\|}{G^{\vartheta}}\Big\}.
\end{equation}
Since $G^{\vartheta-1/2}\leq m^{\vartheta-1/2}$ for $G\geq m$, \eqref{eq:late_triangle} gives, for every nonsingular sample $G\geq m$,
\begin{equation}\label{eq:late_envelope}
\sqrt{Q_G(\btheta_0)}
\leq \frac{\|\bb_R(G)\|}{\sqrt G}
+\omega_m\left(1+\frac{\|\bb_k(G)\|}{\sqrt G}\right).
\end{equation}
Moreover $\omega_m\to0$ almost surely: the first entry of the maximum in \eqref{eq:omega_m} is the tail supremum of a sequence that converges to zero almost surely, and the second is $m^{\vartheta-1/2}\to0$ times two almost surely finite factors, the first of which converges to $\|\bUpsilon\|$.

\emph{(c) A crossing forces a Brownian event.} With $\phi_m=\rho m$ and $t=G/m$, $a_G=\phi_m/(\phi_m+G)=\rho/(\rho+t)$ and $1-a_G=t/(\rho+t)$. For $x\geq0$ and $0<a<1$,
\[
\log(1+x)-\log(1+ax)\leq(1-a)x,
\]
because the difference vanishes at $x=0$ and has derivative $(1-a)/\{(1+x)(1+ax)\}\leq1-a$. Hence, with $x_G=Q_G(\btheta_0)/(G-1)$,
\[
\log\ECR_{G,m}(\btheta_0)=\frac q2\log a_G+\frac G2\big\{\log(1+x_G)-\log(1+a_Gx_G)\big\}
\leq\frac q2\log\frac{\rho}{\rho+t}+\frac G2\,\frac{t}{\rho+t}\,\frac{Q_G(\btheta_0)}{G-1},
\]
so $\ECR_{G,m}(\btheta_0)\geq\alpha^{-1}$ implies
\[
Q_G(\btheta_0)\geq\frac{G-1}{G}\Big(1+\frac\rho t\Big)\Big\{q\log\Big(1+\frac t\rho\Big)+2\log\frac1\alpha\Big\}
=\frac{G-1}{G}\,\mathfrak b_\rho(G/m)
\geq(1-1/m)\,\mathfrak b_\rho(G/m).
\]
Samples at which the process is set to one cannot cross. Fix $\omega\in(0,1/2]$ and let $m\geq1/\omega$, so that $\sqrt{1-1/m}\geq1-1/m\geq1-\omega$. On the event $\{\omega_m\leq\omega\}$, a crossing at some $G\geq m$ therefore implies, by \eqref{eq:late_envelope},
\begin{equation}\label{eq:late_cross}
\frac{\|\bb_R(G)\|}{\sqrt G}+\omega\left(1+\frac{\|\bb_k(G)\|}{\sqrt G}\right)\geq(1-\omega)\sqrt{\mathfrak b_\rho(G/m)} .
\end{equation}
Consequently
\begin{equation}\label{eq:late_split}
P\big(\exists G\geq m:\ \ECR_{G,m}(\btheta_0)\geq\alpha^{-1}\big)
\leq P(\omega_m>\omega)
+P\big(\exists G\geq m:\ \text{\eqref{eq:late_cross} holds}\big).
\end{equation}
The first probability tends to zero as $m\to\infty$ because $\omega_m\to0$ almost surely. The second involves the Brownian motions alone.

\emph{(d) Brownian scaling removes $m$.} Write $G=mt$. The process $t\mapsto\bb_k(mt)/\sqrt m$ is standard $k$-dimensional Brownian motion, and $\bb_R(mt)/\sqrt m=\bUpsilon\bOmega^{1/2}\bb_k(mt)/\sqrt m$, so the rescaled pair has the law of $(\bb_k,\bb_R)$; moreover $\|\bb_R(G)\|/\sqrt G=\|\bb_R(mt)/\sqrt m\|/\sqrt t$. Enlarging the grid $\{G/m:G\geq m\}$ to $[1,\infty)$ can only enlarge the event, so the second probability in \eqref{eq:late_split} is at most
\begin{equation}\label{eq:late_brownian}
P(\mathcal E_\omega),\qquad
\mathcal E_\omega\coloneqq\left\{\exists t\geq1:\
\frac{\|\bb_R(t)\|}{\sqrt t}
+\omega\left(1+\frac{\|\bb_k(t)\|}{\sqrt t}\right)
\geq(1-\omega)\sqrt{\mathfrak b_\rho(t)}\right\},
\end{equation}
which does not depend on $m$. Letting $m\to\infty$ in \eqref{eq:late_split},
\[
\limsup_{m\to\infty}P\big(\exists G\geq m:\ \ECR_{G,m}(\btheta_0)\geq\alpha^{-1}\big)\leq P(\mathcal E_\omega)\qquad\text{for every }\omega\in(0,1/2].
\]

\emph{(e) The tolerance tends to zero.} The events $\mathcal E_\omega$ decrease as $\omega$ decreases, because the left side of the inequality defining them decreases and the right side increases, so $P(\mathcal E_\omega)\downarrow P(\bigcap_{\omega>0}\mathcal E_\omega)$ as $\omega\downarrow0$ by continuity from above. To identify the intersection, first confine the crossings to a compact interval. By the law of the iterated logarithm, $\|\bb_R(t)\|/\sqrt t$ and $\|\bb_k(t)\|/\sqrt t$ are $O(\sqrt{\log\log t})$ almost surely, whereas $\sqrt{\mathfrak b_\rho(t)}\geq\sqrt{q\log(1+t/\rho)}$. Hence there is almost surely a finite random time $t_1\geq1$ with
\[
\frac{\|\bb_R(t)\|}{\sqrt t}+\frac12\left(1+\frac{\|\bb_k(t)\|}{\sqrt t}\right)<\frac12\sqrt{\mathfrak b_\rho(t)}\qquad\text{for all }t\geq t_1,
\]
so that for every $\omega\leq1/2$ the inequality defining $\mathcal E_\omega$ fails after $t_1$. On $\bigcap_{\omega>0}\mathcal E_\omega$ there are therefore, along a sequence $\omega_j\downarrow0$, times $t_j\in[1,t_1]$ at which the inequality defining $\mathcal E_{\omega_j}$ holds. A subsequence converges to some $t^\circ\in[1,t_1]$; the three functions of $t$ in that inequality are continuous, and $\|\bb_k(t)\|/\sqrt t$ is bounded on $[1,t_1]$, so passing to the limit gives $\|\bb_R(t^\circ)\|/\sqrt{t^\circ}\geq\sqrt{\mathfrak b_\rho(t^\circ)}$. Thus, up to a null set,
\[
\bigcap_{\omega>0}\mathcal E_\omega\subseteq\Big\{\exists t\geq1:\ \frac{\|\bb_R(t)\|^2}{t}\geq\mathfrak b_\rho(t)\Big\},
\]
and with (d),
\[
\limsup_{m\to\infty}P\big(\exists G\geq m:\ \ECR_{G,m}(\btheta_0)\geq\alpha^{-1}\big)
\leq P\Big(\exists t\geq1:\ \frac{\|\bb_R(t)\|^2}{t}\geq\mathfrak b_\rho(t)\Big).
\]

\emph{(f) Ville's inequality.} Since $\bb_R$ is standard $q$-dimensional Brownian motion, for each $\bmu\in\R^q$ the process $\exp\{\bmu'\bb_R(t)-t\|\bmu\|^2/2\}$ is a nonnegative martingale equal to one at $t=0$, and mixing over $\bmu\sim\Nor_q(0,\rho^{-1}\bI_q)$ preserves both properties by Tonelli's theorem. The Gaussian integral is
\begin{align*}
E_t^{\mathrm{BM}}&\coloneqq\int\exp\Big\{\bmu'\bb_R(t)-\frac t2\|\bmu\|^2\Big\}\,\Nor_q(0,\rho^{-1}\bI_q)(\dd\bmu)\\
&=\left(\frac{\rho}{\rho+t}\right)^{q/2}
\exp\left\{\frac{\|\bb_R(t)\|^2}{2(\rho+t)}\right\},\qquad t\geq0,
\end{align*}
the continuous-time counterpart of $E_G^S$ in Step~2. Rearranging,
\[
E_t^{\mathrm{BM}}\geq\alpha^{-1}
\iff
\|\bb_R(t)\|^2\geq(\rho+t)\Big\{q\log\Big(1+\frac t\rho\Big)+2\log\frac1\alpha\Big\}
\iff
\frac{\|\bb_R(t)\|^2}{t}\geq\mathfrak b_\rho(t),
\]
so $\mathfrak b_\rho$ is the level-$\alpha$ rejection boundary of this Gaussian mixture martingale. Ville's inequality gives $P(\exists t\geq0:\ E_t^{\mathrm{BM}}\geq\alpha^{-1})\leq\alpha\,\E[E_0^{\mathrm{BM}}]=\alpha$, and restricting to $t\geq1$ can only decrease the probability. With (e) this proves \eqref{eq:late_start}.

\end{proof}

\begin{proof}[Proof of Corollary~\ref{cor:pvalue}]
The reciprocal of the Gaussian comparison process is an exact anytime-valid $p$-value, without needing truncation. The identity
\[
\frac{\log(1/\ECR_G)}{\log(1/E_G^{S\star})}=\frac{\log\ECR_G}{\log E_G^{S\star}}\longrightarrow1\qquad a.s.
\]
shows that the untruncated values satisfy the logarithmic definition; the ratio is stated for the untruncated values because under the null the truncated values eventually equal one, so their logarithms would give $0/0$. Truncation leaves all events $\{\pv_G\leq\alpha\}$, $0<\alpha<1$, unchanged. The late-start probability statement is \eqref{eq:late_start}.
\end{proof}

\begin{proof}[Proof of Corollary~\ref{cor:asympcs}]
$\ECR_G(\btheta)<\alpha^{-1}$ if and only if $(1+x)/(1+a_Gx)<u_G^{-1}$ with $x=Q_G(\btheta)/\nu_G$ and $a_G=\phi/(\phi+G)$, that is, $x(u_G-a_G)<1-u_G$; note $u_G=(\alpha^2a_G^{\,q})^{1/G}<1$. This is the set of Proposition~\ref{prop:common}(b) with strict inequality, and the logarithmic-asymptotic statement follows by inversion. For $\phi_m=\rho m$, the complement of $\{\btheta^\star\in C_{G,m}^{\alpha}\ \text{for all }G\geq m\}$ is the event in \eqref{eq:late_start} at $\btheta_0=\btheta^\star$. Thus \eqref{eq:late_start_coverage} follows from $\liminf_m(1-p_m)=1-\limsup_m p_m$, where $p_m$ is that crossing probability. For $q=1$, $Q_G(\theta)=(\bR\hat\bgamma_G-\theta)^2/(\bR\hat \bV_G\bR')$ and the ellipsoid is the stated interval.
\end{proof}

\subsection{Proof of Corollary~\ref{cor:cr-growth}}
\label{app:cr-growth}
\begin{proof}
Step~5 of the proof of Theorem~\ref{thm:asymptotic} gives $\log\ECR_G(\btheta_0)/G\to r_{\mathrm{CR}}>0$ almost surely, so the first-passage argument in the proof of Proposition~\ref{prop:growth} gives the almost-sure stopping-time limit. To pass to expectations under \eqref{eq:stopping_moments}, we use complete convergence. For vectors, write $\bu_G\to_{\mathrm c}\bu$ if $\sum_G P(\|\bu_G-\bu\|>\epsilon)<\infty$ for every $\epsilon>0$; for matrices use the same definition with the spectral norm. Theorem~1 of \citet[p.~26]{hsu1947complete} states that averages of i.i.d. scalar variables with finite variance converge completely to their expectation. Applying it entrywise, followed by a finite union bound, gives
\[
\frac{\bA_G}{G}\to_{\mathrm c}\bH,\qquad
\frac{\bs_G}{G}\to_{\mathrm c}0,\qquad
\frac1G\sum_{g\leq G}\bpsi_g\bpsi_g'\to_{\mathrm c}\bOmega,\qquad
\frac1G\sum_{g\leq G}\|\bA_g\|^2\to_{\mathrm c}\E\|\bA_g\|^2.
\]
The fourth moments in \eqref{eq:stopping_moments} ensure finite variances for every entry used here. Since $\bH\succ0$, the probabilities of singular $\bA_G$ are summable. Set $\be_G=0$ on those samples for this calculation; on nonsingular samples
$\be_G\coloneqq\hat\bgamma_G-\bgamma^\star=(\bA_G/G)^{-1}(\bs_G/G)\to_{\mathrm c}0$.
The exact score expansion $\hat \bs_g=\bpsi_g-\bA_g\be_G$ and Cauchy--Schwarz give
\begin{align*}
\left\|\frac{\hat \bM_G}{G}-\frac1G\sum_{g\leq G}\bpsi_g\bpsi_g'\right\|
&\leq2\|\be_G\|
\left(\frac1G\sum_{g\leq G}\|\bpsi_g\|^2\right)^{1/2}
\left(\frac1G\sum_{g\leq G}\|\bA_g\|^2\right)^{1/2}\\
&\quad+\|\be_G\|^2\frac1G\sum_{g\leq G}\|\bA_g\|^2.
\end{align*}
Thus $\hat \bM_G/G\to_{\mathrm c}\bOmega$. Continuous mapping in fixed neighborhoods of the positive definite limits, together with the covariance identity in Step~3 of that proof, yields
\[
\hat\btheta_G\to_{\mathrm c}\btheta^\star,\qquad
\hat\bSigma_{R,G}\to_{\mathrm c}\bSigma_R,\qquad
\frac{Q_G(\btheta_0)}G\to_{\mathrm c}c.
\]
The probabilities of singular contrast covariance matrices are also summable. Hence the convention $\ECR_G=1$ at singular samples does not affect complete convergence. Substituting $Q_G/G$ in \eqref{eq:ECR}, with fixed $\phi$, gives
\begin{equation}\label{eq:cr_growth_complete}
\sum_{G=q+1}^\infty
P\!\left(\left|\frac{\log\ECR_G(\btheta_0)}G-r_{\mathrm{CR}}\right|>\eta\right)<\infty
\qquad\text{for every }\eta>0.
\end{equation}
Indeed, viewed as a function of $x=Q_G/G$, the divided logarithm in \eqref{eq:ECR} converges locally uniformly near $c$ to $\tfrac12\log(1+x)$, which is continuous at $c$.

Equation~\eqref{eq:cr_growth_complete} supplies the summable lower-tail deviations required by \eqref{eq:stopping_tail_sum}. Applying that argument with $\ECR_G(\btheta_0)$, $r_{\mathrm{CR}}$ and initial index $q+1$ proves finite expectation, \eqref{eq:cr_expected_stopping} and $L^1$ convergence.
\end{proof}

\subsection{Proof of Proposition~\ref{prop:cr-log-growth}}
\label{app:cr-log-growth}
Fix $\phi_1,\phi_2>0$ and put $x_G=Q_G(\btheta_0)/(G-1)$ on active samples. Subtracting the exact logarithms in \eqref{eq:ECR} cancels the term $(G/2)\log(1+x_G)$:
\begin{equation}\label{eq:cr-precision-ratio}
\begin{split}
\log\frac{E_G^{\mathrm{CR},(\phi_1)}}{E_G^{\mathrm{CR},(\phi_2)}}
={}&\frac q2\log\frac{\phi_1(\phi_2+G)}{\phi_2(\phi_1+G)}\\
&-\frac G2\log
\frac{1+\phi_1x_G/(\phi_1+G)}{1+\phi_2x_G/(\phi_2+G)}.
\end{split}
\end{equation}
By Step~5 of the proof of Theorem~\ref{thm:asymptotic}, $x_G\to c$ almost surely, and samples are eventually active. Since $G\log\{1+\phi x/(\phi+G)\}\to\phi x$ uniformly for $x$ in a compact set,
\[
\log\frac{E_G^{\mathrm{CR},(\phi_1)}}{E_G^{\mathrm{CR},(\phi_2)}}
\longrightarrow
\frac q2\log\frac{\phi_1}{\phi_2}-\frac c2(\phi_1-\phi_2)
\qquad\text{almost surely}.
\]

Bounded $\bA_g$ and $\bpsi_g$ imply exponential concentration of the empirical moments used to form the OLS coefficient and fitted sandwich covariance: these are averages of entries and products of $\bA_g$ and $\bt_g=\bA_g\bgamma^\star+\bpsi_g$. On a fixed small neighborhood of their population values, the fitted matrices are nonsingular, $x_G$ is bounded, and the log ratio in \eqref{eq:cr-precision-ratio} is uniformly bounded. The complement has probability at most $C_0e^{-c_0G}$. For every $x_G\geq0$, the ratio in the second logarithm lies between $1$ and
$\{\phi_1/(\phi_1+G)\}/\{\phi_2/(\phi_2+G)\}$, in either order. Hence the absolute log ratio is bounded by $C(1+G)$ on every active sample, and equals zero on inactive samples. These bounds justify passage to expectations:
\[
\E_P\!\left[\log\frac{E_G^{\mathrm{CR},(\phi_1)}}{E_G^{\mathrm{CR},(\phi_2)}}\right]
=\frac q2\log\frac{\phi_1}{\phi_2}-\frac c2(\phi_1-\phi_2)+o(1).
\]
Maximizing $(q/2)\log\phi-\phi c/2$ gives $\phi_*=q/c$; substituting $(\phi_1,\phi_2)=(\phi_*,\phi)$ proves \eqref{eq:cr-precision-cost}. \qed

\section{Proofs for Section~\ref{sec:rct}}
\label{app:rct}

\begin{proof}[Proof of Proposition~\ref{prop:rct}]
\emph{Positive definiteness.} Independence of $D_i$ and $\bmv_i$ gives, for $a,c\in\R$ and $\bb\in\R^r$,
\[
\E\big[\{a+\bb'\bmv_i+c(D_i-\pi)\}^2\big]
=(a+\bb'\bmu_m)^2+\bb'\bOmega_m \bb+c^2\pi(1-\pi).
\]
Since $\bOmega_m\succ0$ and $\pi\in(0,1)$, this is positive for every nonzero $(a,\bb',c)'$, so $\bOmega_{\bW}=\E(\bw_i\bw_i')\succ0$. The second-moment conditions make $\E(\bw_iy_{ij})$ finite by the Cauchy--Schwarz inequality.

\emph{Separation.} $(\by_i-\bW_i\bgamma)'(\by_i-\bW_i\bgamma)=\sum_{j=1}^q(y_{ij}-\bw_i'\bgamma_j)^2$, so the objective in \eqref{eq:projection} is a sum of $q$ terms each depending on its own $\bgamma_j$, and each is minimized uniquely at $\bgamma_j^\star=\bOmega_{\bW}^{-1}\E(\bw_iy_{ij})$, the population least-squares coefficient of outcome $j$ alone.

\emph{Identification.} Fix $j$ and write $y=y_{ij}$, $y(t)=y_{ij}(t)$, $\dot \bmv=\bmv_i-\bmu_m$. Since $\E[D_i-\pi]=0$, $\E[(D_i-\pi)\bmv_i]=0$ and $\E[\dot \bmv]=0$, the projection equations for the intercept, treatment and centered covariates give
\begin{align*}
\E[y]-\mu_j^\star-\bmu_m'\bzeta_j^\star&=0,\\
\E[(D_i-\pi)y]-\pi(1-\pi)\tau_j^\star&=0,\\
\E[\dot \bmv y]-\bOmega_m\bzeta_j^\star&=0.
\end{align*}
By SUTVA, $y=D_iy(1)+(1-D_i)y(0)$, and by randomization $D_i\perp(y(0),y(1),\bmv_i)$, so $\E[(D_i-\pi)y]=\pi(1-\pi)\{\E[y(1)]-\E[y(0)]\}$. The second equation therefore identifies $\tau_j^\star$ as the ATE, while the third and first give $\bzeta_j^\star$ and $\mu_j^\star$, respectively.

\emph{Scores.} $\bW_i'\bW_i=(\bI_q\otimes \bw_i)(\bI_q\otimes \bw_i')=\bI_q\otimes \bw_i\bw_i'$, so $\bH=\bI_q\otimes\bOmega_{\bW}\succ0$. Block $j$ of $\bpsi_i=\bW_i'(\by_i-\bW_i\bgamma^\star)$ is $(y_{ij}-\bw_i'\bgamma_j^\star)\bw_i=\varepsilon_{ij}\bw_i$, so $\bpsi_i=\beps_i\otimes \bw_i$, whose mean is zero by the first-order conditions. Under the additional score-integrability assumption, its covariance is $\E[(\beps_i\otimes \bw_i)(\beps_i\otimes \bw_i)']=\E(\beps_i\beps_i'\otimes \bw_i\bw_i')$.
\end{proof}

\begin{proof}[Proof of Corollary~\ref{cor:rct}]
(i) The units are i.i.d., $\bH\succ0$ by Proposition~\ref{prop:rct}, $\bSigma_{R_\tau}\succ0$ and $\E\|\bpsi_i\|^s<\infty$ by assumption, and $\E\|\bA_i\|^2=\E\|\bI_q\otimes \bw_i\bw_i'\|^2\leq C\,\E\|\bw_i\|^4<\infty$ because $D_i$ is bounded and the corollary assumes $\E\|\bmv_i\|^4<\infty$. This fourth-moment condition is used here for fitted-score covariance consistency in Lemma~\ref{lem:primitive}, not for the projection identification in Proposition~\ref{prop:rct}. Lemma~\ref{lem:primitive} gives Assumption~\ref{ass:scores}, and Theorem~\ref{thm:asymptotic} and Corollaries~\ref{cor:pvalue} and~\ref{cor:asympcs} apply to $H_0:\bR_\tau\bgamma^\star=\bdelta_0$, which is $H_0:\bdelta^\star=\bdelta_0$ by Proposition~\ref{prop:rct}.

(ii) For $\bV\succ0$ and vectors $\bv,\bc$, the Cauchy--Schwarz inequality in the $\bV^{-1}$ inner product gives $(\bc'\bv)^2\leq(\bv'\bV^{-1}\bv)(\bc'\bV \bc)$. On the event $\{\bdelta^\star\in C_G^\alpha\ \forall G\}$, $\bv=\hat\bdelta_G-\bdelta^\star$ satisfies $\bv'\bV^{-1}\bv<q\kappa_G$ with $\bV=\bR_\tau\hat \bV_G\bR_\tau'$, for every $G$ at which the ellipsoid is finite. Applying this inclusion with $\phi_m=\rho m$ for all $G\geq m$ transfers the $\liminf$ coverage bound in \eqref{eq:late_start_coverage} to all the contrast intervals simultaneously.

\end{proof}

\section{Fully Interacted Regression with Sample Centering}
\label{app:rct_interactions}

The application in Section~\ref{sec:rct} uses regression without interactions. For the fully interacted alternative, at sample size $G$ fit each outcome by OLS using
\begin{equation}\label{eq:rctreg_interacted}
y_{ij}=\mu_j+(\bmv_i-\bar \bmv_G)'\bzeta_j+(D_i-\pi)\tau_j
+D_i(\bmv_i-\bar \bmv_G)'\boldeta_j+\varepsilon_{ij},
\end{equation}
where $\bar \bmv_G=G^{-1}\sum_{i=1}^G \bmv_i$ is the plug-in estimate of the covariate mean. Let $\hat\bdelta_G^{\mathrm{emp}}$ be the vector of fitted treatment coefficients, $\hat{\bm\Lambda}_{\eta,G}=[\hat\boldeta_{1,G},\ldots,\hat\boldeta_{q,G}]$, and $\hat \bV_{\tau,G}^{\mathrm{CR1}}$ the $q\times q$ treatment block of the ordinary CR1 covariance from this fitted regression, with units as clusters. Define
\[
\hat{\bS}_{m,G}=\frac1{G-1}\sum_{i=1}^G(\bmv_i-\bar \bmv_G)(\bmv_i-\bar \bmv_G)'.
\]
For inference on the superpopulation ATE, replace the treatment covariance by
\begin{equation}\label{eq:centering_covariance}
\hat \bV_G^{\mathrm{emp}}=\hat \bV_{\tau,G}^{\mathrm{CR1}}
+\frac1G\hat{\bm\Lambda}_{\eta,G}'\hat{\bS}_{m,G}\hat{\bm\Lambda}_{\eta,G}.
\end{equation}
The added term accounts for estimation of the covariate mean; the ordinary sandwich alone generally omits this variability \citep[Section~5, Remark~(vi), crediting Imbens and Wooldridge (2009)]{lin2013agnostic}. This correction applies to the sample mean of the same $G$ units and is an asymptotic covariance correction.

Assume the randomization and moment conditions of Proposition~\ref{prop:rct} and Corollary~\ref{cor:rct}(i), with the score moment condition applied to the fully interacted population projection centered at $\bmu_m$, and assume that the limiting covariance of $\sqrt G(\hat\bdelta_G^{\mathrm{emp}}-\bdelta^\star)$ is positive definite. Then the asymptotic conclusions of Corollary~\ref{cor:rct} hold with $\hat\bdelta_G^{\mathrm{emp}}$ and $\hat \bV_G^{\mathrm{emp}}$ in place of $\hat\bdelta_G$ and $\bR_\tau\hat \bV_G\bR_\tau'$. In particular, use
\[
Q_G^{\mathrm{emp}}(\bdelta_0)=(\hat\bdelta_G^{\mathrm{emp}}-\bdelta_0)'(\hat \bV_G^{\mathrm{emp}})^{-1}(\hat\bdelta_G^{\mathrm{emp}}-\bdelta_0)
\]
in the same $\ECR_G$ and confidence-ellipsoid formulas, with $\nu_G=G-1$. The late-start coverage conclusion is the $\liminf$ bound \eqref{eq:late_start_coverage}, using the same tuning and monitoring regime as Theorem~\ref{thm:asymptotic}. Thus the fully interacted OLS fit and treatment estimates are retained; the only additional procedural change is the covariance correction \eqref{eq:centering_covariance}. These statements are asymptotic and do not extend the exact Gaussian assertion of Remark~\ref{rem:rctcompare} to empirical centering.

\section{Further Remarks on Optimality and Invariance}
\label{app:further}

\subsection{GROW optimality does not carry over for \texorpdfstring{$q\geq2$}{q >= 2}}
\label{app:grow}
For $q=1$ the group is amenable and the GROW/REGROW statement of \citet{lindon2026anytime} rests on Theorem~2 of \citet{perezortiz2024estatistics}, which assumes amenability. $\GL(q)$ is not amenable for $q\geq2$, and those authors discuss Hotelling's test as an example. Theorem~\ref{thm:main} does not use amenability, so validity is unaffected, but the optimality claim needs to be restated. That a procedure invariant under a smaller group can improve on the best fully invariant one is Stein's example, Example~6.5 of \citet{eaton1989group}: for a Wishart scale hypothesis, the best test invariant under the lower-triangular group dominates the best test invariant under the full general linear group and hence the likelihood-ratio test. Part (b) is the expected-log-growth form of that example.
\begin{enumerate}
\item[(a)] What holds: by Remark~\ref{rem:optimality}, $E_G$ is GRO in the sense of \citet{grunwald2024safe} for every $Q_{\bB,\bSigma}$ at once among $e$-variables that are \emph{invariant} under $\cG$, and hence also GROW and REGROW relative to that class.
\item[(b)] What fails in a nontrivial multivariate problem: $E_G$ is not GROW among \emph{all} $e$-variables against $\{Q_{\bB,\bSigma}\}$ when $q\geq2$ and $\bK_G$ is not proportional to $\bI_N$. The subgroup with $\bC$ lower triangular with positive diagonal is amenable and transitive on $(\bB,\bSigma)$. Its right-Haar prior on the lower-triangular Cholesky factor $\bJ$ of $\bSigma=\bJ\bJ'$ is $\prod_{k=1}^qJ_{kk}^{-(q-k+1)}\dd\bJ\dd\bB$. This is the right-Haar prior of \citet[equation~(9)]{sun2007objective}, there written in terms of the lower-triangular matrix $\bm{\Psi}=\bJ^{-1}$ with $\bSigma^{-1}=\bm{\Psi}'\bm{\Psi}$; its dependence on the triangular decomposition, that is, on the ordering of the coordinates, is the nonuniqueness studied in their Section~5. It gives
\begin{equation}\label{eq:triangular}
E_G^{\mathrm{LT}}=E_G\prod_{k=1}^{q}\left(\frac{t^{(1)}_{kk}}{t^{(0)}_{kk}}\right)^{2k-q-1},
\end{equation}
where $t^{(j)}_{kk}$ are the diagonal entries of the lower Cholesky factor of $\bS_{j,G}$.

The integration is as follows. After removing $\bB$ and mixing $\bDelta$, the covariance kernel is $|\bJ|^{-N}\exp\{-\tr(\bJ^{-1}\bS\bJ^{-T})/2\}$. Write $\bS=\bJ_{\bS}\bJ_{\bS}'$ for its lower Cholesky factorization, with diagonal entries $t_{kk}$, and substitute $\bJ=\bJ_{\bS}\bJ_\circ$, where $\bJ_\circ$ is lower triangular with positive diagonal. The Jacobian on lower-triangular entries is $\prod_k t_{kk}^{k}$, giving
\[
\int |\bJ|^{-N}e^{-\tr(\bJ^{-1}\bS\bJ^{-T})/2}
\prod_kJ_{kk}^{-(q-k+1)}\dd\bJ
=c\prod_k t_{kk}^{-N+2k-q-1}.
\]
The integral is finite for $N\geq q$. Taking the ratio gives \eqref{eq:triangular}. The invariant-likelihood-ratio identity of \citet[Section~4]{andersson1982distributions}, applied to the triangular group as in \citet[Section~4.1]{perezortiz2024estatistics}, or direct integration in the QR decomposition of $\bU$, identifies it as the likelihood ratio on the finer triangular invariant. Thus it is an $e$-variable with null expectation one.

The statistic \eqref{eq:triangular} also factorizes over the coordinates. Write $\by_{g,<k}=(y_{g1},\ldots,y_{g,k-1})'$ and $\bY_{G,<k}$ for the first $k-1$ columns of $\bY_G$. Under \eqref{eq:model} the conditional law of the $k$th coordinate is
\begin{equation}\label{eq:stepdown}
y_{gk}\mid \by_{g,<k},\bx_g,\bz_g\ \sim\ \Nor\big(\bb_k'\bx_g+\bc_k'\by_{g,<k}+\bdelta_k^{\circ\prime}\bz_g,\ J_{kk}^2\big),\qquad k=1,\ldots,q,
\end{equation}
where $\bc_k'=\bJ_{k,<k}\bJ_{<k,<k}^{-1}$ is the regression of $\varepsilon_{gk}$ on $\beps_{g,<k}$, $\bb_k$ is the corresponding combination of the columns of $\bB$, and $\bdelta_k^{\circ}=\bdelta_k-\sum_{l<k}c_{kl}\bdelta_l=J_{kk}\bxi_k$, with $\bdelta_k$ and $\bxi_k$ the $k$th columns of $\bDelta$ and of $\bm\Xi=\bDelta\bJ^{-T}$. Three facts follow. The change of variables $\bc_k=\bJ_{<k,<k}^{-T}\bJ_{k,<k}'$ has $\dd\bJ_{k,<k}=|\bJ_{<k,<k}|\,\dd \bc_k$, so the triangular right-Haar prior is $\prod_kJ_{kk}^{-(q-k+1)}\dd\bJ\dd\bB=\prod_kJ_{kk}^{-1}\dd J_{kk}\dd \bc_k\dd \bb_k$, a product of $q$ scale-location right-Haar priors, one for each conditional regression. The mixture \eqref{eq:mixture} is $\bm\Xi\sim\MN(0,\bPhi^{-1},\bI_q)$, so $\bdelta_k^\circ\mid J_{kk}\sim\Nor_d(0,J_{kk}^2\bPhi^{-1})$ independently over $k$, which is the prior of Corollary~\ref{cor:special}(a) in each regression. And $\bDelta=0$ if and only if $\bdelta_k^\circ=0$ for every $k$. Likelihood, prior and hypotheses therefore factorize, and
\begin{equation}\label{eq:LTproduct}
E_G^{\mathrm{LT}}=\prod_{k=1}^qE_G^{(k)},
\end{equation}
where $E_G^{(k)}$ is the $e$-process of \citet{lindon2026anytime}, that is \eqref{eq:E} with $q=1$ as in Corollary~\ref{cor:special}(a), for the regression of the $k$th column of $\bY_G$ on the nuisance design $[\bX_G,\bY_{G,<k}]$ and the tested design $\bZ_G$, with precision $\bPhi$. Since $\bY_{G,<k}$ is independent of the errors of the $k$th regression, Theorem~\ref{thm:main}(ii) applied conditionally on $[\bX_G,\bY_{G,<k}]$ gives $\E_P[E_G^{(k)}\mid\bY_{G,<k}]=1$, and induction over $k$ gives $\E_P[E_G^{\mathrm{LT}}]=1$ a second time. The triangular statistic is thus the scalar $e$-process of \citet{lindon2026anytime} applied $q$ times, once to each step-down regression, and multiplied. The $k$th factor tests whether $\bz_g$ predicts the $k$th outcome after the first $k-1$ outcomes have been partialled out, so \eqref{eq:LTproduct} is the step-down procedure of \citet[Section~2.3]{roy1958stepdown} for the multivariate linear hypothesis, whose reasoning originates in the independence tests of \citet{roy1958tests} and whose one-sample form is in \citet[Section~2]{subbaiah1978comparison}, with the $F$-test at each step replaced by the $e$-process of \citet{lindon2026anytime}; the coefficients $\bdelta_k^\circ$ are the step-down parameters of his equation~(14); Theorem~1 of \citet{subbaiah1978comparison} identifies the step-down statistics of the one-sample problem as maximal invariants under the group of nonsingular lower-triangular matrices, the counterpart of $\mathsf I_{\mathrm{LT}}$ with the signs removed. Like the step-down procedure \citep[Section~1]{roy1958stepdown}, \eqref{eq:triangular} depends on the ordering of the coordinates. Every ordering, and every orthogonal rotation of the coordinates, gives an $e$-variable with the same expected log-growth, because under both laws the law of $\bU\bSigma^{-1/2}$ is invariant under right multiplication by orthogonal matrices, permutations of its columns included, whereas $E_G$ is invariant under all of them.

To compare growth, let $\mathsf I_{\mathrm{LT}}$ denote that finer invariant and $\mathsf I_{\mathrm{GL}}=\operatorname{col}(\bU)$ the full-group invariant. Since $\mathsf I_{\mathrm{GL}}$ is a function of $\mathsf I_{\mathrm{LT}}$, the chain rule for relative entropy gives
\[
\E_Q\log E_G^{\mathrm{LT}}-\E_Q\log E_G
=\E_{Q^{\mathsf I_{\mathrm{GL}}}}\!\left[\mathrm{KL}\big(Q^{\mathsf I_{\mathrm{LT}}\mid\mathsf I_{\mathrm{GL}}},P^{\mathsf I_{\mathrm{LT}}\mid\mathsf I_{\mathrm{GL}}}\big)\right]\geq0.
\]
The inequality is strict exactly when the finer likelihood ratio is not measurable with respect to $\mathsf I_{\mathrm{GL}}$. For $q\geq2$ and nonscalar $\bK_G$, choose two columns of $\bU$ along eigenvectors of $\bK_G$ with distinct eigenvalues and complete them with further eigenvectors of $\bK_G$ to a matrix of rank $q$. Interchanging those columns preserves $\mathsf I_{\mathrm{GL}}$ and $E_G$ but changes \eqref{eq:triangular}; continuity and positive Gaussian densities make the inequality strict. Both growth expectations are constant across $(\bB,\bSigma)$ by triangular-group invariance and transitivity, so this is also a strict improvement in worst-case growth. If $\bK_G=c\bI_N$, both invariant likelihood ratios equal one and no strict improvement occurs. Under the fixed-alternative conditions of Proposition~\ref{prop:growth}, the Cholesky factors divided by $\sqrt G$ converge to nonsingular limits, so $\log E_G^{\mathrm{LT}}-\log E_G=O_{a.s.}(1)$.
\end{enumerate}

This argument establishes validity of \eqref{eq:triangular} and failure of unrestricted GROW optimality for the full-group statistic in the stated nontrivial cases.

These facts are the reasons for working with $E_G$ in the body of the paper. First, $E_G$ is invariant under the full group, so the evidence does not depend on the ordering or the coordinate system of the outcomes, and by part~(a) it is growth-optimal among all $e$-variables with that property; a triangular statistic is one member of a family indexed by the ordering and rotation of the coordinates, all with the same expected log-growth, and nothing in the problem distinguishes one member. Second, the closed forms on which Sections~\ref{sec:clustered}--\ref{sec:rct} rest are those of $E_G$: the confidence ellipsoid \eqref{eq:ellipsoid}, the simultaneous confidence sequences of Corollary~\ref{cor:rct}(ii), and the identity of Proposition~\ref{prop:common} between the CR1 Wald statistic and Hotelling's $T^2$, through which $\ECR_G$ and Theorem~\ref{thm:asymptotic} extend the exact statistic to clustered designs. For $q\geq2$ the triangular statistic is not determined by the Wald statistic, since the factor in \eqref{eq:triangular} changes under a permutation of the coordinates that leaves $T^2$ fixed; Proposition~\ref{prop:common} therefore identifies $\ECR_G$ with $E_G$ but not with $E_G^{\mathrm{LT}}$. The triangular statistic requires the contrast vector and its full covariance matrix, beyond the scalar Wald statistic: under a common design the two sum-of-products matrices are, up to the common factor $G$, $(G-1)\bR\hat \bV_G\bR'+\bv \bv'$ and $(G-1)\bR\hat \bV_G\bR'+a_G\bv \bv'$ with $\bv=\bR\hat\bgamma_G-\btheta_0$ and $a_G=\phi/(\phi+G)$, so the Cholesky-pivot ratios in \eqref{eq:triangular} are functions of $\bv$ and $\bR\hat \bV_G\bR'$. Its robust asymptotic justification is a separate question. Although \eqref{eq:LTproduct} suggests the product of the $q$ cluster-robust $e$-processes $\ECR_G$ of the step-down regressions, Theorem~\ref{thm:asymptotic} covers one regression at a time and no joint result is proved here (compare Remark~\ref{rem:bonferroni}). Third, the gain is bounded: $\log E_G^{\mathrm{LT}}-\log E_G=O_{a.s.}(1)$ under the conditions of Proposition~\ref{prop:growth}, so both statistics have the growth rate of that proposition and the same power-one conclusion.

\subsection{Invariantly sufficient statistic; matrix angular central Gaussian form}
\label{app:macg}
The induced action on the sufficient statistic is $(\hat\bB,\hDelta,\nu\hSigma)\mapsto(\hat\bB\bC'+\bA,\ \hDelta\bC',\ \nu\bC\hSigma\bC')$, and $\bT_G=\hDelta_G\hSigma_G^{-1}\hDelta_G'$ is a maximal invariant for it: if two values of $\hDelta\hSigma^{-1/2}$ have the same Gram matrix they differ by a $q\times q$ orthogonal matrix $\bm R$, and $\bC=\hSigma_2^{1/2}\bm R'\hSigma_1^{-1/2}$ maps one statistic to the other. Thus $\bT_G$ is an invariantly sufficient statistic in the sense of \citet{hall1965relationship}; their later correction \citep{hall1995correction} requires domination by a member of the family, which holds here because the nonsingular Gaussian laws are mutually absolutely continuous. The density interpretation requires care. At the representative $\bSigma=\bI_q$, put $\bF=\hDelta_G(\nu_G\hSigma_G)^{-1/2}$ with the symmetric square root. Independently, $\nu_G\hSigma_G\sim\mathrm{Wishart}_q(\nu_G,\bI_q)$ and $\hDelta_G\sim\MN(0,\bA_j,\bI_q)$, where
\[
\bA_0=(\tZ_G'\tZ_G)^{-1},\qquad \bA_1=\bPhi^{-1}+\bA_0.
\]
Changing variables $\hDelta_G=\bF(\nu_G\hSigma_G)^{1/2}$ and integrating the Wishart matrix gives the matrix-$t$ density
\[
f_{\bA_j}(\bF)=c_{\nu_G,d,q}|\bA_j|^{-q/2}
\left|\bI_d+\bA_j^{-1}\bF \bF'\right|^{-(\nu_G+d)/2},
\]
with a common normalizing constant; see \citet[Theorem~3.1 and Corollary~3.1]{dickey1967matricvariate}. Its ratio is \eqref{eq:Eols}, since $\bF \bF'=\bT_G/\nu_G$. Under arbitrary $\bSigma$, the law of $\bF \bF'$ is unchanged by whitening, but the law of the oriented $\bF$ need not be. Consequently it is the ratio on the Gram-matrix invariant, or equivalently on the orientation after whitening by the true $\bSigma$ as in Appendix~\ref{app:evalue}, rather than a nuisance-free distribution of the unwhitened oriented matrix, that justifies the result.

On the full sample space, the maximal invariant is the column space of $\bU$, a point of the Grassmannian of $q$-planes in $\R^{G-p}$, uniformly distributed under the null. The proof of Theorem~\ref{thm:main}(ii) in Appendix~\ref{app:evalue} whitens the columns and uses the matrix angular central Gaussian density of the orientation under the alternative; that density is invariant under right orthogonal rotations, so it descends to the Grassmannian and equals \eqref{eq:hK}. Under the alternative with nonidentity column covariance, the unwhitened polar orientation $\bU(\bU'\bU)^{-1/2}$ need not have the same Stiefel density; under the null it is uniform for every $\bSigma$. The subspace law is the nuisance-free object. For $q=1$ the expression is the angular central Gaussian density studied by \citet{tyler1987statistical} on the sphere at $\bO_G\bY_G/\|\bO_G\bY_G\|_2$. This explains the closed form for the Gaussian mixture, whereas point alternatives for the scalar-outcome unit vector lead to the hypergeometric expressions discussed in \citet{lindon2026anytime}.

\end{document}